\documentclass[11pt]{article}
\usepackage{jheppub}

\usepackage{epsfig}
\usepackage{amssymb}
\usepackage{amsmath}
\usepackage{multirow}

\usepackage{ulem}

\usepackage{graphicx, subfigure, array, placeins, float}
\usepackage{epstopdf}
\usepackage{extarrows}

\usepackage{dsfont,bbm}
\usepackage{tabularx}

\usepackage{stackengine}

\newcommand{\pvint}{=\!\!\!\!\!\!\!\int}

\makeatletter \@addtoreset{equation}{section} \makeatother

\newcommand{\be}{\begin{equation}}
\newcommand{\ee}{\end{equation}}
\newcommand{\bea}{\begin{eqnarray}}
\newcommand{\eea}{\end{eqnarray}}

\allowdisplaybreaks[4]

\title{Two-Meson Form Factors in Unitarized \\ Chiral Perturbation Theory}

\author[a]{Yu-Ji Shi,}
\emailAdd{shiyuji92@126.com}
\author[a]{Chien-Yeah Seng,}
\emailAdd{cseng@hiskp.uni-bonn.de}
\author[b,c]{Feng-Kun Guo,}
\emailAdd{fkguo@itp.ac.cn}
\author[a]{Bastian Kubis,}
\emailAdd{kubis@hiskp.uni-bonn.de}
\author[a,d,e]{Ulf-G. Mei{\ss}ner}
\emailAdd{meissner@hiskp.uni-bonn.de}
\author[f]{\\ and Wei Wang}
\emailAdd{wei.wang@sjtu.edu.cn}
\affiliation[a]{Helmholtz-Institut f\"ur Strahlen- und Kernphysik and Bethe Center for Theoretical Physics, Universit\"at Bonn, D-53115 Bonn, Germany}
\affiliation[b]{CAS Key Laboratory of Theoretical Physics, Institute of Theoretical Physics, Chinese Academy of Sciences,
Beijing 100190, China}
\affiliation[c]{School of Physical Sciences, University of Chinese Academy of Sciences, Beijing 100049, China}
\affiliation[d]{Institute for Advanced Simulation, Institut f{\"u}r Kernphysik and J\"ulich Center for Hadron Physics,
Forschungszentrum J{\"u}lich, D-52425 J{\"u}lich, Germany}
\affiliation[e]{Tbilisi State University, 0186 Tbilisi, Georgia}
\affiliation[f]{INPAC, Shanghai Key Laboratory for Particle Physics and Cosmology, MOE Key Lab for Particle  Astrophysics and
Cosmology, School of Physics and Astronomy, Shanghai Jiao Tong University, Shanghai, 200240, China }

\abstract{
We present a comprehensive analysis of  form factors for two light pseudoscalar mesons induced by  scalar,
vector, and tensor  quark operators. The theoretical  framework is based on a combination
of unitarized chiral perturbation theory and dispersion relations. The low-energy constants in chiral
perturbation theory are fixed by a global fit to the available data of the two-meson scattering phase shifts.
Each form factor derived from
unitarized chiral perturbation theory is improved by iteratively applying a  dispersion relation.
This study updates the  existing results  in the literature and explores those that
have not been systematically studied previously, in particular the two-meson tensor form factors
within unitarized chiral perturbation theory. We also discuss the applications of these form factors as
mandatory inputs for  low-energy phenomena, such as the semi-leptonic decays $B_s\to \pi^+\pi^-\ell^+\ell^-$
and the $\tau$ lepton decay $\tau\rightarrow\pi^{-}\pi^{0}\nu_{\tau}$, in searches for physics
beyond the Standard Model. 
}

\stackMath
\begin{document}

\maketitle

\setcounter{footnote}{0}

\section{Introduction}

The study of multi-meson systems is an interesting problem as they are universal  in
various physical processes. 
An example of this is the  $B \to K^{*}(\to K\pi)  \mu^+ \mu^-$ decay that is induced by the flavor-changing
neutral current. Such a process is highly suppressed in the  Standard Model (SM), and thus sensitive to physics
beyond the Standard Model (BSM).  As a result it   offers a large number of observables
to test the SM  ranging from differential decay widths and polarizations to  a full  analysis of angular
distributions of the final-state particles. Recent experimental studies have led to some hints for
moderate deviations from the SM~\cite{Aaij:2013qta,Aaij:2017vbb,Abdesselam:2019wac}. Note that this process is in
fact a four-body decay since the $K^{*}$ meson is reconstructed from the $K\pi$ final state. Therefore,
to handle such decay processes, the narrow-width approximation is usually assumed in phenomenological
studies. However, this assumption may lead to sizable
systematic uncertainties as it captures only part of the $K\pi$ final-state interactions.  

To solve this problem, one should use a complete factorization analysis that can systematically
separate the low-energy final-state interaction from the short-ranged weak transition. In
semi-leptonic processes like  $B \to M_1 M_2  \ell^+ \ell^-$, the final two-meson state decouples from the
leptons to a good approximation. Thus, it is guaranteed by the Watson--Madigal
theorem~\cite{Watson:1952ji,Migdal:1956tc}
that the phase of the hadronic transition matrix element is equal to the phase of $M_1 M_2$ elastic
scattering below the first inelastic threshold. More explicitly, as
pointed out, e.g., in Ref.~\cite{Meissner:2013hya}, the decay matrix element is proportional to
a two-meson form factor,
\begin{eqnarray}
\langle M_1M_2 |\bar q(0) \Gamma q(0)|0\rangle~, \label{eq:current-matrix-element}
\end{eqnarray} 
where the Dirac matrices $\Gamma=1, \gamma_\mu, \sigma_{\mu\nu}$ correspond to the scalar, vector, and antisymmetric
tensor currents, respectively. The choice depends on whether $M_1$ and $ M_2$ are in relatively $S$-wave or $P$-wave.  The relation between vacuum-to-two-meson form factors as in Eq.~\eqref{eq:current-matrix-element} and those appearing in heavy-meson decays can occasionally be sharpened based on chiral-symmetry relations~\cite{Albaladejo:2016mad}.
 
One of the standard approaches to calculating these two-meson form factors is using chiral perturbation
theory (ChPT), which is a low-energy effective theory of quantum chromodynamics (QCD) that describes the
interaction among light
mesons and baryons. The next-to-leading-order (NLO) ChPT calculation for the $\pi\pi$ scalar form factor
was firstly given in Ref.~{\cite{Gasser:1984gg}}. Its two-loop representation and some unitarization schemes
were discussed in Ref.~\cite{Gasser:1990bv}.
After that, Refs.~{\cite{Meissner:2000bc,Lahde:2006wr}}
performed more complete studies of the scalar form factors in unitarized chiral perturbation theory (uChPT),
where the results of the NLO ChPT were extended to a higher energy scale around $1\ \text{GeV}$, which was realized
by involving the channel coupling between the $\pi\pi$ and the $K\bar K$ systems to impose unitarity
constraints on the form factors. 
The reconstruction of the scalar $\pi\pi$ and $K\bar K$ form factors based on a Muskhelishvili--Omn\`es representation, relying on phenomenological phase shift input, has by now a long history~\cite{Donoghue:1990xh,Moussallam:1999aq,DescotesGenon:2000ct,Hoferichter:2012wf,Daub:2015xja}, which includes several dedicated applications in the context of BSM physics searches~\cite{Daub:2012mu,Celis:2013xja,Monin:2018lee,Winkler:2018qyg}.  Extensions beyond $1\ \text{GeV}$ with a new formalism including further inelastic channels were discussed in Ref.~\cite{Ropertz:2018stk}.
Studies of the $\pi\eta$ isovector scalar form factor (also coupled to $K\bar K$) are much rarer~\cite{Albaladejo:2015aca,Albaladejo:2016mad,Lu:2020qeo}, largely due to far less experimental information on $\pi\eta$ scattering.
The $K\pi$, $K\eta$ scalar form factors up to $2\ \text{GeV}$ were given
in Ref.~\cite{Jamin:2001zq,Jamin:2000wn,Jamin:2006tj,Bernard:2009ds} with a coupled-channel dispersive analysis.
The two-meson
vector form factors for $K\pi$ were first derived in ChPT~{\cite{Gasser:1983yg}}, while it was mostly obtained
by fitting to the data of semi-leptonic $\tau$ decays in Refs.~\cite{Boito:2011tp,Boito:2010hq,Boito:2010ak}.  There is a 
number of works for the pion vector form factor based on the Omn\`es dispersive
representation~{\cite{Guerrero:1997ku,DeTroconiz:2001rip,Pich:2001pj,deTroconiz:2004yzs,Guo:2008nc,Hanhart:2012wi,Dumm:2013zh,Colangelo:2018mtw,Gonzalez-Solis:2019iod,Colangelo:2020lcg}}, ChPT calculations~\cite{Gasser:1990bv,Bijnens:1998fm,Bijnens:2002hp,Oller:2000ug}, a model based on analyticity~\cite{Bruch:2004py}, and in the large-$N_c$ limit~\cite{Pich:2010sm}.
Throughout the present study, we will work in the isospin limit, although also isospin-violating scalar and vector form factors have been studied in uChPT or using dispersive methods, such as in the context of $a_0$--$f_0$ mixing~\cite{Hanhart:2007bd} or for studies of second-class currents in $\tau$-decays~\cite{Descotes-Genon:2014tla,Escribano:2016ntp}.

Unlike the scalar and vector cases, naturally the coupling with the antisymmetric tensor current does not
exist in the SM and conventional ChPT (the energy-momentum tensor  is symmetric, and has been built into ChPT up
to NLO~\cite{Donoghue:1991qv,Kubis:1999db}).  However, in terms of the research for BSM physics, for
example in the Standard Model Effective Field Theory (SMEFT), a number of high-dimensional operators including
the tensor current are necessary.  Besides the conventional ChPT Lagrangian, additional terms with an antisymmetric
(antisymmetric is implicit in the following discussion of the tensor part) tensor source was first given in Ref.~\cite{Cata:2007ns},
which is crucial to calculating the tensor form factors.  Recently, dispersive analyses of tensor form factors in specific channels ($\pi\pi$~\cite{Miranda:2018cpf}, $\pi K$~\cite{Cirigliano:2017tqn}, and for the nucleon~\cite{Hoferichter:2018zwu}) have been carried out.

In this work, we will perform a study of all three kinds of two-meson form factors
based on uChPT and dispersion relations. 
Section~\ref{sec:2} gives a brief introduction to ChPT and its unitarization, where we will discuss how
unitarized meson--meson scattering amplitudes can be obtained by the inverse amplitude method (IAM). 
The coupled-channel IAM~\cite{GomezNicola:2001as} is modified by removing the imaginary parts of the $t$-
and $u$-channel loops in order to restore unitarity in coupled-channel systems, which is otherwise violated
in particular around the $\rho$-meson region in the isospin-1 sector.
In Sect.~\ref{sec:3}, we will calculate the two-meson scalar form factors, which are then unitarized
by the IAM. There,  unphysical sub-threshold singularities, related to the so-called Adler zeros, will show up. 
To eliminate these defects, an iteration procedure based on dispersion relations is performed for each form
factor, so that the improved form factors behave well in a wide energy range $0\ldots1.2\ \text{GeV}$. In
Sects.~\ref{sec:4} and~\ref{sec:5}, we will apply the same procedure
to the calculation of unitarized vector and tensor form factors, respectively. 
Some of the form factors obtained in this work are compared  with those that have appeared
in earlier works.  For each kind of form factor, we will also briefly introduce
their applications in corresponding phenomenological studies. This includes the application of the
$\pi\pi$ form factor for the $S$-wave-dominated decay $B_s\to f_0(980)(\to\pi^+\pi^-) \mu^+ \mu^-$,
the application of two-meson vector form factors
in the two-body hadronic decays of a charged lepton $l\to\phi\phi'\nu$, where $\phi$ and $\phi'$ denote light pseudoscalar mesons, and the application of
two-meson tensor form factors for the BSM effects in two-body $\tau$ decays $\tau\to\phi\phi'\nu_{\tau}$.
Finally, Sect.~\ref{sec:conclusions}
contains a summary. Various technicalities are relegated to the appendices.

\section{Framework}
\label{sec:2}

\subsection{Chiral Perturbation Theory and Its Unitarization } 

ChPT~\cite{Weinberg:1978kz,Gasser:1983yg,Gasser:1984gg} provides a systematic framework to investigate the
strong interaction at low energies.
It is based on the spontaneous breaking of the global $G=\text{SU}(3)_{L}\times \text{SU}(3)_{R}$ symmetry of the QCD Lagrangian,
in the limit of  vanishing $u$, $d$, and $s$ quark masses, down to the global $H=\text{SU}(3)_{V}$  symmetry of the QCD vacuum.
This spontaneous symmetry breaking gives rise to eight pseudo-Nambu--Goldstone bosons (pNGBs), which are
the relevant degrees of freedom at low energies. In fact, it can be proven that the eight pNGBs construct a
space which is isomorphic to the quotient space  $G/H=\text{SU}(3)_{L}\times \text{SU}(3)_{R}/\text{SU}(3)_{V}$~\cite{Coleman:1969sm}.
This isomorphism enables one to parametrize any of these quotient elements $U \in G/H$ by the eight pNGBs as
\begin{eqnarray}
 U = {\rm exp} \left[\frac{i\phi}{F_0}\right],
\end{eqnarray}
where $F_0$ is the pion decay constant in the chiral limit, and
\begin{eqnarray}
 \phi=\phi^a\lambda^a = \left( \begin{array}{ccc}
    \pi^{0} +\eta/\sqrt3 & \sqrt{2}\pi^{+} & \sqrt{2}K^{+} \\ 
    \sqrt{2}\pi^{-} & -\pi^{0} +\eta/\sqrt3 & \sqrt{2}K^{0} \\ 
    \sqrt{2}K^{-} & \sqrt{2}\bar K^{0} & -2
 \eta/\sqrt3 \\ 
  \end{array}\right)
\end{eqnarray}
contains the pNGB octet. Here, exact isospin symmetry is assumed, which turns off the $\pi^0$-$\eta$
mixing for simplicity. We use the convention that under $\text{SU}(3)_{L}\times \text{SU}(3)_{R}$ transformations $U$ behaves as $U
\to R U L^{\dagger}$, with $R \in \text{SU}(3)_{R}$ and $L \in \text{SU}(3)_{L}$.
With $U$ as the building block, the leading-order (LO) effective  Lagrangian of ChPT is constructed  as
\begin{eqnarray}
  \mathcal{L}^{(2)} = \frac{F_0^2}{4} \langle D_{\mu}U D^{\mu}U^{\dagger}\rangle
  +  \frac{F_0^2}{4} \langle\chi U^{\dagger} + \chi^{\dagger} U\rangle~,\label{eq:Op2}
\end{eqnarray}
where $\langle \ldots \rangle$ denotes the trace in SU(3) flavor space, $\chi = 2B_{0} (M+s)$, with $M$
the quark mass matrix, $D_{\mu} U \equiv \partial_{\mu} U - i r_{\mu} U+iU l_{\mu}$, and $s, l_{\mu}, r_{\mu}$ are
the scalar, the left-handed, and the right-handed external sources. The parameter $B_{0}$ is proportional to
the QCD quark condensate, $3 F_0^2 B_0 =-\langle {\bar u} u+\bar d d+\bar s s\rangle$.

Applying Eq.~\eqref{eq:Op2} at one-loop produces ultraviolet (UV) divergences that can be regulated
using dimensional regularization and then reabsorbed into the low-energy constants (LECs) in the
next-to-leading-order (NLO) Lagrangian~\cite{Gasser:1983yg,Gasser:1984gg}:
\begin{eqnarray}
  {\cal L}^{(4)} &=& L_{1} \langle D_{\mu}U D^{\mu}U^{\dagger}\rangle^{2}
  + L_{2}   \langle D_{\mu}U \left(D_{\nu}U\right)^{\dagger}\rangle\langle D^{\mu}U (D^{\nu}U)^{\dagger} \rangle 
  +  L_{3}  \langle D_{\mu}U (D^{{\mu}}U)^{\dagger} D_{{\nu}}U (D^{\nu}U)^{\dagger}\rangle \nonumber\\&&
  +  L_{4}  \langle D_{\mu}U (D^{{\mu}}U)^{\dagger}\rangle \langle\chi^{\dagger}U + U\chi^{\dagger}\rangle 
  +  L_{5}  \langle D_{\mu}U D^{{\mu}}U^{\dagger} (\chi^{\dagger}U + U\chi^{\dagger})\rangle \nonumber\\&&
  +  L_{6}  \langle\chi^{\dagger}U + U\chi^{\dagger}\rangle^{2}  
  +  L_{7}  \langle\chi^{\dagger}U - U\chi^{\dagger}\rangle^{2}  + L_{8} \langle\chi U^{\dagger}\chi U^{\dagger}+ U\chi^{\dagger}U\chi^{\dagger} \rangle \nonumber\\
  && -i L_{9} \langle R_{\mu\nu} D^{\mu}U (D^{\nu} U)^{\dagger} + L_{\mu\nu}(D^{\mu}U)^{\dagger} D^{\nu} U\rangle+ L_{10} \langle L_{\mu\nu}U R^{\mu\nu}U^{\dagger}\rangle~,
\end{eqnarray}
where $L_{\mu\nu}$ and $R_{\mu\nu}$ are field-strength tensors of the external sources
\begin{eqnarray}
 L_{\mu\nu} = \partial_{\mu} l_{\nu }-\partial_{\nu} l_{\mu} - i[l_{\mu}, l_{\nu}]~,~~~~
 R_{\mu\nu} = \partial_{\mu} r_{\nu }-\partial_{\nu} r_{\mu} - i[r_{\mu}, r_{\nu}]~. 
\end{eqnarray}

The UV-finite, scale-dependent renormalized LECs $\{L_i^r\}$ are defined as $L_{i}^{r} =  L_{i} -\Gamma_{i} \lambda$, 
with the UV-divergent parts proportional to
\begin{eqnarray}
\lambda= \frac{1}{32\pi^{2}} \left[\frac{2}{d-4} -\ln (4\pi)+\gamma-1\right],
\end{eqnarray}
and  the nonzero values for their coefficients $\Gamma_{i}$ relevant to this work are
\begin{eqnarray}
 \Gamma_{1} =\frac{3}{32},\;\; \Gamma_{2} = \frac{3}{16},\;\; \Gamma_{4} = \frac{1}{8},\;\; \Gamma_{5} =\frac{3}{8},\;\; 
\Gamma_{6} = \frac{11}{144},\;\; \Gamma_{8} = \frac{5}{48}, \;\;
 \Gamma_{9} =\frac{1}{4},\;\; \Gamma_{10} = -\frac{1}{4}.
\end{eqnarray}
The scale dependence of these LECs is given by 
\begin{eqnarray}
 L_i^{r}(\mu_2) &=& L_i^{r}(\mu_1) -\frac{\Gamma_i}{32\pi^2}\ln\left(\frac{\mu_2^2}{\mu_1^2}\right). 
\end{eqnarray}
Some details of the loop functions occurring in the one-loop ChPT calculations can be found in
Appendix~\ref{sec:loopfunc}.
Table~\ref{tab:NLOLEC} collects  the numerical results for the $L_i^r$ at the scale $\mu=M_{\rho}$
that were obtained previously. The first column corresponds to the analysis up to ${\cal O}(p^4)$ in
ChPT~\cite{Gasser:1983yg,Bijnens:1994ie}, and the second refers to the fit with ${\cal O}(p^6)$
corrections~\cite{Bijnens:2014lea}.  The third column corresponds to the previous  fit of
meson--meson scattering phase shifts and inelasticities in the coupled-channel
IAM~\cite{GomezNicola:2001as}, which we will discuss further below. 
\begin{table}
\begin{center}
  \caption{Low-energy constants $L_i^r$ (in units of $10^{-3}$) at $\mu=M_{\rho}$. For the description of
    the fits given in the last two columns, we refer to Sect.~\ref{sec:GloabFit}. }
\label{tab:NLOLEC}
\setlength{\tabcolsep}{1.5mm}{
\begin{tabular} {lcccccccc}
\hline
\hline
&$L_i^r$	 &  ${\cal O}(p^4)$ ChPT~\cite{Gasser:1983yg,Bijnens:1994ie}    & ${\cal O}(p^6)$ ChPT~\cite{Bijnens:2014lea}  & uChPT~\cite{GomezNicola:2001as} & uChPT Fit~1  & uChPT Fit~2   \\\hline
&$L_{1}^{r}$   & $0.4\pm0.3$   & $1.11\pm0.10$  & $0.56\pm0.10$ & $0.52\pm0.01$ & $0.55\pm0.02$  \\ 
&   $L_{2}^{r}$  & $1.35\pm0.3$   & $1.05\pm0.17$&  $1.21\pm0.10$ &  $1.08\pm0.01$ & $1.07\pm0.03$ \\
&  $L_{3}$  & $-3.5\pm1.1$   & $-3.82\pm0.30$&  $-2.79\pm0.24$ & $-2.72\pm0.01$& $-2.76\pm0.03$\\
& $L_{4}^{r}$   & $-0.3\pm0.5$  & $1.87\pm0.53$& $-0.36\pm0.17$ & $-0.27\pm0.01$ & $-0.20\pm0.02$ \\
& $L_{5}^{r}$ & $1.4\pm0.5$ & $1.22\pm0.06$& $1.4\pm0.5$  & $1.36\pm0.19$  & $1.13\pm0.61$  \\
& $L_{6}^{r}$  & $-0.2\pm0.3$  &$1.46\pm0.46$& $0.07\pm0.08$ & 0.07 (fixed)& 0.07 (fixed)\\
& $L_{7}$  & $-0.4\pm0.2$  &$-0.39\pm0.08$& $-0.44\pm0.15$ & $-0.60\pm0.04$ & $-0.52\pm0.12$ \\
& $L_{8}^{r}$  & $0.9\pm0.3$ & $0.65\pm0.07$& $0.78\pm0.18$ & $0.89\pm0.09$ & $0.68\pm0.29$ \\
& $L_{9}^{r}$  & $6.9\pm0.7$ &--& --&--& -- \\
& $L_{10}^{r}$  & $-5.5\pm0.7$ &--& --&--& -- \\ 
\hline
\end{tabular}}
\end{center}
\end{table}

The power counting of ChPT is organized according to the increasing  power of the ratio $p/\Lambda_\chi$ given in terms of
a typical small pNGB momentum $p$ of the order of the pNGB mass and the chiral symmetry breaking scale $\Lambda_\chi
\sim 4\pi F_\pi$~\cite{Manohar:1983md},
where $F_\pi\approx 92.1\ \text{MeV}$ is the physical pion decay constant. Therefore, the perturbative
expansion in ChPT is expected to break down when $p/\Lambda_\chi\sim 1$. Moreover, a perturbative expansion in
powers of momenta to any finite order cannot describe the physics of resonances, which are given by poles of the
$S$-matrix on unphysical Riemann sheets. Thus, the masses of the lowest resonances in each meson--meson
scattering channel limit the applicability region of ChPT in the corresponding sector.

Unitarization (or resummation) is a systematic prescription intended to extend the applicability of ChPT to
higher energies, say $1\ \text{GeV}$, by modifying the perturbative expression such that it satisfies the full instead
of only the perturbative unitarity requirement of quantum field theory. 
Since unitarity is nonperturbative in its nature, in this way the lowest meson resonances may also be described.
Note, however, that unitarization comes with a price: as there are various unitarization schemes, some scheme-dependence
is introduced. Also, crossing symmetry is often broken in such approaches.

Let us start with a simple example. In the case of $2 \to 2$ multi-channel
scattering with the momenta of initial and final particles as $p_1, p_2$ and $p_3, p_4$, respectively, we can define the
partial-wave amplitude $T_{J}(s)$ with total angular momentum $J$ from the full scattering amplitude
$T(s,\cos\theta)$ by
\begin{align}
T_{J}(s)=\frac{1}{2} \int_{-1}^{+1} P_{J}(\cos \theta) T(s, \cos\theta) d \cos \theta~,
\end{align}
where  $s=(p_1+p_2)^2=(p_3+p_4)^2$ is one of the usual Mandelstam variables, $\theta$ is the angle between ${\vec p}_3$ and
${\vec p}_1$ in the center-of-mass (c.m.)\ frame, and $P_{J}(\cos \theta)$ is the Legendre polynomial of order $J$. If
we consider only two-particle
intermediate states, then the partial-wave amplitudes should satisfy the following unitarity relation:   
\begin{align}
  \operatorname{Im}\left[T_{J}(s)\right]_{i j} =\sum_{k k^{\prime}}\left[T_{J}(s)\right]_{i k} \delta_{k k^{\prime}}
  \frac{\left|\vec{p}_{k}\right|}{8 \pi \sqrt{s}}\Theta(s-s_\mathrm{th}^i)\left[T_{J}(s)\right]_{k^{\prime} j}^{*},
  \label{OpticalRela}
\end{align}
where time reversal invariance is assumed.  The indices $i, j, k$, and $k^{\prime}$
denote different scattering channels, $\left|\vec{p}_{k}\right|$ is the modulus of the c.m.\ 3-momentum in the $k$th channel,
and $s_\mathrm{th}^i=(M_{a_i}+M_{b_i})^2$ is the production threshold of the $i$th channel particles.
Equation~(\ref{OpticalRela}) can be written in matrix form:
\begin{align}
  \operatorname{Im}\left[T_{J}(s)\right]=T_{J}(s) \Sigma(s) T_{J}^{*}(s)~,~~~\left[\Sigma(s)\right]_{i j}
  \equiv \delta_{i j} \frac{\left|\vec{p}_{i}\right|}{8 \pi \sqrt{s}}\Theta(s-s_\mathrm{th}^i)~.
  \label{MatrOpticalRela}
\end{align}
Multiplying $T_J^{-1}$ and $T_J^{*-1}$ on both sides leads to
\begin{align}
  \frac1{2i}\left[T_J^{*-1}(s) - T_J^{-1}(s) \right] =
  -\operatorname{Im}T_{J}^{-1}(s) =\Sigma(s). 
  \label{MatrOpticalRela1}
\end{align}

In a fixed-order ChPT calculation, the relation~\eqref{OpticalRela} is only satisfied
perturbatively. For instance, substituting the ChPT expression of $T_J(s)$ at $\mathcal{O}(p^4)$
to both sides will result in the breaking of the equality at the $\mathcal{O}(p^6)$ level. There are many
ways to recover exact unitarity, and in this work we mainly adopt
the multi-channel IAM outlined in Ref.~\cite{GomezNicola:2001as}. The procedure
is as follows. First, the partial-wave amplitude $T_J$ can be written as
\begin{equation}
T_{J}=\left(\mathrm{Re}[T_{J}^{-1}]-i\Sigma\right)^{-1}.\label{eq:TJinverse}
\end{equation}
Since $\Sigma$ is a known matrix, we just need to calculate $\mathrm{Re}[T_{J}^{-1}]$,
i.e., the real part of the inverse matrix $T_{J}^{-1}$ in order to
obtain the full $T_{J}$. The IAM is a way to calculate $\mathrm{Re}[T_{J}^{-1}]$
approximately~\cite{Truong:1988zp,Dobado:1996ps,GomezNicola:2001as}.
The idea is to start with the perturbative expansion of $T_{J}$: $T_{J}  =  T_{J}^{(2)}+T_{J}^{(4)}+\ldots$,
where the superscripts $(2),(4)$ denote the order of the amplitude, which are ${\cal O}(p^2)$
and ${\cal O}(p^4)$, respectively. The corresponding inverse matrix can be perturbatively expanded as
\begin{eqnarray}
T_{J}^{-1} =  T_{J}^{(2)-1}\left(1-T_{J}^{(4)}T_{J}^{(2)-1}+\ldots\right).
\end{eqnarray}
Next, using the fact that $T_{J}^{(2)}$ is real (again assuming time reversal invariance), we have
\begin{equation}
\mathrm{Re}[T_{J}^{-1}]=T_{J}^{(2)-1}\left(1-\mathrm{Re}[T_{J}^{(4)}]T_{J}^{(2)-1}+\ldots\right).
\end{equation}
Since both $T_{J}^{(2)-1}$ and $\mathrm{Re}[T_{J}^{(4)}]$ are
calculable in ChPT, we have already achieved our goal.
However, we can further simplify the expression by plugging it into Eq.~(\ref{eq:TJinverse}),
which results in
\begin{eqnarray}
  T_{J} & = & \left[T_{J}^{(2)-1}\left(1-\mathrm{Re}[T_{J}^{(4)}]T_{J}^{(2)-1}+\ldots\right)-i\Sigma\right]^{-1}
  \nonumber \\
  & = & T_{J}^{(2)}\left(T_{J}^{(2)}-\mathrm{Re}[T_{J}^{(4)}]-iT_{J}^{(2)}\Sigma T_{J}^{(2)}+\ldots\right)^{-1}T_{J}^{(2)}
  \nonumber \\
  & = & T_{J}^{(2)}\left(T_{J}^{(2)}-\mathrm{Re}[T_{J}^{(4)}]-i\mathrm{Im}[T_{J}^{(4)}]+\ldots\right)^{-1}T_{J}^{(2)}
  \nonumber \\
 & \approx & T_{J}^{(2)}\left(T_{J}^{(2)}-T_{J}^{(4)}\right)^{-1}T_{J}^{(2)}~.\label{eq:TJIAM}
\end{eqnarray}
In the third line we have used the perturbative unitarity relation, which can be obtained by power
expansion of Eq.~(\ref{MatrOpticalRela}).
The last line is the central equation for  the IAM at NLO. It effectively implies that
we can use perturbation theory to calculate $T_{J}^{(2)}$ and $T_{J}^{(4)}$,
and resum their effects to all orders. An advantage of this method is that the
resummation is a simple matrix inversion, and does not involve
any extra integral equation.
\begin{figure}
\begin{center}
\includegraphics[width=1.0\columnwidth]{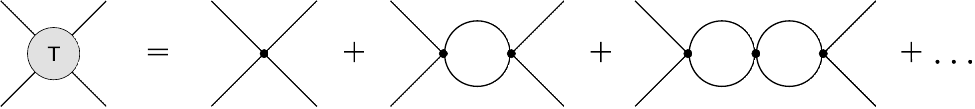} 
\caption{A Feynman diagram interpretation for the IAM method: 
  summing up all the $s$-channel ${\cal O}(p^{4})$ scattering amplitudes to all orders.    }
\label{fig:uchpt}
\end{center}
\end{figure} 
There is also a simple Feynman diagram interpretation of Eq.~(\ref{eq:TJIAM}).
As shown in Fig.~\ref{fig:uchpt}, we simply sum up all the $s$-channel ${\cal O}(p^{4})$ corrections to all orders:
\begin{eqnarray}
T & = & T^{(2)}+T^{(4)}+T^{(4)}(T^{(2)})^{-1}T^{(4)}+T^{(4)}(T^{(2)})^{-1}T^{(4)}(T^{(2)})^{-1}T^{(4)}+\ldots\nonumber \\
 & = & T^{(2)}+\left[1+T^{(4)}(T^{(2)})^{-1}+T^{(4)}(T^{(2)})^{-1}T^{(4)}(T^{(2)})^{-1}+\ldots\right]T^{(4)}\nonumber \\
 & = & T^{(2)}+\left(1-T^{(4)}(T^{(2)})^{-1}\right)^{-1}T^{(4)}\nonumber \\
 & = & \left(1-T^{(4)}(T^{(2)})^{-1}\right)^{-1}\left[\left(1-T^{(4)}(T^{(2)})^{-1}\right)T^{(2)}+T^{(4)}\right]\nonumber \\
 & = & T^{(2)}\left(T^{(2)}-T^{(4)}\right)^{-1}T^{(2)},
\end{eqnarray}
which gives exactly the IAM result. The explicit expressions for the scattering amplitudes classified by definite
isospin states can be found in Appendix~\ref{sec:ScatAmp}.

An obvious shortcoming of the IAM formula is that it leads to a peak when the determinant
$\mathrm{det}\left[T^{(2)}-T^{(4)}\right]$ approaches a minimum. This peak may be unphysical and, in terms
of dispersion relations, is due to the failure to incorporate the pole contributions from the so-called Adler
zero of the partial wave in the sub-threshold region. This problem can be satisfactorily resolved in the case of the single-channel IAM~\cite{GomezNicola:2007qj} but not for coupled channels, see Appendix~\ref{sec:Adler}
for a brief explanation of the procedure. In Sect.~\ref{disperImprov}, we will introduce an effective solution
based on dispersion relations for coupled-channel systems. 


\subsection{Unitarity}
\label{sec:unitarity}

\begin{figure}[tb]
  \centering
  \includegraphics[width=\textwidth]{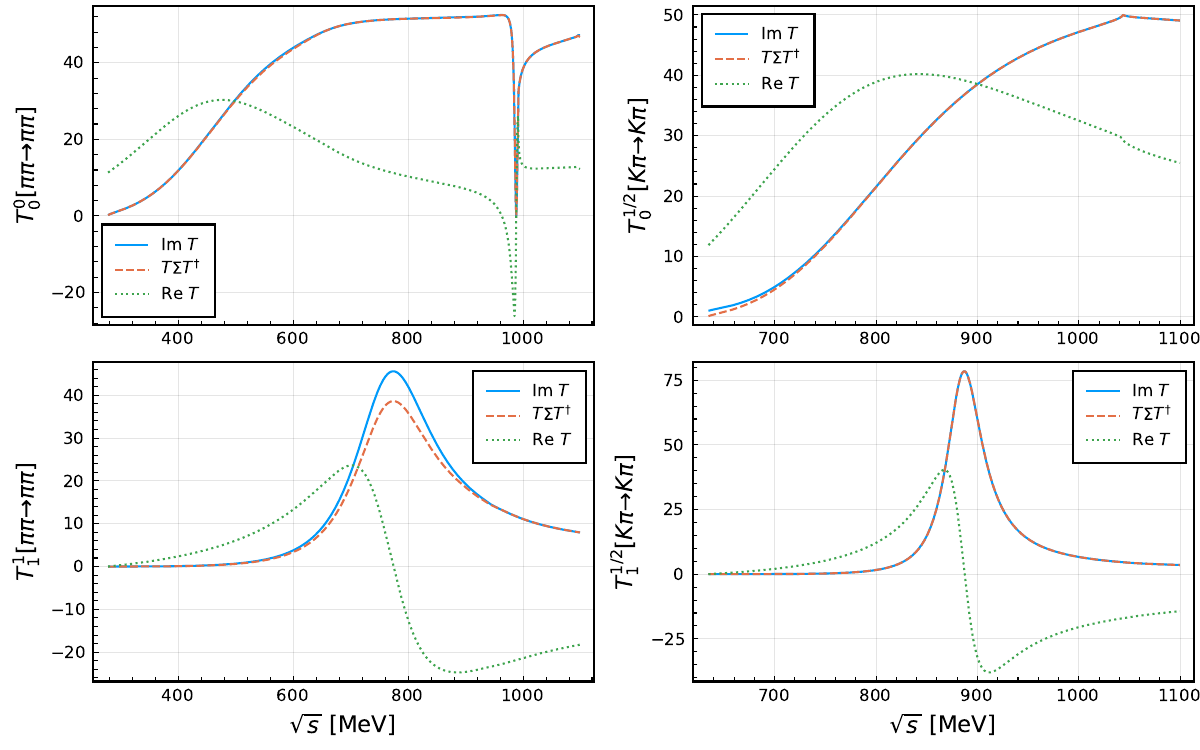}
  \caption{Check of the unitarity relations for several $T$-matrix elements in coupled-channel uChPT, with the LECs
    taking the central values given in Ref.~\cite{GomezNicola:2001as}. The indices of $T_J^I$ refer to the isospin
    $I$ and angular momentum $J$ partial wave.}
  \label{fig:uni_nomod}
\end{figure}

Although the uChPT was constructed to fulfill the unitarity relation $\mathrm{Im}T=T\Sigma T^\dag$,
the one-step IAM solution of the partial waves actually satisfies the exact relation only above the highest
threshold. In general,  the unitarity relation below the highest threshold 
is broken due to the mixing between the left-hand and right-hand cuts
of different matrix elements in $T$ during the process of matrix inversion. This phenomenon occurs since all the
particles in the initial and final states of the $T^{(4)}$-matrix are treated as on shell~\cite{Du:2018gyn}. 
Such a problem is well-known and also exists in other methods of
unitarization~\cite{Iagolnitzer:1973fq,Badalian:1981xj,Xiao:2001pt,Xiao:2001sw,GomezNicola:2001as}.
However, depending on the values of LECs, such unitarity violation is usually very small and would not cause
any real problem in practical applications of IAM results. With the LEC values reported in
Ref.~\cite{GomezNicola:2001as}, the only exception is the $I=1,J=1$ channel, see Fig.~\ref{fig:uni_nomod}.
The imaginary part of the partial-wave amplitude in this channel is peaked at $\sqrt{s}\approx 0.77\ \text{GeV}$ due
to the existence of the $\rho$-resonance, and it turns out that the IAM approach leads to a breaking of the
unitarity relation by as much as 20\% around the $\rho$-peak.

As discussed in Ref.~\cite{GomezNicola:2001as}, this problem can in principle be solved by adopting a multi-step
unitarization approach, namely to take
the dimension of the $T$-matrix as a function of $s$, which changes by one unit every time $s$ crosses a
threshold. By doing so one explicitly avoids the mixing of left-hand and right-hand cuts between different
matrix elements below the largest thresholds, and thus the unitarity relation is exactly satisfied
in all regions. One disadvantage of this approach is that one cannot study the scattering amplitudes
(and their associated form factors) below their respective production thresholds because their
corresponding matrix elements simply do not exist in the scattering matrix. It is therefore
highly desirable to search for a prescription that allows for a simultaneous study of form factors
below and above thresholds, and at the same time avoids the mixing of left-hand and right-hand cuts as much
as possible. In the following, we demonstrate a simple procedure that satisfies both requirements.

\begin{figure}[tb]
  \centering
  \includegraphics[width=\textwidth]{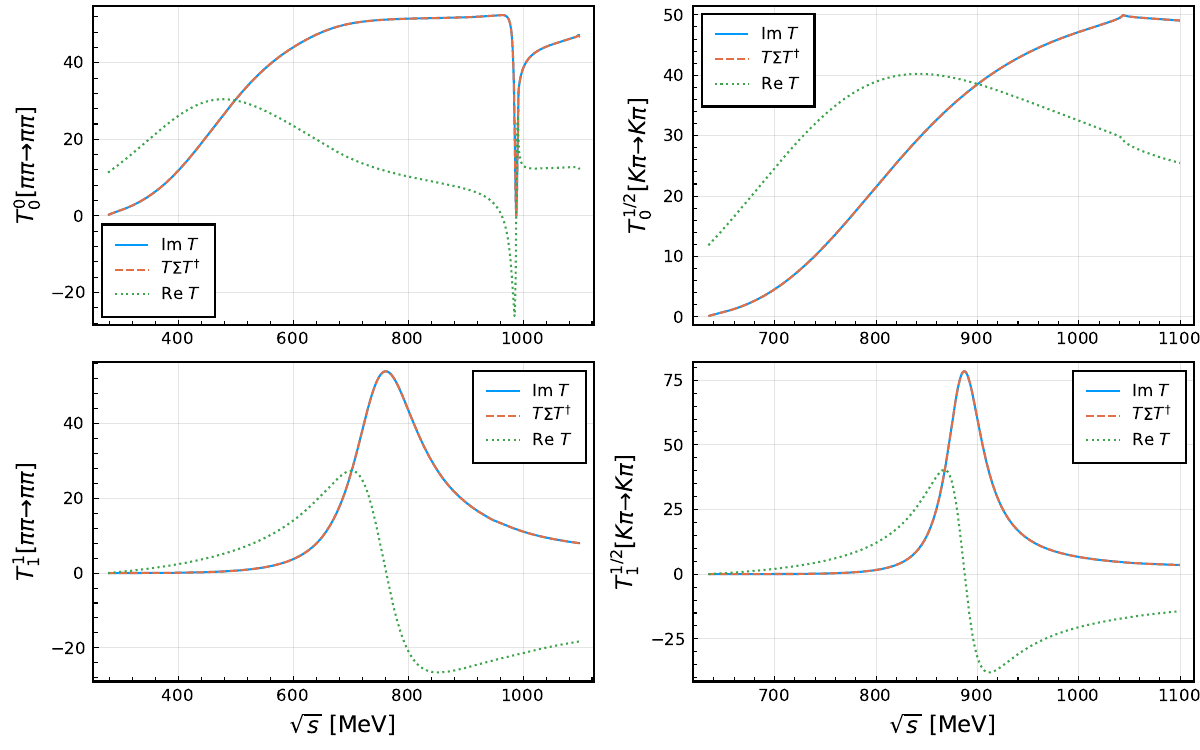}
  \caption{Unitarity relations are fulfilled exactly after removing the imaginary parts of the troublesome
    $t$- and $u$-channel loops. Here, the LECs take the central values given in Ref.~\cite{GomezNicola:2001as}.}
  \label{fig:uni_mod}
\end{figure}
Let us focus on the $(I,J)=(1,1)$ $\pi\pi$ scattering, which couples to the $K\bar K$ channel. At NLO ChPT, the $K\bar K\to K\bar K$ scattering receives contributions from $t$- and $u$-channel meson loops, where the
Mandelstam variables $t$ and $u$ are defined as $t=(p_1-p_3)^2$ and $u=(p_1-p_4)^2$. If the initial and final
kaons are on shell, the branch point for the $K\bar K\to K\bar K$ partial wave due to the $\pi\pi$ $t$-channel loop is at $s = 4(M_K^2 - M_\pi^2)$,
which is below the $K\bar K$ threshold, but above the $\pi\pi$ threshold. 
Similarly, the branch point of the $t$-channel $\pi\eta$ loop occurs at $s=4M_K^2-(M_\pi+M_\eta)^2$, again
above the $\pi\pi$ threshold.
If the kaons are off shell, such singularities would not be on the physical Riemann sheet of $\pi\pi$
scattering~\cite{Du:2018gyn}, and would not cause any problem. However, in the IAM treatment, the kaons
are on shell, leading to an overlap between such left-hand cuts with the $\pi\pi$ right-hand cut and
thus a violation of unitarity.
We find that if we remove the imaginary part of the troublesome $t$- and $u$-channel loops, unitarity can be
exactly maintained, see Fig.~\ref{fig:uni_mod}, where the curves for ${\rm Im} T$ and $T\Sigma T^{\dagger}$ coincide.
These loops include the $t$-channel $\pi\pi$ and $\pi\eta$ loops in $K\bar K\to K\bar K$ (both $I=0$ and $I=1$),
the $t$- and $u$-channel $\pi\pi$ loops in $\eta\eta\to\eta\eta$, the $t$- and $u$-channel $\pi K$ loops in
$K\bar K\to \eta\eta$, and the $t$-channel $\pi\pi$ and $u$-channel $\pi K$ loops in $K\eta\to K\eta$.

\subsection{Global Fit of LECs}
\label{sec:GloabFit}

\begin{figure}[tbhp]
  \centering
  \includegraphics[width=\textwidth]{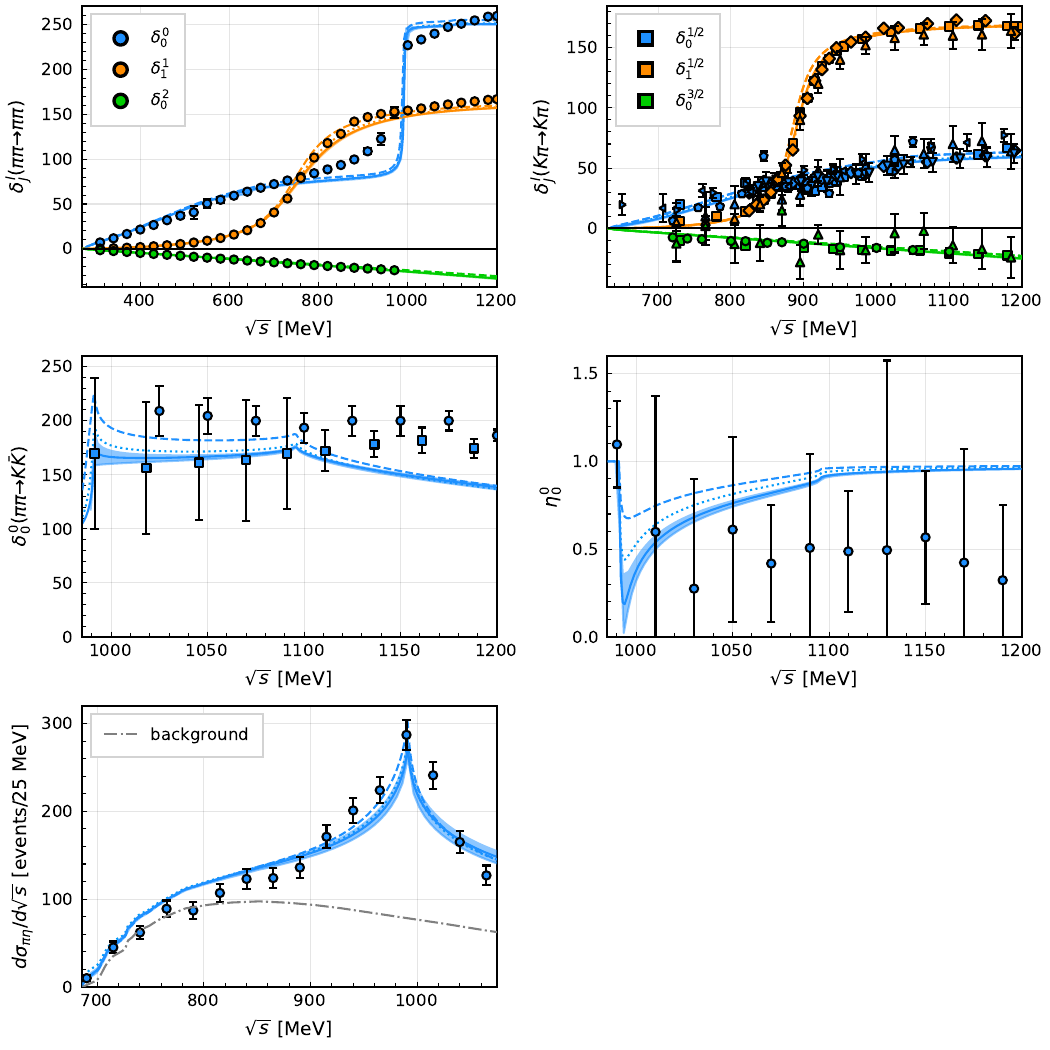}
  \caption{After removing the unitarity-violating contributions, the LECs are fitted to data. The narrow
    bands are the results from Fit~2 with errors of all data pointed increased by adding 5\% of the central values.
    The dotted lines are from the best-fit values of Fit~1, i.e., fitting to data with the original data errors.
    The results computed using the central values of LECs given in Ref.~\cite{GomezNicola:2001as} are shown as
    dashed lines. The $\pi\pi$ phase shift data are from Ref.~\cite{GarciaMartin:2011cn}. 
    For the $K\pi$ phase shifts, the data are taken from Refs.~\cite{Mercer:1971kn} (up triangles),
    \cite{Bingham:1972vy} (down triangles), \cite{Linglin:1973ci} (circles), \cite{Baker:1974kr} (pentagons),
    \cite{Estabrooks:1977xe} (rectangles), \cite{Aston:1987ir} (diamonds), \cite{delAmoSanchez:2010fd} (left triangles),
    and~\cite{Ablikim:2015mjo} (right triangles). 
    The $\pi\pi\to K\bar K$ data are from Refs.~\cite{Martin:1979gm} (rectangles) and~\cite{Cohen:1980cq} (circles).
    The $\eta_0^0$ data are from Ref.~\cite{Kaminski:1996da}.
    The data for $d\sigma_{\pi\eta}/d\sqrt{s}$, as well as the corresponding background, are taken from
    Ref.~\cite{Armstrong:1991rg}.   }
  \label{fig:fit}
\end{figure}

After the modification given above, the values of the LECs need to be refixed. We fit to the following data
sets using the MINUIT function minimization and error analysis package~\cite{James:1975dr,iminuit1,iminuit2}: the $\pi\pi$
scattering phase shifts are taken from the dispersive analysis compiled in Ref.~\cite{GarciaMartin:2011cn}
(which is perfectly compatible with the alternative Roy equation analyses of
Refs.~\cite{Ananthanarayan:2000ht,Colangelo:2001df,Caprini:2011ky} at the level of accuracy aimed for with the
IAM);\footnote{The table in Appendix~D of Ref.~\cite{GarciaMartin:2011cn} gives the $\pi\pi$ scattering phase
shifts up to $970\ \text{MeV}$. The $\delta_0^0$ data points above this energy were read off from the band in Fig.~15
of this reference, and those of $\delta_1^1$ were taken from Ref.~\cite{Kaminski:2006qe}. The latter reference
also provides an analysis of $\delta_0^2$, which, however, does not match exactly to the data given in
Ref.~\cite{GarciaMartin:2011cn}. Thus, for $\delta_0^2$, we only use those of Ref.~\cite{GarciaMartin:2011cn}
up to $970\ \text{MeV}$.}
the data for the inelasiticity $\eta_0^0$ are taken from the analysis of Ref.~\cite{Kaminski:1996da};
the $\pi\pi\to K\bar K$ data are from Refs.~\cite{Martin:1979gm,Cohen:1980cq} (cf.\ also
Refs.~\cite{Buettiker:2003pp,Pelaez:2018qny}); the $K\pi$ phase shifts are taken from
Refs.~\cite{Mercer:1971kn,Bingham:1972vy,Linglin:1973ci, Baker:1974kr, Estabrooks:1977xe, Aston:1987ir, delAmoSanchez:2010fd, Ablikim:2015mjo} (cf.\ also the corresponding dispersive analyses~\cite{Buettiker:2003pp,Pelaez:2016tgi,Pelaez:2020gnd}); 
the data for the $\pi\eta$ invariant mass distribution are taken from Ref.~\cite{Armstrong:1991rg},
and the background is extracted from the corresponding curve in that reference. 

We notice that in the NLO ChPT amplitudes for $\pi\pi\to \pi\pi$, $K\pi\to K\pi$, and $K\bar K\to K\bar K$,
$L_6^r$ and $L_8^r$ always appear as the same linear combination $2L_6^r+L_8^r$. Since most of the available data
are on these channels, it is difficult to fix $L_6^r$ and $L_8^r$ independently. Thus, we fix $L_6^r$ to the
central value given in Ref.~\cite{GomezNicola:2001as}, and fit the other parameters to the above data.
The $\pi\eta$ invariant mass distribution is fitted with the following expression~\cite{Oller:1998hw,GomezNicola:2001as}:
\begin{equation}
  \frac{d\sigma_{\pi\eta}}{d\sqrt{s}} = c \, q_{\rm cm} |(T_0^1)_{\pi\eta\to\pi\eta}|^2 + \text{background},
\end{equation}
where $c$ is a normalization constant to be fitted, $q_{\rm cm}$ is the $\pi\eta$ c.m.\ momentum, and the background
is extracted from the experimental analysis~\cite{Armstrong:1991rg}.
A direct fit to all these data sets leads to a value of $\chi^2/\text{dof}=7.76$ with the LECs given in
the column ``Fit~1'' in Table~\ref{tab:NLOLEC}. The large $\chi^2/\text{dof}$ value is due to the inconsistency among the data sets.
Following Refs.~\cite{Oller:1998zr,GomezNicola:2001as}, we increase the errors of the data points
by hand, and find an additional error of $5\%$ (of the central values) to all data points leads to
$\chi^2/\text{dof}=1.21$. The LECs from such a fit are listed in the last column in Table~\ref{tab:NLOLEC},
labelled as ``Fit~2''.
A comparison of these fits, as well as the results using the central values of LECs in Ref.~\cite{GomezNicola:2001as},
to the data is shown in Fig.~\ref{fig:fit}.
The errors propagated from the data in Fit~2 are plotted as bands, which are rather narrow.
One sees that the increase of $\delta_0^0$ around the $K\bar K$ threshold is more abrupt in uChPT than
that from the dispersive analysis~\cite{GarciaMartin:2011cn}. Other than that, the data are well described
using these different sets of LEC values.

An additional remark is in order. We find that when the LECs take certain values, $T^{(2)}-T^{(4)}$ in the $(I,J)=(0,0)$
channel can have zeros in the physical region using this modified version of coupled-channel IAM. These are not
the Adler zeros in the single-channel scattering amplitudes, and can lead to sharp kinks in phase shifts and
other observables at the zeros. 
Such unphysical singularities also exist in the original coupled-channel IAM. However, in that case, due to the
presence of nonvanishing imaginary parts from the unphysical left-hand cuts, the singularities are in the complex
$s$-plane, and thus lead to smoother kinks.\footnote{See the kink around $800\ \text{MeV}$ in the solid line of the $\delta_0^0$
plot in Fig.~2 of Ref.~\cite{GomezNicola:2001as}.} 
Nevertheless, we checked that the best-fit LECs in both Fit~1 and Fit~2, as well as those from
Ref.~\cite{GomezNicola:2001as}, do not have that problem. We will use the central values of these fits
(the three last columns in Table~\ref{tab:NLOLEC}) in the study of form factors in the following to estimate
the uncertainty of this method.

In what follows, we shall apply the idea above to calculate the two-meson scalar, vector, and tensor
form factors. We shall also consider a dispersion-theoretical improvement that will get rid of
the unphysical sub-threshold singularities due to Adler zeros.

\section{Scalar form factors}
\label{sec:3}

\begin{figure}
\begin{center}
\includegraphics[width=0.7\columnwidth]{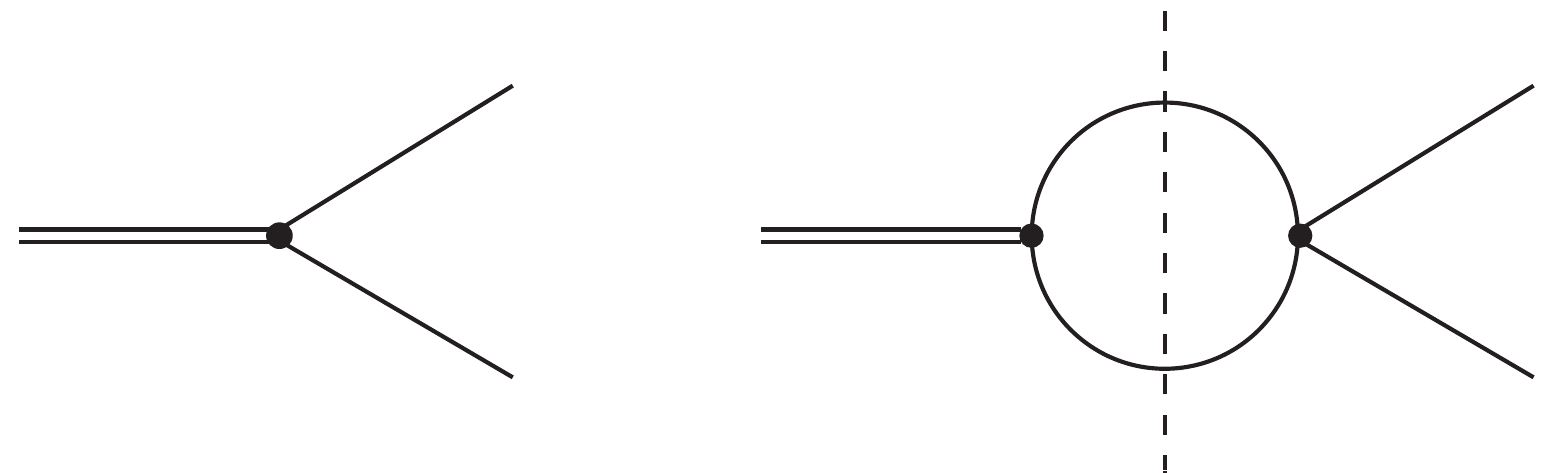} 
\caption{The imaginary part of a form factor is caused by the on shell configuration (cutting along the dashed
  line) of the intermediate states. In the elastic region, the phase of a form factor is the same as that of
  the scattering amplitude.}
\label{fig:watson}
\end{center}
\end{figure} 

In this section, we will give a systematic calculation for the two-meson scalar form factors in uChPT, where the
IAM approach is applied. Note that the unitarization of the two-loop scalar (and vector) pion form factor
was already discussed in Ref.~\cite{Gasser:1990bv}. The scalar form factor of a two-meson system is
defined by the matrix element
\begin{equation}
B_0F_{S,i}^{\bar{q}'q}(s)\equiv\left\langle a_{i}(p_{a_{i}})b_{i}(p_{b_{i}})\right|\bar{q}'q\left|0\right\rangle,
\end{equation}
where the subscript $i$ is again the channel index, $\{a_{i},b_{i}\}$
are the two mesons in channel $i$, and $s=(p_{a_{i}}+p_{b_{i}})^{2}$. The unitarity relation of the scalar form factor reads 
\begin{equation}
  2\mathrm{Im}F_{S,i}^{\bar{q}'q}(s)=\sum_{j}\frac{|\vec{p}_{j}|}{4\pi\sqrt{s}}\left[T_{0}^{*}(s)\right]_{ji}
  \Theta(s-s_\mathrm{th}^j)F_{S,j}^{\bar{q}'q}(s)~,\label{URelaS}
\end{equation}
where the subscript ``0'' at the partial-wave amplitude denotes the
$J=0$ component.  A sketch is shown in Fig.~\ref{fig:watson}, where the imaginary part of the form
factor is caused by the on-shell configuration of the intermediate states.
Furthermore,  time reversal invariance leads to  $(T_{0})_{ij}=(T_{0})_{ji}$. Now, we can simplify the expression
above by taking $F^{\bar{q}'q}(s)$ as a column vector in the channel space and write Eq.~(\ref{URelaS}) as a matrix
equation
\begin{equation}
\mathrm{Im}F^{\bar{q}'q}_S(s)=T_{0}^{*}(s)\Sigma(s)F^{\bar{q}'q}_S(s)~.\label{eq:SFFexactunitarity}
\end{equation}
This is the exact unitarity relation of the multi-channel scalar form
factor. We can also derive the perturbative unitarity relation by
expanding $F^{\bar{q}'q}(s)$ according to the chiral power counting
\begin{equation}
F_S=F_S^{(0)}+F_S^{(2)}+\ldots~,
\end{equation}
where the superscript $\bar{q}'q$ and the argument $s$ are suppressed for simplicity. Here we
define the leading term of the expansion to be $\mathcal{O}(p^0)$ because it is not suppressed
by the chiral expansion parameter. Meanwhile, $T_{0}=T_{0}^{(2)}+T_{0}^{(4)}+\ldots$.
Therefore, the perturbative unitarity relation reads
\begin{equation}
\mathrm{Im}F_S^{(0)} = 0, \qquad
\mathrm{Im}F_S^{(2)} = T_{0}^{(2)}\Sigma F_S^{(0)}, \qquad
\ldots . \label{eq:SFFperturbunitarity}
\end{equation}

\subsection{ChPT Result}

We have to first compute the ChPT results for the scalar form factor up to ${\cal O}(p^2)$ as input to
the IAM formula. For that purpose, we need to express the scalar current in terms of ChPT fields.
The scalar current in QCD is defined as
\begin{equation}
S_{ij}(x)\equiv\bar{q}_{i}(x)q_{j}(x)~,\label{SqcdCurr}
\end{equation}
where $q=(\begin{array}{ccc}
u & d & s\end{array})^{T}$. To obtain the ChPT version of this current, we start with the QCD Lagrangian
and promote its quark mass matrix $M$ to a general matrix $X_{q}$,
\begin{equation}
  \mathcal{L}_{QCD}=\bar{q}_{L}iD\!\!\!\!/\:q_{L}+\bar{q}_{R}iD\!\!\!\!/\:q_{R}-\bar{q}_{R}X_{q}q_{L}
  -\bar{q}_{L}X_{q}^{\dagger}q_{R}-\frac{1}{4}G_{\mu\nu}^{a}G^{a\mu\nu}.
\end{equation}
The scalar current defined in Eq.~(\ref{SqcdCurr}) is then obtained by taking the partial
derivative of the Lagrangian with respect to matrix elements of $X_{q}$, which results in
\begin{equation}
  S_{ij}=-\left.\left(\frac{\partial\mathcal{L}_{QCD}}{\partial(X_{q})_{ij}}
  +\frac{\partial\mathcal{L}_{QCD}}{\partial(X_{q}^{\dagger})_{ij}}\right)\right|_{X_{q}=M}.
\end{equation}
We can now derive the scalar current in ChPT by applying the formula above to the chiral Lagrangian. 
We can write the scalar current as $S_{ij}=S_{ij}^{(2)}+S_{ij}^{(4)}+\ldots$,
where $S_{ij}^{(n)}$ is defined as the scalar current derived from the chiral Lagrangian $\mathcal{L}^{(n)}$.
The outcome for the scalar currents, up to $S^{(4)}$, is
\begin{eqnarray}
S_{ij}^{(2)} & = & -\frac{F_0^2B_{0}}{2}[U^{\dagger}+U]_{ji},\nonumber\\
S_{ij}^{(4)} & = &-2B_{0}L_{4}\langle\partial_{\mu}U\partial^{\mu}U^{\dagger}\rangle\big[U^{\dagger}+U\big]_{ji}-2B_{0}L_{5}\big[U^{\dagger}\partial_{\mu}U\partial^{\mu}U^{\dagger}+\partial_{\mu}U\partial^{\mu}U^{\dagger}U\big]_{ji}\nonumber\\
 &  & -8B_{0}^{2}L_{6}\langle MU^{\dagger}+UM^{\dagger}\rangle\big[U^{\dagger}+U\big]_{ji}-8B_{0}^{2}L_{7}\langle MU^{\dagger}-UM^{\dagger}\rangle\big[U^{\dagger}-U\big]_{ji}\nonumber\\
 &  & -8B_{0}^{2}L_{8}\big[UM^{\dagger}U+U^{\dagger}MU^{\dagger}\big]_{ji}.\label{eq:Scurrent}
\end{eqnarray}
In particular, the components of $S^{(2)}$ are
\begin{eqnarray}
\bar{u}u & = & B_{0}\bigg[-F_0^2+K^{+}K^{-}+\frac{\pi^{0}\eta}{\sqrt{3}}+\frac{1}{6}\eta^{2}+\frac{1}{2}(\pi^{0})^2+\pi^{+}\pi^{-}\bigg]+\ldots,\nonumber \\
\bar{d}d & = & B_{0}\bigg[-F_0^2+K^{0}\bar{K}^{0}-\frac{\pi^{0}\eta}{\sqrt{3}}+\frac{1}{6}\eta^{2}+\frac{1}{2}(\pi^{0})^{2}+\pi^{+}\pi^{-}\bigg]+\ldots,\nonumber \\
\bar{s}s & = & B_{0}\bigg[-F_0^2+K^{0}\bar{K}^{0}+K^{+}K^{-}+\frac{2}{3}\eta^{2}\bigg]+\ldots,\nonumber \\
\bar{u}d & = & B_{0}\bigg[K^{0}K^{-}+\sqrt{\frac{2}{3}}\eta\pi^{-}\bigg]+\ldots,\nonumber \\
\bar{u}s & = & B_{0}\bigg[\bar{K}^{0}\pi^{-}-\frac{K^{-}\eta}{\sqrt{6}}+\frac{K^{-}\pi^{0}}{\sqrt{2}}\bigg]+\ldots,\nonumber \\
\bar{d}s & = & B_{0}\bigg[-\frac{\bar{K}^{0}\eta}{\sqrt{6}}-\frac{\bar{K}^{0}\pi^{0}}{\sqrt{2}}+K^{-}\pi^{+}\bigg]+\ldots.\label{eq:ScurrentCompo}
\end{eqnarray}

The ChPT prediction for scalar form factors is then obtained by calculating the matrix elements for
the currents in Eq.~\eqref{eq:ScurrentCompo} with respect to two-meson states up to one loop,
as shown in Fig.~\ref{fig:feynmanFormFactor}. The full analytical results can be found in Appendix~\ref{sec:FFoneloop}.

\begin{figure}
  \begin{center}
    \includegraphics[width=0.9\columnwidth]{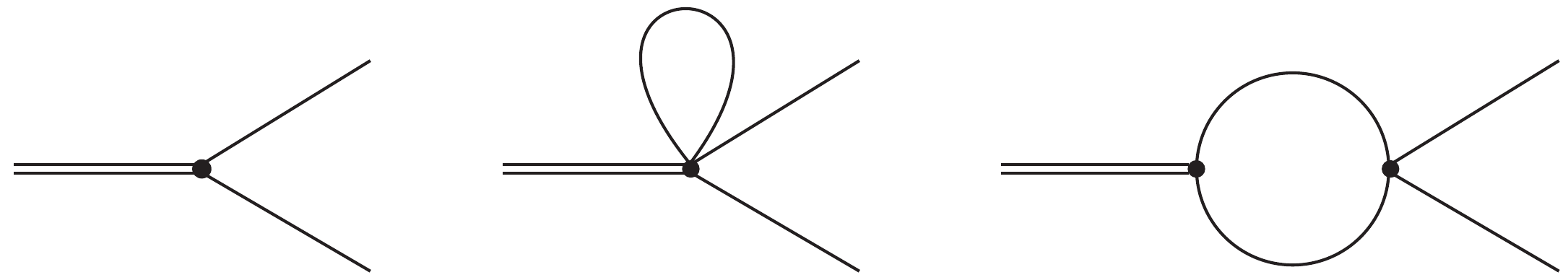} 
    \caption{Feynman diagrams for a generic  form factor at tree and one-loop level.
      The wave function renormalization diagrams are not shown here.   }
    \label{fig:feynmanFormFactor}
  \end{center}
\end{figure} 

\subsection{Unitarization}

If we restrict ourselves to one single channel, the IAM unitarization formula for  form
factors (scalar, vector, and tensor) can be derived rigorously from a dispersion relation in complete
analogy to the derivation of the single-channel IAM formula for partial waves, see Appendix~\ref{sec:Adler}
for more details (for an early application of the single-channel IAM to scalar and vector form factors,
see Ref.~\cite{Truong:1988zp}). For coupled-channel form factors, a dispersive derivation of the IAM is not available,
so here we offer a more empirical derivation of the unitarization formula, which we expect at
least to work well above the highest production threshold of the coupled channels considered here,
which is sufficient for the applications to most of the interesting processes we mentioned in the Introduction.

From Eq.~\eqref{eq:SFFexactunitarity}, one notices that a possible solution to this unitarity relation for
the scalar form factor is $F_S=T_0 A$,
where $A$ is a real vector. The proof is simple:
\begin{equation}
T_0^*\Sigma F_S=T_0^*\Sigma T_0A=(\mathrm{Im}T_0)A=\mathrm{Im}F_S.
\end{equation} 
In the second equality we have used the unitarity relation for $T_0$. So the question is how
to choose the form of the real vector $A$ such that its expansion reproduces the ChPT result
up to $\mathcal{O}(p^4)$. The correct choice turns out to be
\begin{equation} 
A=(T^{(2)}_0)^{-1}\left[F^{(0)}_S+F^{(2)}_S-T_0^{(4)}(T_0^{(2)})^{-1}F^{(0)}_S\right].
\end{equation} 
With this choice and a bit of algebra, we obtain the IAM formula for a unitarized scalar form factor
\begin{equation}
F_S=F_S^{(0)}+T_0^{(2)}\left(T_0^{(2)}-T_0^{(4)}\right)^{-1}F_S^{(2)}.\label{eq:IAMSFF}
\end{equation}
Remember that as argued in Sect.~\ref{sec:unitarity},  the LHCs of $T_0^{(4)}$ will be transferred into $F_S$. Thus we also need to remove the imaginary part of the troublesome $t$- and $u$-channel loops of $T_0^{(4)}$ in this formula manually.

\subsection{Improvement by Dispersion Relation} \label{disperImprov}

\begin{figure}
\begin{center}
\includegraphics[width=0.6\columnwidth]{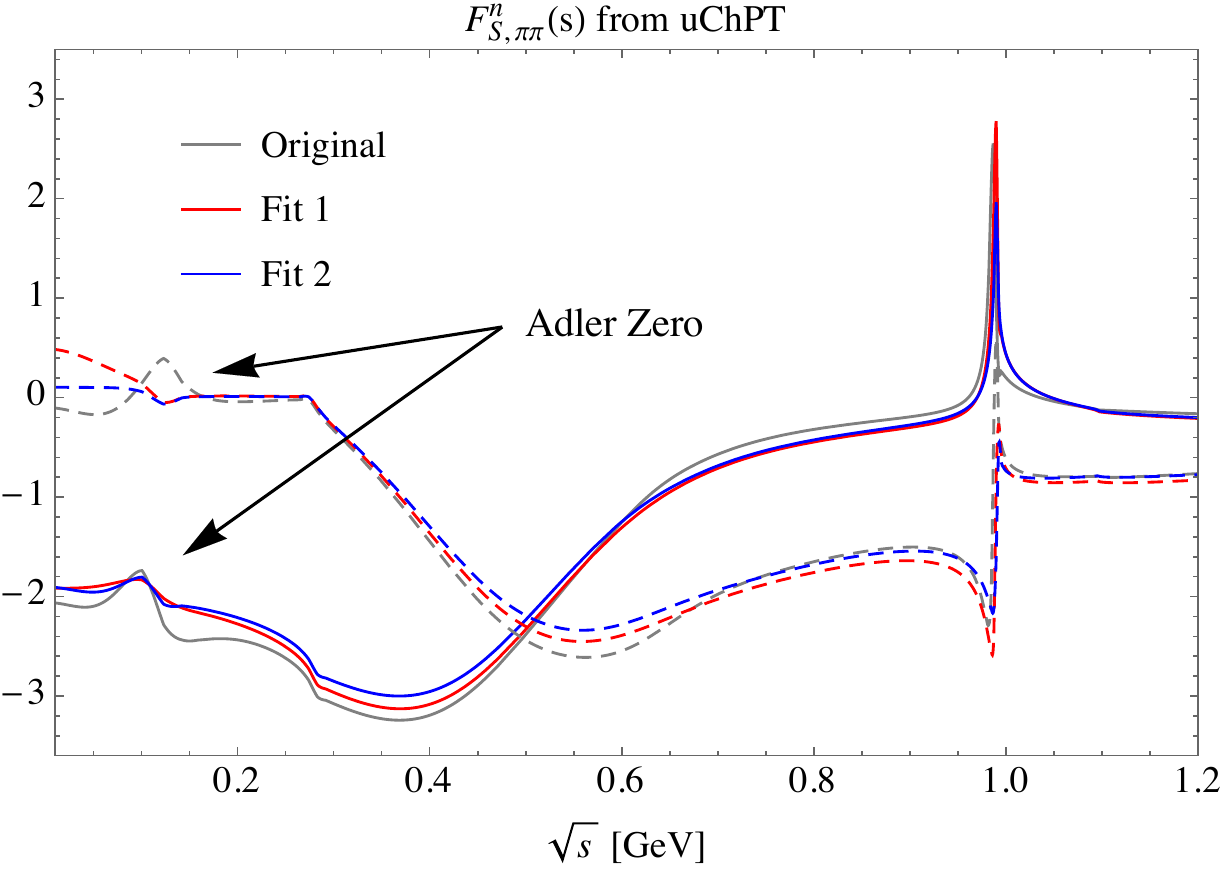} 
\caption{Real (solid lines) and imaginary (dashed lines) parts of  $F_{S,\pi\pi}^{n}$ from the
  original IAM calculation.  The LECs are taken from the original work~\cite{GomezNicola:2001as} (gray line),
  Fit~1 (red line) and Fit~2 (blue line), respectively. In the sub-threshold region there exist unphysical peaks
  due to Adler zeros in the scattering amplitude (pointed out by the black arrows) and nonvanishing imaginary parts below the lowest threshold.}
\label{fig:Sn_ChPT} 
\end{center}
\end{figure}

It is well known that the IAM generates spurious structures such as peaks that do
not correspond to any physical resonance. This happens in particular in the region below the lightest
two-meson production threshold. In fact, since $F_S=T_0 A$ is only a possible solution for
the unitarity equation~\eqref{eq:SFFexactunitarity} above threshold (more rigorously, above the highest
two-meson threshold since we are using a one-step unitarization for a coupled-channel problem), it
is natural that the outcome can only be trusted above threshold. For example, Fig.~\ref{fig:Sn_ChPT}
shows the scalar nonstrange current form factor from the IAM calculation, where the LECs are taken
from Ref.~\cite{GomezNicola:2001as} (gray line), Fit~1 (red line), and Fit~2 (blue line), respectively.
Obviously it suffers from sub-threshold irregularities, such as unphysical peaks (the tiny peaks near $0 \ \text{GeV}$ in
Fig.~\ref{fig:Sn_ChPT}) and nonvanishing imaginary parts, in all channels. 

The unphysical sub-threshold singularities due to the IAM are studied extensively in terms of
dispersion relation for the case of single-channel scattering amplitudes~\cite{Hannah:1997sm, GomezNicola:2007qj}. There,
the existence of spurious poles in the scalar partial wave is identified as a consequence of the failure
to include the effect of the so-called Adler zero in the dispersion integral of the inverse amplitude.
This problem can be solved by appropriately adding back such contributions in the IAM formula.
Unfortunately, a similar solution is not available in the coupled-channel case
because there is so far no dispersive derivation of the coupled-channel IAM formula.

On the other hand, the dispersion relations of the form factors themselves are much more straightforward.
It simply takes the following form:
\begin{equation}
{\rm Re}F(s)=\frac{1}{\pi}\pvint_{s_{\rm{th}}}^{\infty}dz\frac{{\rm Im}F(z)}{z-s}~,
\end{equation}
where $\pvint$ means the principal-value integration, and $s_{\rm{th}}$ denotes the lowest threshold.
Here it is sufficient to employ an unsubtracted dispersion relation because $F(s)$ falls off
as $1/s$ or faster at large $s$ as suggested by perturbative QCD~\cite{Brodsky:1973kr}. This integral
equation suggests
a better way to proceed: use the IAM-predicted imaginary part of the form factor as the input
to the dispersion integral, and obtain an improved real part of the form factor. The obtained real
part is then used to modify the imaginary part, which will be the input of a subsequent dispersive analysis. Such a procedure can be iterated until the curves of both the real and imaginary
parts of the form factor are stable. In the following we shall depict the actual procedure and outline
some of the details of such iterations.

First, we use $F^{[n]}$ to denote the form factor after $n$  iterations (to avoid confusion with
the chiral order denoted by superscripts with parentheses, here we use square brackets). Obviously, $F^{[0]}$
then represents the original IAM
result without undergoing any dispersive correction. To start the iteration process, in the first step we
set the imaginary part of $F^{[1]}$ as
\begin{equation}
\mathrm{Im}F^{[1]}\equiv T^*\Sigma F^{[0]},
\label{eq:iterated-unitarity}
\end{equation}
which will  be used later as an input to the dispersion integral to obtain $\mathrm{Re}F^{[1]}$.
However, we have to apply one extra modification before evaluating the dispersion integral: the IAM result $F^{[0]}$, which certainly does not apply to arbitrarily large $s$ values, fails to reproduce the asymptotic $1/s$-behavior. We therefore need to
introduce a smooth transition between the IAM-predicted $\mathrm{Im}F$ at small $s$ and
the expected $1/s$ at large $s$ by hand. This can be achieved by defining a modified imaginary
part $\mathrm{Im}\tilde{F}^{[1]}$ as
\begin{equation}
\mathrm{Im}\tilde{F}^{[1]}(s)\equiv \left[1-\sigma(s)\right]\mathrm{Im}F^{[1]}(s)+\sigma(s)\frac{\alpha^{[1]}}{s},
\end{equation}
where $\alpha^{[1]}$ is a constant to be determined later, and $\sigma(s)$ is a monotonically increasing
activation function that satisfies $\sigma(-\infty)=0$ and $\sigma(+\infty)=1$. A simple choice of
such a function is
\begin{equation}
 \sigma(s)=\frac{1}{2}\left(\tanh\left\{\frac{4(s-s_0)}{\delta s}\right\}+1\right).
\end{equation}
This activation function is centered at $s_0$ and has a width of $\delta s$. This means
that $\mathrm{Im}\tilde{F}^{[1]}(s)$ can smoothly transform from $\mathrm{Im}F^{[1]}(s)$ to $\alpha^{[1]}/s$
in the region $(s_0-\delta s/2,s_0+\delta s/2)$. 
 
Now, we should use the modified imaginary part $\mathrm{Im}\tilde{F}^{[1]}$, instead of $\mathrm{Im}F^{[1]}$,
in the dispersion integral to ensure the convergence at infinity. The unknown constant $\alpha^{[1]}$ can
be fixed by requiring that $F(s)$ reproduces the NLO ChPT result at $s=0$, i.e., $F^{[1]}(0)=\mathrm{Re}F^{[1]}(0)
=F_{\mathrm{ChPT}}(0)$. Once $\alpha^{[1]}$ is fixed, everything in the dispersion integral is known and we can
use it to numerically determine $\mathrm{Re}F^{[1]}(s)$. The dispersion relation guarantees that
the outcome makes sense both below and above threshold.   
 
The whole procedure above can be iterated until a stable result is obtained. At the
second step, we define $\mathrm{Im}F^{[2]}\equiv \mathrm{Re}\{T^*\Sigma F^{[1]}\}$ where
$F^{[1]}=\mathrm{Re}\tilde{F}^{[1]}+i\,\mathrm{Im}\tilde{F}^{[1]}$, and then modify its UV-behavior by
constructing $\mathrm{Im}\tilde{F}^{[2]}(s)$. Notice that this construction will involve a new unknown
constant $\alpha^{[2]}$ which is in general different from $\alpha^{[1]}$ so that it has to be re-determined.
After that, we can plug $\mathrm{Im}\tilde{F}^{[2]}$ into the dispersion integral to obtain $\mathrm{Re}F^{[2]}$.
This procedure will be iterated for several times so that we can obtain a series of increasingly
refined form factors $F^{[3]},F^{[4]},\ldots$, which will eventually stabilize. Finally, the unphysical peaks
such as those in Fig.~\ref{fig:Sn_ChPT} are completely wiped out after such a dispersive improvement. 

It is worthwhile to stress that the Watson's theorem is still fulfilled perturbatively at the fixed point of the iteration procedure, and this dispersive treatment is applicable to all scalar, vector, and tensor form factors. 
  
\subsection{Numerical Results}

\begin{figure}
\includegraphics[width=0.48\columnwidth]{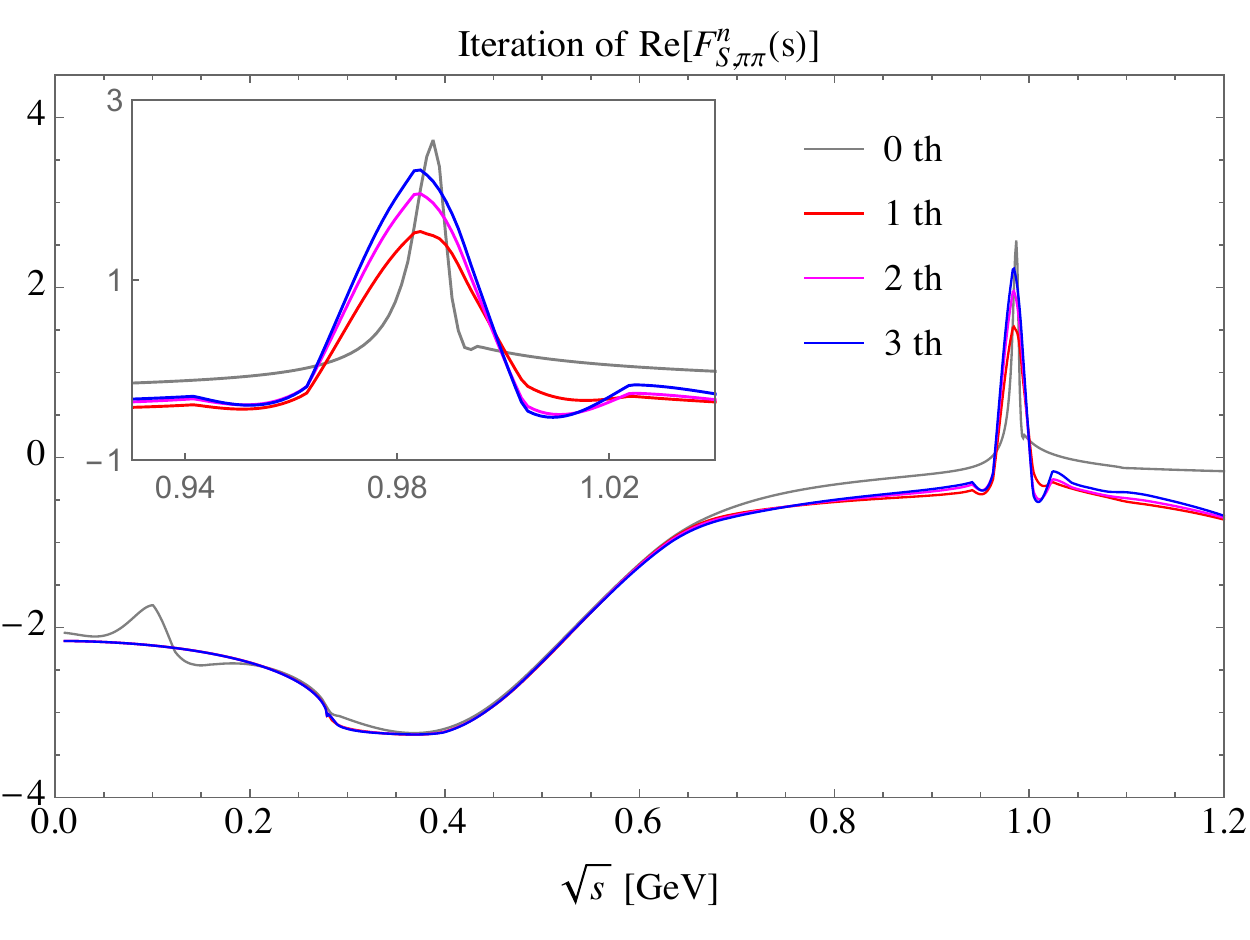} 
\includegraphics[width=0.48\columnwidth]{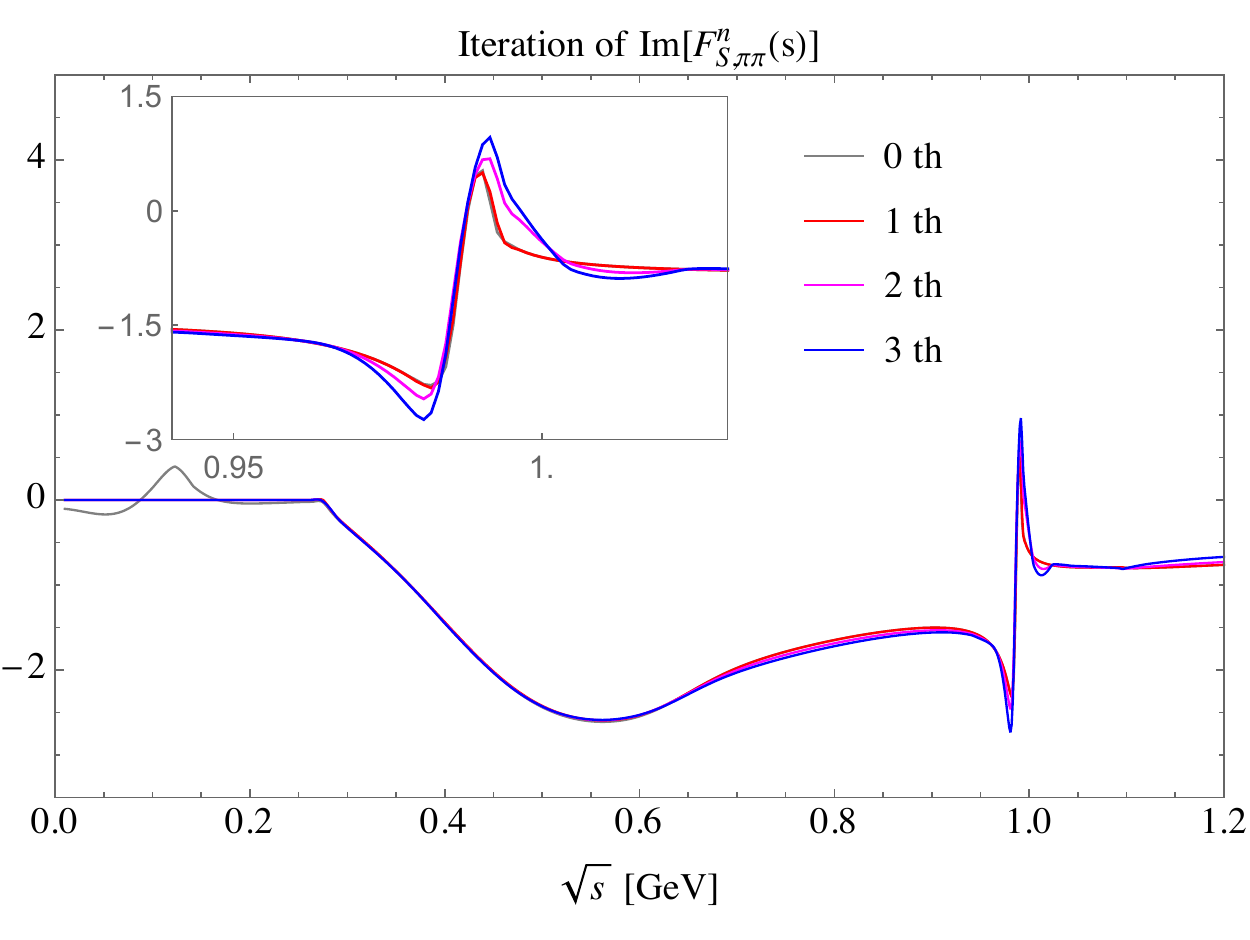}
\caption{Four iterations for the real and imaginary parts of $F_{S,\pi\pi}^{n}$ with the LECs taking the central values
  of those given in Ref.~\cite{GomezNicola:2001as}. Different colors distinguish various steps of the iteration and
  the sub-figures zoom in to show the details. }
\label{fig:SnIteration} 
\end{figure}
\begin{figure}
\begin{center}
\includegraphics[width=0.49\columnwidth]{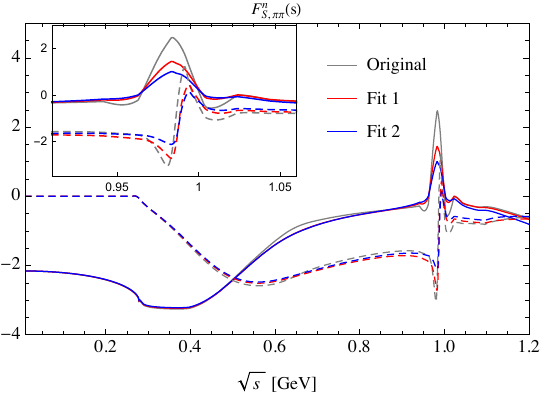} 
\includegraphics[width=0.49\columnwidth]{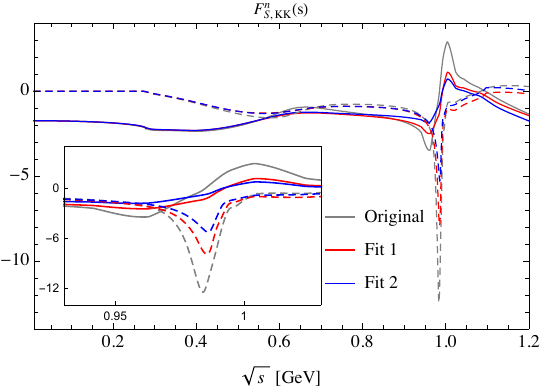}
\includegraphics[width=0.49\columnwidth]{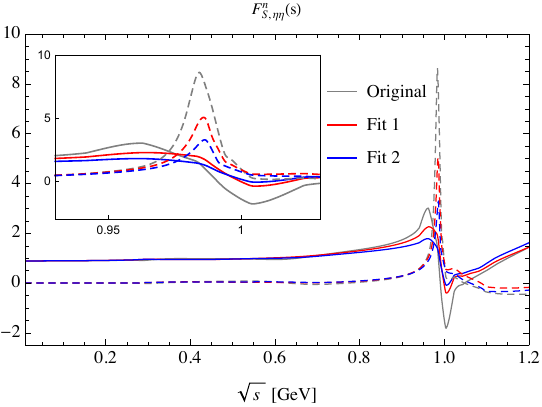} 
\caption{Real (solid lines) and imaginary (dashed lines) parts of  $F_{S,\pi\pi}^{n}$, $F_{S,K\bar{K}}^{n}$, and $F_{S,\eta\eta}^{n}$, respectively.  Three sets of LECs are used to plot the form factors: the original one from Ref.~\cite{GomezNicola:2001as} (gray lines), one from Fit~1 (red lines) and one from Fit~2 (blue lines).}
\label{fig:Sn} 
\end{center}
\end{figure}
\begin{figure}
\begin{center}
\includegraphics[width=0.49\columnwidth]{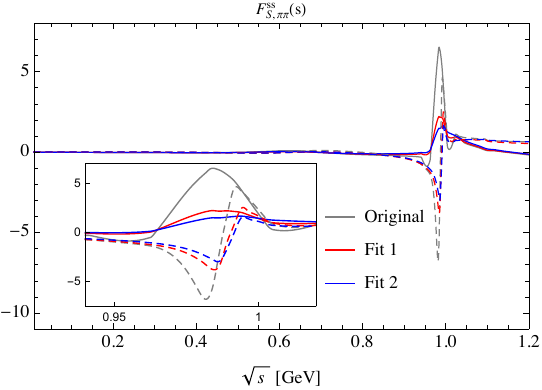} 
\includegraphics[width=0.49\columnwidth]{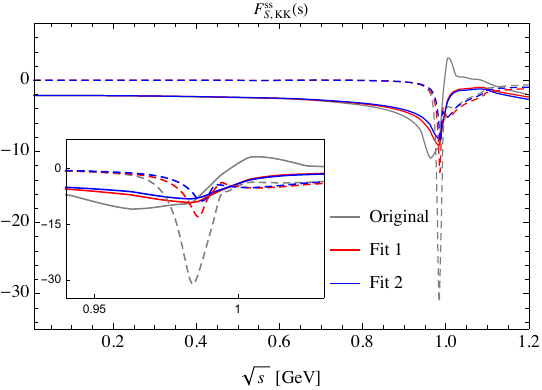}
\includegraphics[width=0.49\columnwidth]{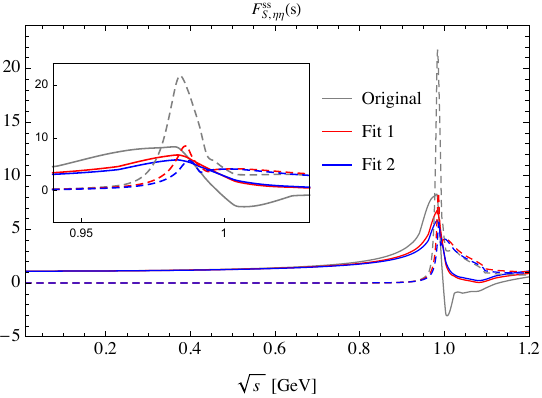} 
\caption{Real and imaginary parts of  $F_{S,\pi\pi}^{\bar s s}$, $F_{S,K\bar{K}}^{\bar s s}$, and $F_{S,\eta\eta}^{\bar s s}$, respectively. The meaning of different types of lines is the same as that in Fig.~\ref{fig:Sn}.}
\label{fig:Sss} 
\end{center}
\end{figure}
\begin{figure}
\includegraphics[width=0.49\columnwidth]{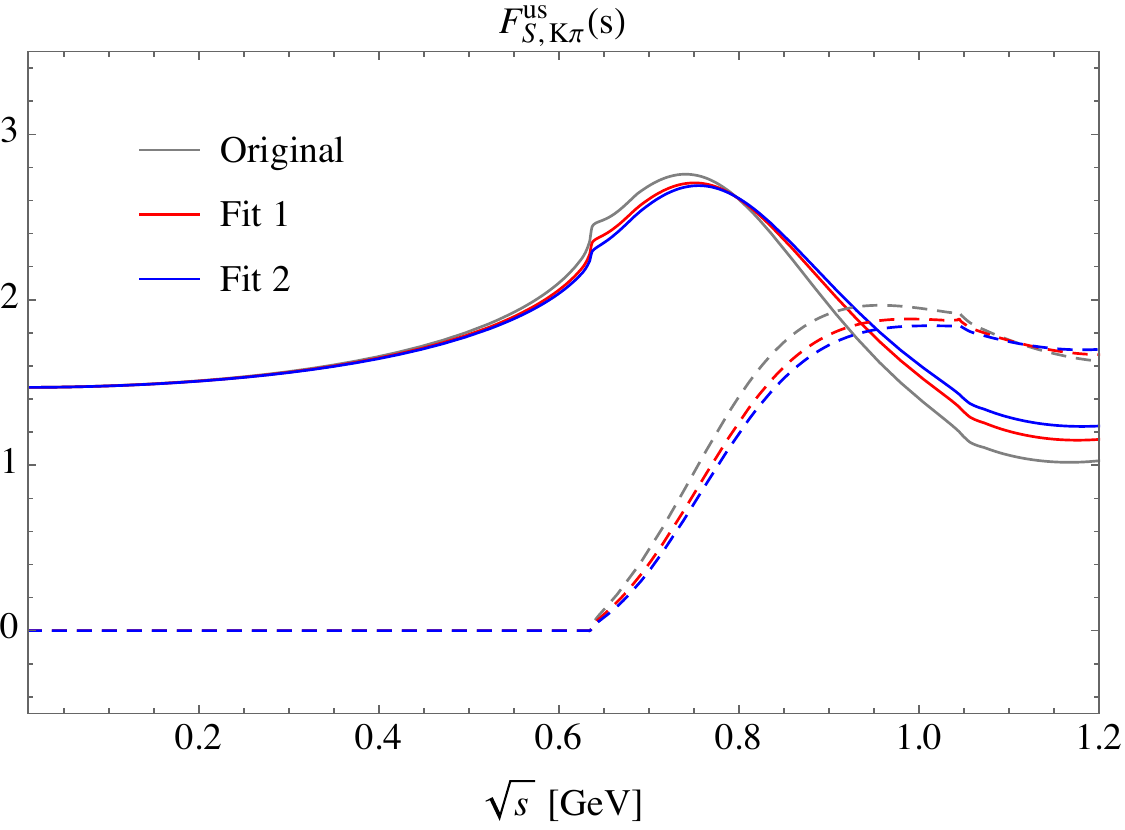} 
\includegraphics[width=0.49\columnwidth]{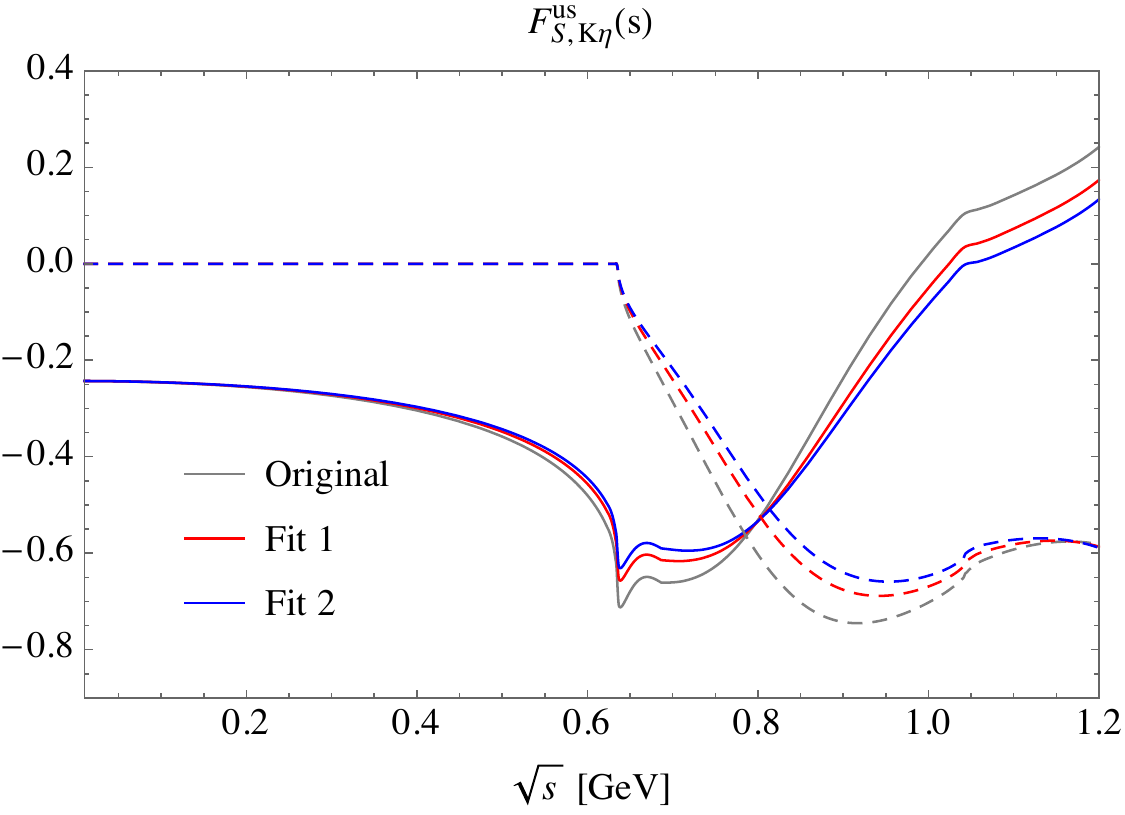}
\caption{Real and imaginary parts of  $F_{S,K\pi}^{\bar u s}$ and $F_{S,K\eta}^{\bar u s}$, respectively.
  For notations, see
  Fig.~\ref{fig:Sn}.}
\label{fig:Sus} 
\end{figure}
\begin{figure}
\includegraphics[width=0.49\columnwidth]{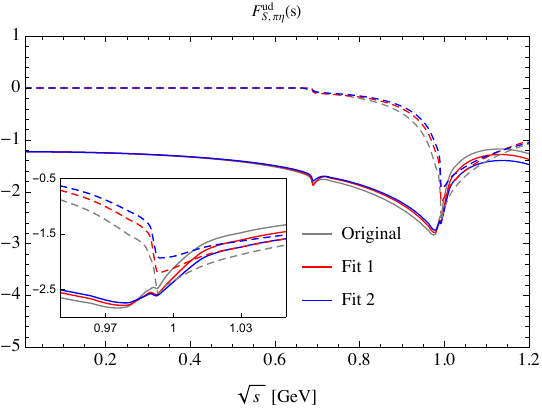} 
\includegraphics[width=0.49\columnwidth]{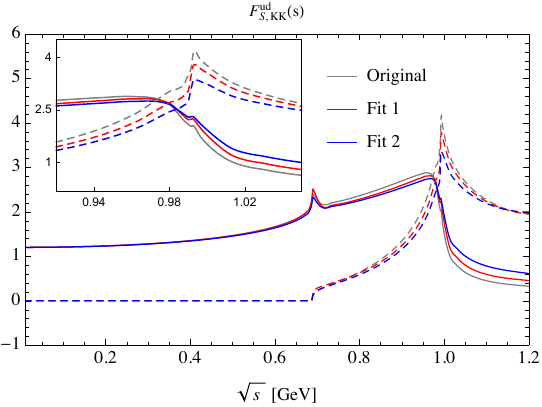}
\caption{Real and imaginary parts of  $F_{S,\pi\eta}^{\bar u d}$ and $F_{S,K\bar{K}}^{\bar u d}$, respectively.
  For notations, see
  Fig.~\ref{fig:Sn}.}
\label{fig:Sud} 
\end{figure}

In this section we show the numerical results of the unitarized scalar form factors after the dispersive improvement. 
To show the convergence of the iteration, as an example, Fig.~\ref{fig:SnIteration} gives the first four
iterations for the real and imaginary parts of $F_{S,\pi\pi}^{n}$, with the LECs taken from Ref.~\cite{GomezNicola:2001as}.
It can be seen that the iteration successfully removes the kink due to the Adler zeros of the scattering amplitudes,
and the imaginary part vanishes below the lowest threshold.   The curves of all the scalar form factors
after the iteration are plotted in 
Figs.~\ref{fig:Sn}--\ref{fig:Sud} 
within the plot region
$0\ \text{GeV}<\sqrt{s}<1.2\ \text{GeV}$. Here, we use three sets of LECs: the original one from Ref.~\cite{GomezNicola:2001as}
(gray lines), the one from Fit~1 (red lines), and the one from Fit~2 (blue lines) to plot the form factors.
The solid and dashed lines correspond to the real and imaginary parts, respectively.  The parameters for
the activation function $\sigma(s)$
are taken to be $s_0=1.8\ \text{GeV}^2$ and $\delta s=0.6\ \text{GeV}^2$, which implies a smooth
transition between the IAM result and the $1/s$-behavior within the range $1.1\ \text{GeV}<\sqrt{s}<1.6\ \text{GeV}$. 


In the $n=(\bar{u}u+\bar{d}d)/\sqrt{2}$ channel and the $\bar{s}s$ channel, the real parts of the form
factors generate sharp peaks around $\sqrt{s}\sim 0.99\ \text{GeV}$ due to the simultaneous
existence of the $f_0(980)$ resonance and the $K\bar{K}$-threshold within a narrow region. Our results
for $F_{S,\pi\pi}^{n}$, $F_{S,\pi\pi}^{\bar{s}s}$, and $F_{S,K\bar{K}}^{\bar{s}s}$ are consistent with those presented in
Ref.~\cite{Doring:2013wka} (barring differences in overall normalization), the latter were obtained by a
slightly different version of algebraic unitarization formula, incorporating only the $s$-channel cuts of
the partial waves. However, the $f_0(980)$ region shown in Fig.~\ref{fig:Sn} has a narrower structure compared with that given in Refs.~\cite{Daub:2015xja,Ropertz:2018stk}. Other disagreement appears in the $F_{S,K\bar{K}}^{n}$ form factor. In particular,
the outcome of Ref.~\cite{Doring:2013wka} does not match the NLO ChPT value at $s=0$. Our result, on
the other hand, guarantees such a matching as it is implemented during the determination of the
coefficient $\alpha^{[i]}$. Also, we present the $\eta\eta$ form factors that were not calculated in
that paper. 

\begin{figure}
\begin{center}
\includegraphics[width=0.6\columnwidth]{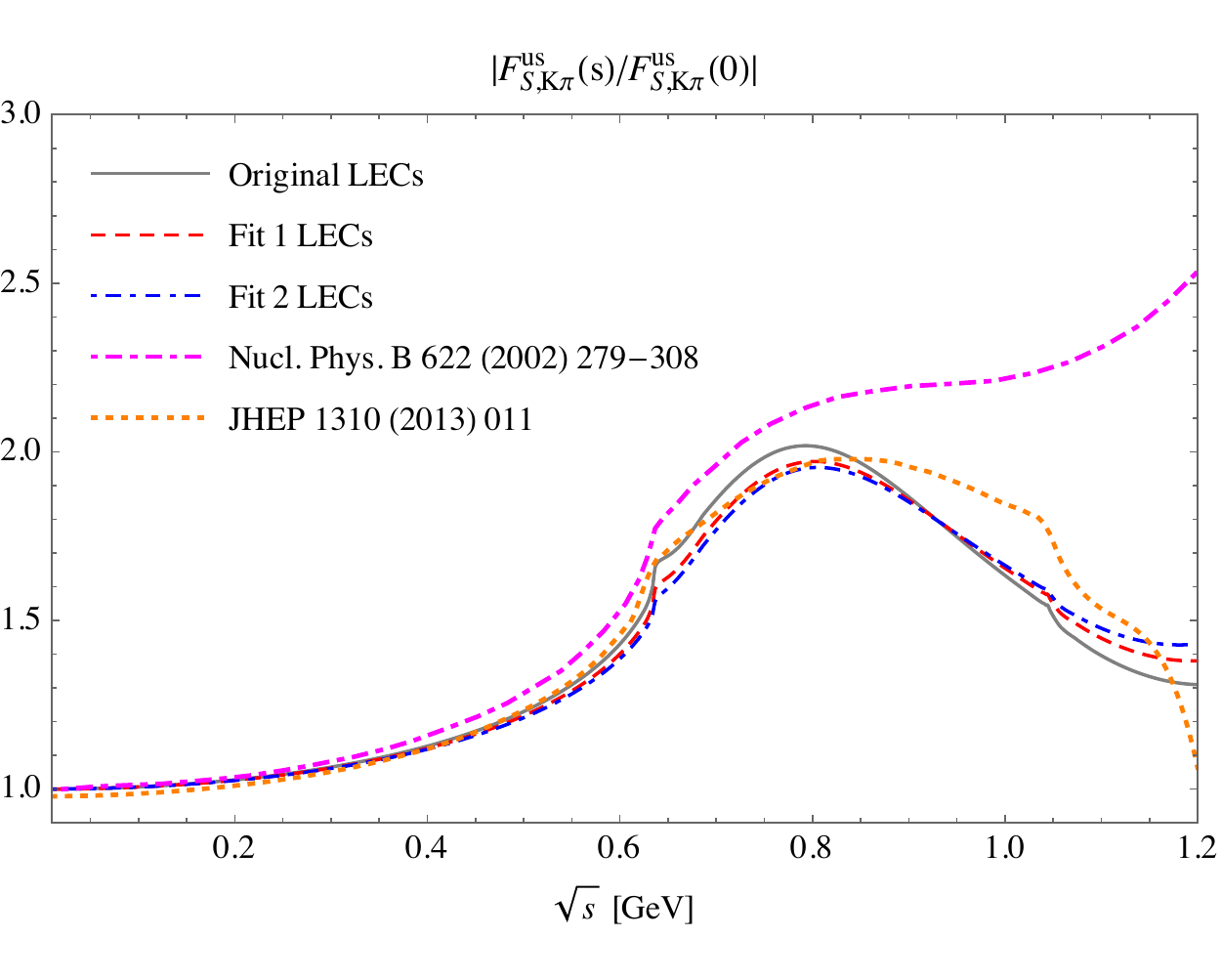} 
\caption{Comparison of $F_{S,K\pi}^{\bar u s}$ in this work using the original LECs in Ref.~\cite{GomezNicola:2001as}
  (gray solid line), Fit~1 LECs (red dashed line) and Fit~2 LECs (blue dot-dashed line), respectively, with
  that from a coupled-channel dispersion relation analysis~\cite{Jamin:2001zq} (magenta dot-dashed line)
  as well as that from a Muskhelishvili--Omn\`es solution~\cite{Doring:2013wka} (orange dotted line). }
\label{fig:CompareWithEarlywork} 
\end{center}
\end{figure}

In the $\bar{u}s$ channel, the strategy adopted in Ref.~\cite{Doring:2013wka} therein is 
computationally involved as one has to first discretize $s\rightarrow\{s_i\}$ and solve the
dispersion relation by inverting a huge rank matrix (in the $s$-space) to obtain the discretized form
factor $F(s_i)$. Our approach is much simpler because we are simply taking the IAM results above
threshold as the input of the dispersion integral, and the outcomes quickly stabilize after two or
three iterations. In Fig.~\ref{fig:CompareWithEarlywork}, our results are compared with those
from Refs.~\cite{Jamin:2001zq,Doring:2013wka}. All of them coincide with each other below $0.8\ \text{GeV}$.
The mismatch at larger $\sqrt{s}$ can be
understood because the convergence of ChPT becomes weaker at higher energies, which inevitably causes
more uncertainties. 
One notices that there is a cusp at the $K\pi$ threshold, signaling the opening of the $K\pi$ channel
in the $S$-wave. The second cusp appears at the $K\eta$ threshold, which enters through the coupled-channel treatment.\footnote{The strength of the $K\eta$ threshold cusp in the $K\pi$ $S$-wave is likely overestimated in uChPT, as the phenomenological inelasticity is very small below $K\eta'$ threshold~\cite{Pelaez:2016tgi}; see also Refs.~\cite{Noel:2020,vonDetten:2020}.}

\subsection{Applications of the Scalar Form Factors} 

We end this section by discussing applications of the two-meson scalar form factors, especially its
$s$-dependence.  Let us consider for example the decay $B_s\to f_0(980)(\to\pi^+\pi^-) \mu^+ \mu^-$,
which is a four-body decay dominated by the $S$-wave contribution $f_0(980)\to\pi^+\pi^-$. To study
this decay process, the main task is to evaluate the  $B_s\to (\pi^+\pi^-)_S$ transition matrix elements,
which are parameterized as
 \begin{align}
 \langle (\pi^+\pi^-)_S|\bar s \gamma_\mu\gamma_5 b|\bar B_s
 \rangle 
 &=  \frac{-i }{M_{B_s}} \Bigg\{ \bigg[P_{\mu}
 -\frac{M_{B_s}^2-M_{\pi\pi}^2}{q^2} q_\mu \bigg] {\cal F}^{B_s\to \pi\pi}_{1}(M_{\pi\pi}^2, q^2) \nonumber \\ & \qquad \qquad
 +\frac{M_{B_s}^2-M_{\pi\pi}^2}{q^2} q_\mu  {\cal F}^{B_s\to \pi\pi}_{0}(M_{\pi\pi}^2, q^2)  \Bigg\},
 \nonumber\\
\langle (\pi^+\pi^-)_S|\bar s \sigma_{\mu\nu} q^\nu \gamma_5 b|
 \bar B_s \rangle  & = \frac{{\cal F}^{B_s\to \pi\pi}_T(M_{\pi\pi}^2,
 q^2)}{M_{B_s}(M_{B_s}+M_{\pi\pi})} \bigg[ ({M_{B_s}^2-M_{\pi\pi}^2}) q_\mu - q^2
 P_{\mu}\bigg],
 \label{eq:generalized_form_factors}
\end{align}
where $M_{\pi\pi}^2$ is the invariant mass square of the two-pion system.
These $B_s\to \pi^+\pi^-$ form factors can be calculated
by light-cone sum rules (LCSR) and expressed in terms of the $\pi^+\pi^-$ light-cone distribution amplitudes
(LCDAs)~\cite{Mueller:1998fv,Diehl:1998dk,Polyakov:1998ze,Kivel:1999sd,Diehl:2003ny,Hagler:2002nh,Pire:2008xe}. 

According to the Watson--Migdal theorem, since the $B_s\to \pi^+\pi^-$ transition totally decouples from the
leptonic part at leading order, its amplitude must share the same phase as that of the $ \pi^+\pi^-$ scalar
form factor $F_{S,\pi\pi}^{\bar s s}(M_{\pi\pi}^2)$ below the lowest inelastic threshold (which is that of $K\bar K$
since the four-pion channel is not considered, and the inelasiticity is known to be negligible below about $1\ \text{GeV}$). 
Accordingly, in the framework of LCSR, the $S$-wave $\pi^+\pi^-$  LCDAs are defined almost the same as those of a single
scalar meson $f_0$ but with the normalization factor taken as the scalar $\pi\pi$ form
factor~\cite{Diehl:1998dk,Polyakov:1998ze,Kivel:1999sd,Diehl:2003ny,Meissner:2013hya}.
For example, the twist-2 LCDA is defined as
\begin{eqnarray}
 \langle {(\pi^+\pi^-)_S}|\bar s (x)\gamma_\mu s(0)|0\rangle &=&  F_{S,\pi\pi}^{\bar s s}(M_{\pi\pi}^2)B_0\ 
 p_{{\pi\pi},\mu}
 \int_0^1 du \, e^{i up_{\pi\pi}\cdot x}\phi_{\pi\pi}(u)~,
\end{eqnarray}
where $F_{S,\pi\pi}^{\bar s s}(M_{\pi\pi}^2)B_0$ stands for the original meson decay constant $f_{f_0}$
in the definition of the $f_0$ LCDAs.
The definition of all the twist-2 and 3 $S$-wave $\pi^+\pi^-$  LCDAs  as well as the explicit form
of the resulting form factors~\cite{Meissner:2013hya} can be found in Appendix~\ref{sec:LCDAs}. 
As a result,  the form factors take the form
\begin{eqnarray}
 {\cal F}_i(M_{\pi\pi}^2, q^2)&=& B_0 F_{S,\pi\pi}^{\bar s s}(M_{\pi\pi}^2) \overline F_i(M_{\pi\pi}^2,q^2)~.\label{Fpipifactor}
\end{eqnarray}
Generally,  the $\overline F_i$ depend on both $M_{\pi\pi}^2$ and $q^2$. However, in the case of
$B \to K^{*}(\to K\pi)$,  as shown by Fig.~3 of Ref.~\cite{Meissner:2013hya}, the $\overline F_i$'s have a
much weaker dependence on $M_{K\pi}^2$ than that on $q^2$. Such behavior is similar to the case of $B_s\to \pi^+\pi^-$.
This enables one to approximately suppress the $M_{\pi\pi}^2$ dependence of $\overline F_i$~, which leads to a
factorized form
\begin{eqnarray}
 {\cal F}_i(M_{\pi\pi}^2, q^2)&\approx& B_0 F_{S,\pi\pi}^{\bar s s}(M_{\pi\pi}^2) \overline F_i(q^2)~,\label{FpipifactorApprox}
\end{eqnarray}
so that $\overline F_i$ can be described by a suitable
parametrization~\cite{Cheng:2005nb,Wang:2015paa,Shi:2015kha,Shi:2017pgh}. Practically, one can first
fix  $M_{\pi\pi}=M_{f_0(980)}$ to extract
$\overline F_i(q^2)$, and then multiply it with $F_{S,\pi\pi}^{\bar s s}(M_{\pi\pi}^2)$ again in the form of
Eq.~\eqref{FpipifactorApprox} to recover the complete transition form factor $ {\cal F}_i(M_{\pi\pi}^2, q^2)$.
The detailed calculation can be found in Ref.~\cite{Shi:2015kha}.
On the other hand, instead of the $S$-wave $\pi^+\pi^-$  LCDAs, one can firstly use the LCDAs of  the $f_0$
to get the transition form factors $\tilde F_i(q^2)$ of $B_s\to f_0$, then due to the approximation leading
to Eq.~\eqref{FpipifactorApprox}, one can write the total form factor as
\begin{eqnarray}
  {\cal F}_i(M_{\pi\pi}^2, q^2)&\approx& \frac{B_0}{f_{f_0}} F_{S,\pi\pi}^{\bar s s}(M_{\pi\pi}^2) \tilde F_i(q^2)~.
  \label{FpipifactorApprox2}
\end{eqnarray} 
Equations~\eqref{FpipifactorApprox} and~\eqref{FpipifactorApprox2} explicitly
reflect that the $\pi\pi$ distribution of the $S$-wave-dominated decay $B_s\to f_0(980)(\to\pi^+\pi^-)
\mu^+ \mu^-$ is determined by the distribution of the $\pi\pi$ scalar form factor. 

All these scalar form factors are also required in the study of three-body $B$ decays, which have
been studied using dispersively reconstructed form factors or phenomenological parametrizations, see for instance
Refs.~\cite{Daub:2015xja,Ropertz:2018stk,Chen:2002th,Wang:2015uea,Ma:2017idu,Boito:2017jav,Liang:2018szw,Xing:2019xti} and many references therein. 

\section{Vector form factors}
\label{sec:4}
 
Next we discuss the vector form factors. The matrix element of a vector current with respect to a
two-meson system can be parametrized as
\begin{equation}
  \label{eq:vectorFFdef}\left\langle a_{i}(p_{a_{i}})b_{i}(p_{b_{i}})\right|\bar{q}'\gamma^{\mu}q\left|0\right\rangle
  \equiv F_{V+,i}^{\bar{q}'q}(s)(p_{a_{i}}-p_{b_{i}})^{\mu}+F_{V-,i}^{\bar{q}'q}(s)(p_{a_{i}}+p_{b_{i}})^{\mu}.
\end{equation}
Notice that we label the form factors $F_{V\pm}$ as above because they are more commonly defined in the
$t$-channel, where $p_{b_i}$ will switch sign. From the equation of motion
$\partial_{\mu}(\bar{q}'\gamma^{\mu}q)=i(m_{q'}-m_{q})\bar{q}'q$, one sees that the form factor
$F_{V-,i}$  is not independent since it can be expressed in terms of $F_{V+,i}$ and the scalar
form factor $F_{S,i}$ according to
\begin{equation}
\label{eq:sFFtovFF}
F_{V-,i}^{\bar{q}'q}(s)=\frac{1}{s}\left[B_0(m_{q'}-m_{q})F_{S,i}^{\bar{q}'q}(s)
  -(m_{a_{i}}^{2}-m_{b_{i}}^{2})F_{V+,i}^{\bar{q}'q}(s)\right].
\end{equation}
Therefore, it is sufficient to concentrate only on $F_{V+,i}^{\bar{q}'q}$. 

The unitarity relation of $F_{V+,i}^{\bar{q}'q}$ is most conveniently expressed in terms of
\begin{eqnarray}
\tilde{F}_{V+}^{\bar{q}'q}(s) & \equiv &\mathbb{P} F_{V+}^{\bar{q}'q}(s)~,
\end{eqnarray}
where $\mathbb{P}_{ij}\equiv |\vec{p}_i|\delta_{ij}$~.
One can then straightforwardly express the unitarity relation as
\begin{eqnarray}
  \mathrm{Im}\tilde{F}^{\bar{q}'q}_{V+} & = & T_{1}^{*}\Sigma\tilde{F}^{\bar{q}'q}_{V+}~.
  \label{eq:VFFexactunitarity}
\end{eqnarray}
Notice that $\tilde{F}$ is associated with the $J=1$ partial-wave
scattering amplitude. However, the relation above is rigorously true only above the highest threshold,
where all elements of $\mathbb{P}$ are real. When $s$ is between the lowest and highest thresholds,
a more rigorous form of the unitarity relation is
\begin{equation}
  \mathrm{Im}F^{\bar{q}'q}_{V+}=(\mathbb{P}^{-1})^*T_1^*\Sigma \mathbb{P}F^{\bar{q}'q}_{V+}.
  \label{eq:VFFunitarityEq}
\end{equation}
In particular, at the right-hand side of the equation above we have $(\mathbb{P}^{-1})^*$ instead of
$\mathbb{P}^{-1}$ so that the kinematical imaginary part of $T_1^*$ (i.e., the imaginary part due to
$|\vec{p}_i|$) below the highest threshold can be canceled by that of $(\mathbb{P}^{-1})^*$.

\subsection{ChPT Result}
\label{sec:vector}

There are two kinds of vector currents: the SU(3)-octet current
$V_{\mu}^{a}$ ($a=1,..,8$) and the singlet current $V_{\mu}^{0}$ due
to the $\text{SU}(3)_{V}$ and $\text{U}(1)_{B}$ ($B$ stands for baryon number)
symmetry, respectively. They can be defined as
\begin{eqnarray}
V_{\mu}^{0}  =  \bar{q}\gamma_{\mu}q,~~~V_{\mu}^{a} & = & \bar{q}T^{a}\gamma_{\mu}q,
\end{eqnarray}
respectively. The easiest way to obtain such currents from the QCD
Lagrangian is to first promote $\text{SU}(3)_{V}$ and $\text{U}(1)_{B}$
to local symmetries by introducing external fields $v_\mu=T^av_\mu^a$
and $v_{\mu}^{(s)}$:
\begin{equation}
\label{2:6:lqcds}
{\cal L}={\cal L}^0_{\rm QCD}+{\cal L}_{\rm ext}
={\cal L}^0_{\rm QCD}+\bar{q}\gamma_\mu \bigg(v^\mu +\frac{1}{3}v^\mu_{(s)}
+\gamma_5 a^\mu \bigg)q
-\bar{q}(s-i\gamma_5 p)q.
\end{equation}
Taking the derivative of the Lagrangian with respect to the external fields gives the currents
\begin{eqnarray}
V_{\mu}^{0} = \left.3\frac{\partial\mathcal{L}}{\partial v_{(s)}^{\mu}}\right|_{v=v_{(s)}=0} , \qquad
V_{\mu}^{a} = \left.\frac{\partial\mathcal{L}}{\partial v^\mu_a}\right|_{v=v_{(s)}=0}.
\end{eqnarray}
 
Again, we can apply the formulae above to obtain vector currents in
ChPT. The strict SU(3) symmetry of the ChPT Lagrangian up to ${\cal O}(p^4)$ leads to a vanishing $V_{\mu}^{0}$.
It should be noted that at ${\cal O}(p^6)$ a certain SU(3) breaking term can be introduced so that $V_{\mu}^{0}$
no longer vanishes. However, at ${\cal O}(p^4)$ we will not consider this effect. 
For the octet currents, we have
\begin{eqnarray}
V_{a\mu}^{(2)} & = & -\frac{iF_0^2}{4}\langle\lambda^{a}[U,\partial_{\mu}U^{\dagger}]\rangle~,\nonumber \\
V_{a\mu}^{(4)} & = & -2iL_{1}\langle\partial_{\nu}U\partial^{\nu}U^{\dagger}\rangle\langle\lambda^{a}[U,\partial_{\mu}U^{\dagger}]\rangle \nonumber\\
&&
-iL_{2}\big\{\langle\partial_{\mu}U\partial_{\nu}U^{\dagger}\rangle\langle\lambda^{a}[U,\partial^{\nu}U^{\dagger}]\rangle
+\langle\partial_{\nu}U\partial_{\mu}U^{\dagger}\rangle\langle\lambda^{a}[U^{\dagger},\partial^{\nu}U]\rangle\big\}
 \nonumber\\
&& -iL_{3}\langle\big([\lambda^{a},U]\partial_{\mu}U^{\dagger}+\partial_{\mu}U[\lambda_{a},U^{\dagger}]\big)\partial_{\nu}U\partial^{\nu}U^{\dagger}\rangle
-2iB_{0}L_{4}\langle\lambda^{a}[U,\partial_{\mu}U^{\dagger}]\rangle\langle MU^{\dagger}+UM^{\dagger}\rangle
\nonumber \\&  & 
 -iB_{0}L_{5}\langle([\lambda^{a},U]\partial_{\mu}U^{\dagger}+\partial_{\mu}U[\lambda_{a},U^{\dagger}])(MU^{\dagger}+UM^{\dagger})\rangle
 \nonumber\\&&
+iL_{9}\langle\lambda^{a}\partial_{\nu}(\partial^{\nu}U\partial_{\mu}U^{\dagger}-\partial_{\mu}U\partial^{\nu}U^{\dagger})\rangle
~.
\end{eqnarray}
In particular, the components of $V^{(2)}_{a\mu}$ in ChPT are (making use of the fact that
$V_{\mu}^{0}=\bar{u}\gamma_{\mu}u+\bar{d}\gamma_{\mu}d+\bar{s}\gamma_{\mu}s=0$
in the meson sector):
\begin{eqnarray}
\bar{u}\gamma^{\mu}d & = & V_{1}^{\mu}+iV_{2}^{\mu}= -iK^{0}\overleftrightarrow{\partial^{\mu}}K^{-}+i\sqrt{2}\pi^{0}\overleftrightarrow{\partial^{\mu}}\pi^{-}+\ldots,\nonumber \\
\bar{u}\gamma^{\mu}s & = & V_{4}^{\mu}+iV_{5}^{\mu}= i\pi^{-}\overleftrightarrow{\partial^{\mu}}\bar{K}^{0}-i\sqrt{\frac{3}{2}}K^{-}\overleftrightarrow{\partial^{\mu}}\eta-\frac{i}{\sqrt{2}}K^{-}\overleftrightarrow{\partial^{\mu}}\pi^{0}+\ldots,\nonumber \\
\bar{d}\gamma^{\mu}s & = & V_{6}^{\mu}+iV_{7}^{\mu}= -iK^{-}\overleftrightarrow{\partial^{\mu}}\pi^{+}+i\sqrt{\frac{3}{2}}\eta\overleftrightarrow{\partial^{\mu}}\bar{K}^{0}-\frac{i}{\sqrt{2}}\pi^{0}\overleftrightarrow{\partial^{\mu}}\bar{K}^{0}+\ldots,\nonumber \\
\bar{u}\gamma^{\mu}u & = & \frac{1}{\sqrt{3}}V_{8}^{\mu}+V_{3}^{\mu}= iK^{-}\overleftrightarrow{\partial^{\mu}}K^{+}+i\pi^{-}\overleftrightarrow{\partial^{\mu}}\pi^{+}+\ldots,\nonumber \\
\bar{d}\gamma^{\mu}d & = & \frac{1}{\sqrt{3}}V_{8}^{\mu}-V_{3}^{\mu}= -iK^{0}\overleftrightarrow{\partial^{\mu}}\bar{K}^{0}-i\pi^{-}\overleftrightarrow{\partial^{\mu}}\pi^{+}+\ldots,\nonumber \\
\bar{s}\gamma^{\mu}s & = & -\frac{2}{\sqrt{3}}V_{8}^{\mu}= iK^{0}\overleftrightarrow{\partial^{\mu}}\bar{K}^{0}-iK^{-}\overleftrightarrow{\partial^{\mu}}K^{+}+\ldots.
\end{eqnarray}

The one-loop ChPT results for the vector form factors are given in Appendix~\ref{sec:FFoneloop}.

\subsection{Unitarization, Dispersive Improvement and Numerical Results}
 
The IAM formula for the unitarized vector form factors can be obtained directly from Eq.~\eqref{eq:IAMSFF} by
the replacements $T_0\rightarrow T_1$ and $F_S\rightarrow \tilde{F}_{V+}$:
\begin{eqnarray}
 \tilde{F}_{V+}&=&\tilde{F}_{V+}^{(0)}+T_1^{(2)}(T_1^{(2)}-T_1^{(4)})^{-1}\tilde{F}_{V+}^{(2)}\nonumber\\
 \Rightarrow F_{V+}&=&F_{V+}^{(0)}+\mathbb{P}^{-1}T_1^{(2)}(T_1^{(2)}-T_1^{(4)})^{-1}\mathbb{P}F_{V+}^{(2)}~.\label{eq:VFFIAM}
\end{eqnarray}
This result is also required to be improved by a dispersion relation. The whole procedure is identical
to that of the scalar form factors, except that now the imaginary parts of the vector form factors that enter
the dispersion integrals should be taken as
\begin{equation}
 \mathrm{Im}F^{[i]}_{V+}\equiv(\mathbb{P}^{-1})^*T_1^*\Sigma \mathbb{P}F^{[i-1]}_{V+}~,\label{eq:ImVFFsubs}
\end{equation}
where $i$ is the number of iteration following the unitarity relation of the vector form factors.

\begin{figure}
\begin{center}
\includegraphics[width=0.49\columnwidth]{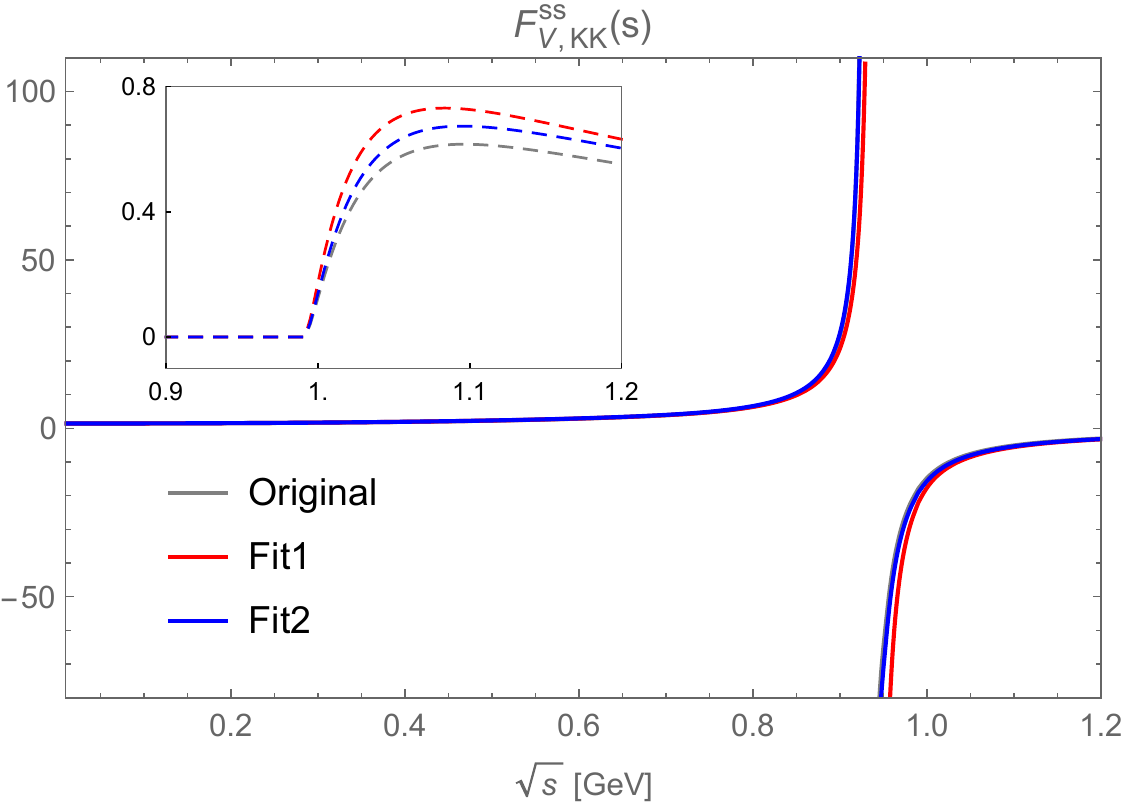} 
\caption{Real and imaginary parts of  $F_{V+,K\bar{K}}^{\bar{s}s}$, respectively.
 The notations are the same as that of 
  Fig.~\ref{fig:Sn} but here the imaginary parts are presented in the inset panel.}
\label{fig:Vsbars} 
\end{center}
\end{figure}

\begin{figure}
\begin{center}
\includegraphics[width=0.49\columnwidth]{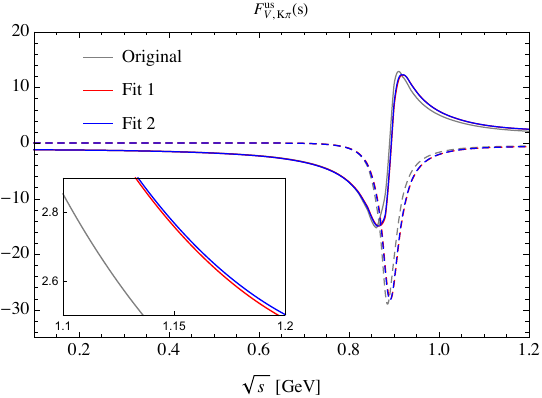} 
\includegraphics[width=0.49\columnwidth]{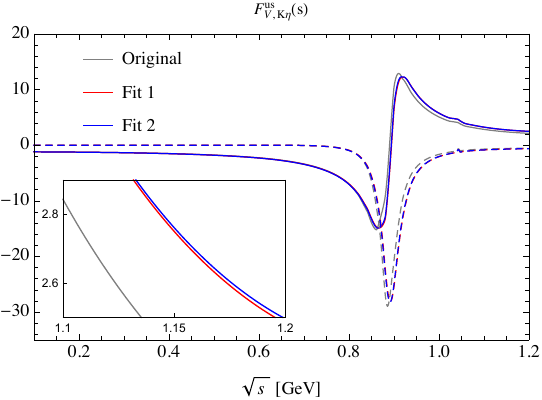} 
\caption{Real and imaginary parts of  $F_{V+,K\pi}^{\bar{u}s}$ and $F_{V+,K\eta}^{\bar{u}s}$, respectively.
  For notations, see Fig.~\ref{fig:Sn}.
  The insets show that the Fit 1 and Fit 2 curves are very close to each other.}
\label{fig:Vubars} 
\end{center}
\end{figure}

\begin{figure}
\begin{center}
\includegraphics[width=0.49\columnwidth]{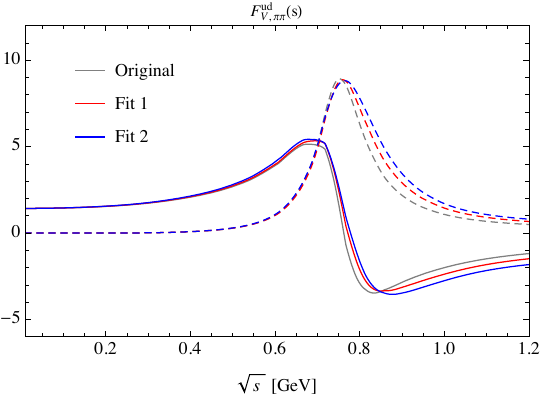} 
\includegraphics[width=0.49\columnwidth]{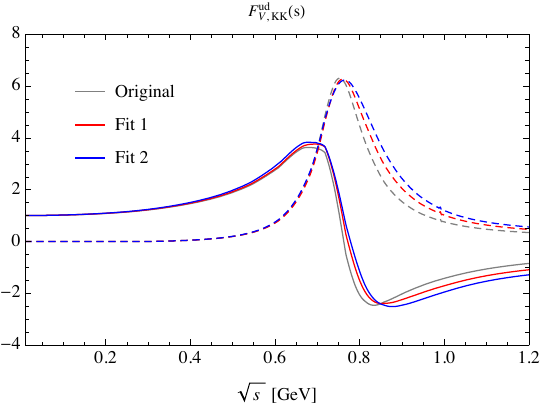} 
\caption{Real and imaginary parts of  $F_{V+,\pi\pi}^{\bar{u}d}$ and $F_{V+,K\bar{K}}^{\bar{u}d}$, respectively.
  For notations, see Fig.~\ref{fig:Sn}.
}
\label{fig:Vubard} 
\end{center}
\end{figure}
Our final results for the vector form factors are summarized in Figs.~\ref{fig:Vsbars}--\ref{fig:Vubard}. In the
$\bar{s}s$ channel, there is a pole on the real axis below the $K\bar K$ threshold, which corresponds to the $\omega_8$ resonance as pointed out in Ref.~\cite{GomezNicola:2001as}. Its mass is between the masses of $\omega(782)$ and $\phi(1020)$, which is reasonable since the $\omega_8$ is a mixture of $\omega$ and $\phi$. Note that for this channel we have removed the unphysical left-hand cuts of the $\pi\pi$ and $\pi\eta$ loops in $T_{I=0,J=1}^{(4)}$ as we did for the $I=1$, $J=1$ channel in Sect.~\ref{sec:unitarity}. Since we do not consider $3\pi$ intermediate states here, this sub-threshold resonance must be a bound state with zero width. Keeping the left-hand cuts of the scattering amplitude results in an unphysical imaginary part of the form factor below the $K\bar K$ threshold, which is due to the on-shell approximation of the IAM, see Sect.~\ref{sec:unitarity}. On the other hand, due to the absence of Adler zeros 
and with the unphysical imaginary parts cut off, further improvement by dispersive iteration is unnecessary for this channel. 

In the $\bar{u}s$ channel, it turns out that the IAM result is almost invariant under
the dispersive treatment; the imaginary parts of the $K\pi$ and $K\eta$ form factors peak
around $\sqrt{s}\approx 0.89\ \text{GeV}$, reflecting the existence of the $K^*(892)$ resonance. 

As explained in Sect.~\ref{sec:unitarity}, there is a breaking of the unitarity relation below the highest
threshold due to the mixing between the left-hand and right-hand cuts using the coupled-channel IAM as
in Ref.~\cite{GomezNicola:2001as}. Such a violation is especially serious for the $I=1,J=1$ channel,
see Fig.~\ref{fig:uni_nomod}, and thus must happen as well when calculating  the $u\bar{d}$ vector form factors.
It is found that only after the modification proposed in Sect.~\ref{sec:unitarity}, with the imaginary part
of the troublesome $t$- and $u$-channel loops removed, the iteration procedure for  the $u\bar{d}$ vector
form factors converges. The result is shown in Fig.~\ref{fig:Vubard}, and agrees well with the
existing literature (see the discussion below).

\begin{figure}
\includegraphics[width=0.49\columnwidth]{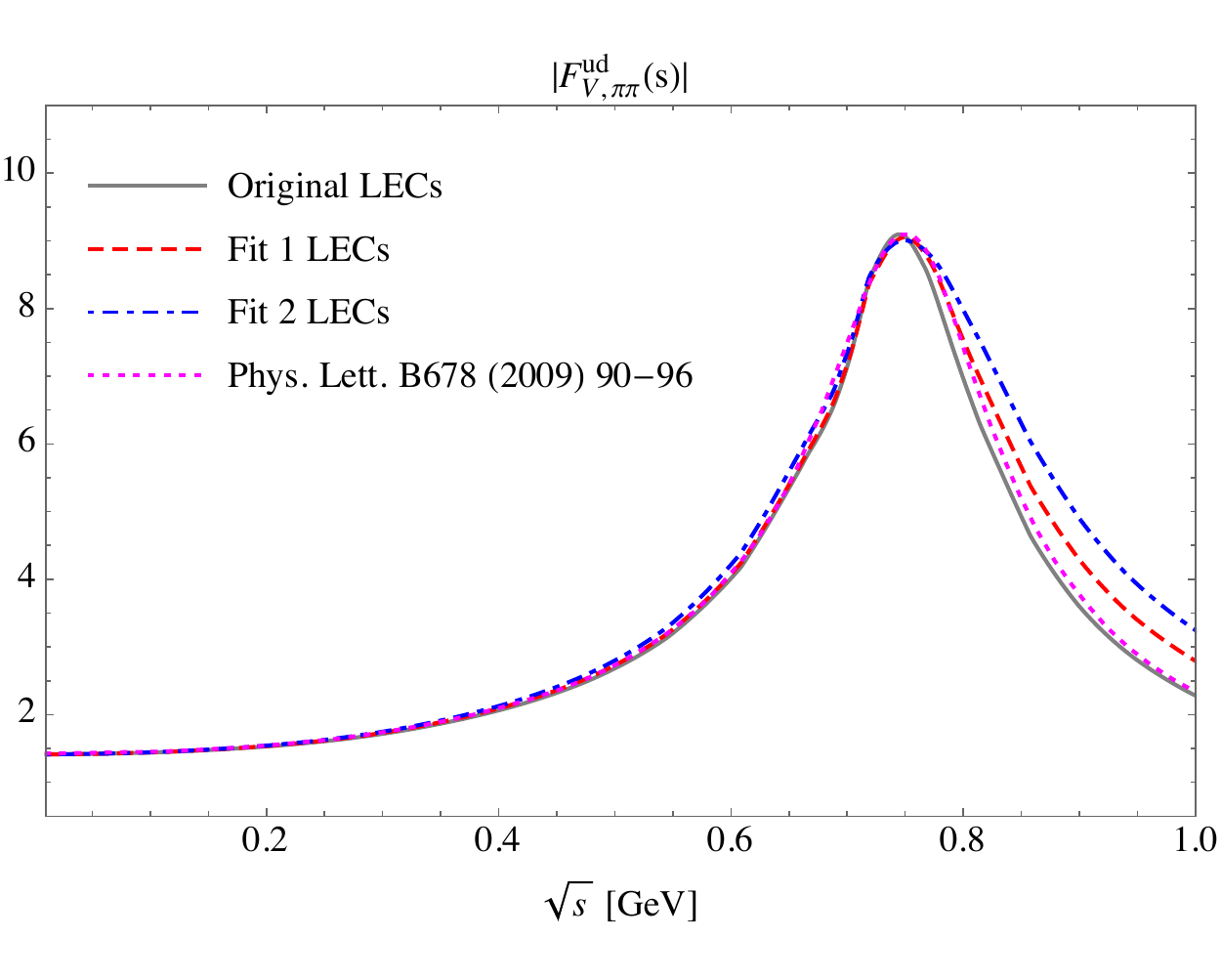} 
\includegraphics[width=0.49\columnwidth]{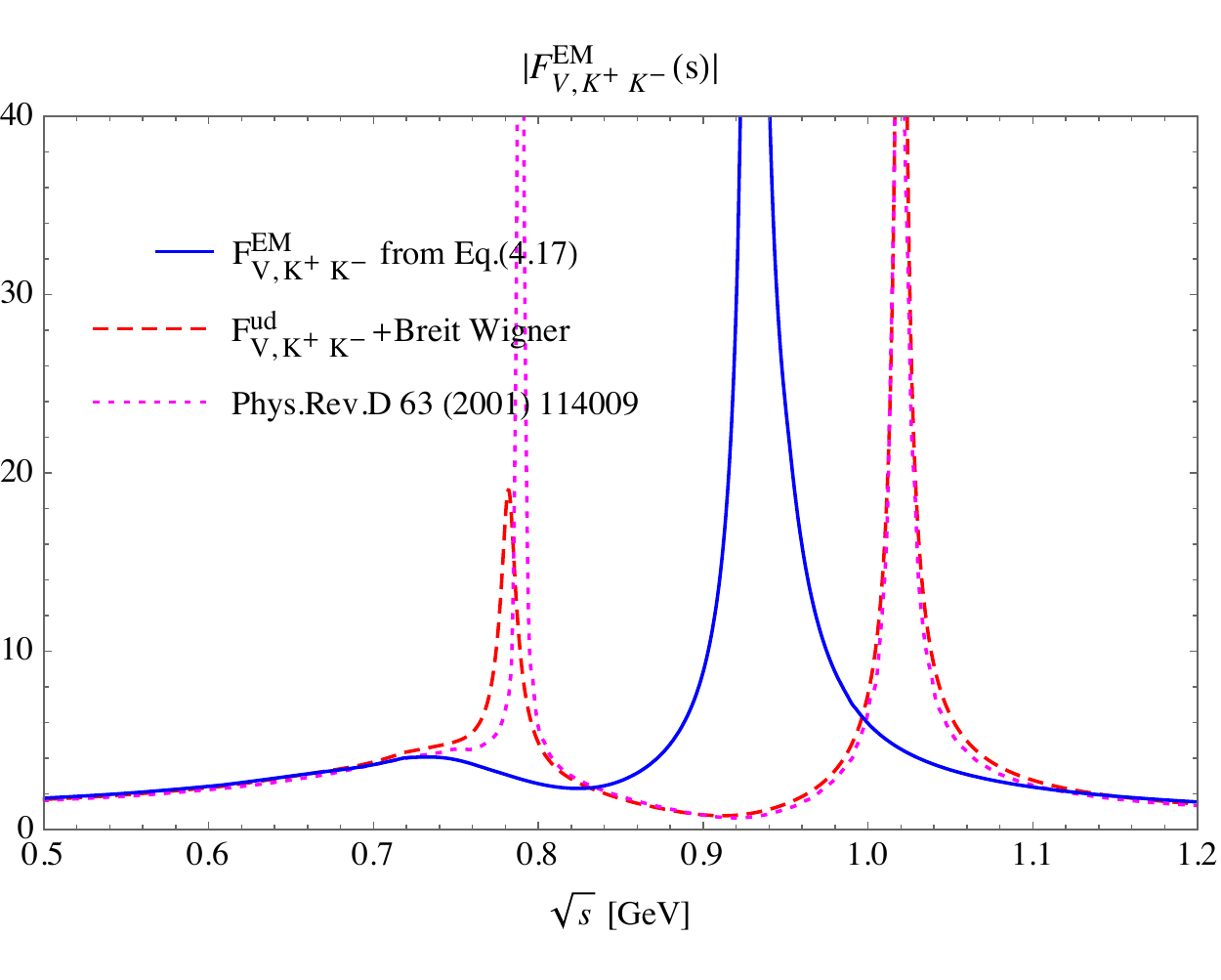} 
\caption{Comparison between the $\pi\pi$ vector form factor derived in this work and that from the Omn\`{e}s
  representation~\cite{Guo:2008nc} (left), comparison between the $K^{+}K^{-}$ EM form factor derived in this work
  and that from an earlier uChPT work~\cite{Oller:2000ug} (right). }
\label{fig:CompareWithOm} 
\end{figure}

The $\pi\pi$ vector form factor is calculated in  Ref.~\cite{Guo:2008nc} through the Omn\`{e}s
representation
\begin{align}
  F(s)=\exp \left(P_{1} s+\frac{s^{2}}{\pi} \int_{4 M_{\pi}^{2}}^{\infty} ds'
  \frac{\delta_{1}^1(s')}{(s')^{2}(s'-s-i \epsilon)}\right),
\end{align}
where $P_{1}=\left\langle r^{2}\right\rangle / 6$ with $\left\langle r^{2}\right\rangle$ the pion radius
squared (see Refs.~\cite{Hanhart:2016pcd,Ananthanarayan:2017efc,Colangelo:2018mtw} for recent determinations of the pion charge radius from data), and a twice-subtracted dispersion relation is applied. The Omn\`{e}s representation
relates the form factor with the corresponding phase shift $\delta_{1}^1(s)$, which can be extracted in a rough approximation from
the mass and decay width of the $\rho$ meson using a Breit--Wigner parametrization. The scattering
amplitude dominated by the $s$-channel $\rho$ resonance reads
\begin{align}
  a_1^1(s)=\frac{c}{s-M_{\rho}^{2}-i M_{\rho} \Gamma_{\mathrm{tot}}(s)}~,\ \ \ \Gamma_{\mathrm{tot}}
  =\frac{g_{\rho \pi \pi}^{2} p^{3}}{6 \pi M_{\rho}^{2}}=\frac{g_{\rho \pi \pi}^{2}\left(\frac{s}{4}
    -M_{\pi}^{2}\right)^{3 / 2}}{6 \pi M_{\rho}^{2}}~,
\end{align}
where $c$ is an irrelevant constant. The phase shift is derived as
\begin{align}
  \delta_1^1(s)=\arctan \frac{\operatorname{Im} a_1^1(s)}{\operatorname{Re} a_1^1(s)}
  =\arctan \frac{M_{\rho} \Gamma_{\operatorname{tot}}(s)}{M_{\rho}^{2}-s}~.
\end{align}
The left panel in Fig.~\ref{fig:CompareWithOm} 
shows a comparison between the $\pi\pi$ vector
form factor derived in this work and that from the Omn\`{e}s representation, and we observe good agreement. In the right panel, instead of $F_{V,K^{+}K^{-}}^{\bar u d}$ we compare
our result for the kaon electromagnetic (EM) form factor with that from the earlier literature also using uChPT~\cite{Oller:2000ug}, since the explicit result for $F_{V,K^{+}K^{-}}^{\bar u d}$ is not available from that source. The EM form factor is defined as
\begin{equation}
  \left\langle K^{+}(p)K^{-}(p^{\prime})\right|\sum_{i}e_i\bar{q}_i'\gamma^{\mu}A_{\mu}q_i\left|\gamma\right\rangle
  \equiv F_{V,\ K^{+}K^{-}}^{EM}(s)\epsilon\cdot(p-p^{\prime}),
\end{equation}
where $i=u,d,s$ and $e_i=2/3,-1/3,-1/3$.
Generally, due to isospin symmetry, our vector form factors derived above are related to this EM form factor according to
\begin{eqnarray}
F_{V,K^{+}K^{-}}^{EM}(s)=\frac{1}{2}F_{V+,K\bar{K}}^{\bar{u}d}(s)+\frac{1}{3\sqrt{2}}F_{V+,K\bar{K}}^{\bar{s}s}(s)-\frac{1}{6\sqrt{2}}F_{V+,K\bar{K}}^{\bar{u}u+\bar{d}d}(s).\label{EMFF1}
\end{eqnarray}
Note that due to charge conservation we must have $F_{V,K^{+}K^{-}}^{EM}(0)=1$, which can be checked by our results
(using the U(1) symmetry constraint: $\bar u\gamma_{\mu}u+\bar d\gamma_{\mu}d=-\bar s\gamma_{\mu}s$). The EM form factor from Eq.~\eqref{EMFF1} is shown by the blue solid line in the right panel of Fig.~\ref{fig:CompareWithOm}, where the divergent bound state peak around 935~${\rm MeV}$ corresponds to the $\omega_8$.
However, in Ref.~\cite{Oller:2000ug}, instead of the $\omega_8$ two physical resonances $\omega$ and $\phi$ were introduced manually, which are mixtures of $\omega_1$ and $\omega_8$. Thus, to reproduce these two distinct resonances for better comparison, we replace the last two terms in Eq.~(\ref{EMFF1}) with the standard Breit--Wigner distributions of the $\omega$ and $\phi$ :
\begin{equation}
  \frac{1}{\sqrt{2}}\frac{g}{g_{\phi}}\frac{M_{\phi}^{2}}{s-M_{\phi}^{2}+i\ M_{\phi}\Gamma_{\phi}}
  -\frac{1}{2}\frac{g}{g_{\omega}}\frac{M_{\omega}^{2}}{s-M_{\omega}^{2}+i\ M_{\omega}\Gamma_{\omega}},
\end{equation}
where $g$ is the SU(3)-symmetric vector-to-two-pseudoscalars coupling constant, and $g_{\phi}$, $g_{\omega}$ refer to the coupling constants of $\phi$ and $\omega$ to the  electromagnetic
current. Their explicit definitions  can be found in Refs.~\cite{Meissner:1987ge,Klingl:1996by},
where  $g_{\phi}=-12.89,\ g_{\omega}=17.05$ and $g=6.05$. 
The decay widths are taken as $\Gamma_{\phi}=4$~MeV and
$\Gamma_{\omega}=8.5$~MeV. The corresponding curve is the red dashed line shown in the right plot of Fig.~\ref{fig:CompareWithOm}. Finally, the magenta dotted line in the right panel is the EM form factor extracted from Fig.~4 of Ref.~\cite{Oller:2000ug}. From this comparison, we find that regardless the region where the un-mixed resonant $\omega_8$ or the physical mixtures $\omega$ and $\phi$ emerge, the agreement is good at both low (below 0.7~GeV) and high (above 1.1~GeV) energies. 

\subsection{Applications of the Vector Form Factors}
 
We end this section by discussing an application of the two-meson vector form factors
in the two-body hadronic decays of a charged lepton $l\to\phi\phi'\nu$. The leading contribution
to $l\to\phi\phi'\nu$ is due to a single exchange of a $W$-boson, which, at low energies, can
be approximated by the Fermi interaction. The corresponding amplitude is
\begin{equation}
  iM=-\frac{iG_F}{\sqrt{2}}V_{qq'}^*\bar{u}_\nu\gamma^\mu(1-\gamma_5)u_l\left\langle \phi\phi'
  |\bar{q}'\gamma_\mu q|0\right\rangle,
\end{equation} 
where $G_F$ is Fermi's constant and $V_{qq'}$ is the Cabibbo--Kobayashi--Maskawa (CKM) matrix element. Since
the axial vector component
in the hadronic matrix element vanishes due to parity, the matrix element can be expressed in terms of
the vector form factors defined in Eq.~\eqref{eq:vectorFFdef}. Considering only the spin-averaged decay,
the differential decay width is a function of two kinematic variables. They can be chosen as
$s=(p_\phi+p_{\phi'})^2$ and $\theta$, which is the angle between $\vec{p}_\phi$ and $\vec{p}_l$ in the c.m.\
frame of $\phi\phi'$. After carrying out the phase space integration, one obtains
\begin{equation}
  \frac{d\Gamma}{ds\:d\cos\theta}=G_F^2|V_{qq'}|^2\frac{(m_l^2-s)|\vec{p}_\phi|}{(8\pi)^3m_l^3\sqrt{s}}\left[A(s)
    +\frac{|\vec{p}_\phi|}{\sqrt{s}}B(s)\cos\theta+\frac{|\vec{p}_\phi|^2}{s}C(s)\cos^2\theta\right],
\end{equation}   
where the $\phi\phi'$ subscript and $\bar{q}'q$ superscript in the form factors have been suppressed
for notational simplicity, and
\begin{eqnarray}
  A(s)&\equiv&\frac{2(m_l^2-s)}{s^2}\Big[m_l^2s\big(s|F_{V-}|^2+2(M_\phi^2-M_{\phi'}^2)\mathrm{Re}\{F_{V+}^*F_{V-}\}\big)
    \nonumber\\
    &&+|F_{V+}|^2\left(M_\phi^4(m_l^2+s)-2m_\phi^2(M_{\phi'}^2m_l^2+M_{\phi'}^2s+s^2)+M_{\phi'}^4(m_l^2+s)-2M_{\phi'}^2s^2+s^3\right)\Big],
  \nonumber\\
  B(s)&=&-\frac{8m_l^2(m_l^2-s)}{s}\left(|F_{V+}|^2(M_\phi^2-M_{\phi'}^2)+s\,\mathrm{Re}\{F_{V+}^*F_{V-}\}\right),
  \nonumber\\
  C(s)&=&8|F_{V+}|^2(m_l^2-s)^2.
\end{eqnarray}

Two-body hadronic decay of charged leptons is one of the commonly studied processes in the extraction of the
CKM matrix elements, for example  $\tau\to K\pi\nu_\tau$ for $V_{us}$~\cite{Amhis:2019ckw}. Therefore,
an improved understanding of the $s$-dependence in the vector form factors may improve on the $V_{us}$
precision~\cite{Bernard:2011ae} and lead to a better reconciliation of the same quantity measured in other processes such as
the kaon leptonic/semi-leptonic decays.

 \section{Tensor form factors} 
\label{sec:5}

Next, we study the tensor form factors of a two-meson system, which were so far only investigated in limited channels.
We define the tensor form factors through the following matrix elements:
\begin{equation}
  \label{eq:tensorFFdef}\left\langle a_{i}(p_{a_{i}})b_{i}(p_{b_{i}})\right|\bar{q}'\sigma^{\mu\nu}q\left|0\right\rangle
  \equiv i\frac{\Lambda_2}{F_\pi^2}(p_{a_{i}}^{\mu}p_{b_{i}}^{\nu}-p_{a_{i}}^{\nu}p_{b_{i}}^{\mu})F_{T,i}^{\bar{q}'q}(s)~,
 \end{equation}
where $\Lambda_2$ is an LEC that appears when introducing external tensor sources to the chiral
Lagrangian, which we shall discuss later. The unitarity relation obeyed by the tensor form factor is
identical with that of the vector form factor $F_{V+}$~\cite{Hoferichter:2018zwu}:
\begin{equation}
\mathrm{Im}F^{\bar{q}'q}_{T}=(\mathbb{P}^{-1})^*T_1^*\Sigma \mathbb{P}F^{\bar{q}'q}_{T}~.\label{eq:TFFunitarityEq}
\end{equation}
Again, it is associated with the $J=1$ partial-wave amplitude. 
 
 \subsection{ChPT Result}
 
The derivation of tensor currents in ChPT requires the introduction of an antisymmetric Hermitian tensor
source $\bar{t}_{\mu\nu}$ into the QCD Lagrangian:
\begin{equation}
\mathcal{L}=\bar{q}\sigma^{\mu\nu}\bar{t}_{\mu\nu}q~.
\end{equation}
The corresponding effective field theory was first investigated in Ref.~\cite{Cata:2007ns}.
The LO chiral Lagrangian coupled to the tensor source scales as ${\cal O}(p^4)$ and is given by
\begin{eqnarray}
 {\cal L}_{T}^{(4)} = -i \Lambda_{2} \langle t_{+}^{\mu\nu} u_{\mu} u_{\nu}\rangle~,
 \end{eqnarray}
where 
 \begin{eqnarray}
  u_{\mu} =  i \left[ u^{\dagger} (\partial_{\mu} -ir_{\mu}) u - u(\partial_{\mu} -il_{\mu}) u^{\dagger}\right]~,~~~
  t_{\pm}^{\mu\nu} = u^{\dagger } t^{\mu\nu} u^{\dagger } \pm u\ t^{\mu\nu\dagger } u~.
 \end{eqnarray}
$t^{\mu\nu}$ and  $t^{\mu\nu\dagger}$ are given as
\begin{eqnarray} 
 t^{\mu\nu} = \frac{1}{4} \big(g^{\mu\lambda} g^{\nu\rho} - g^{\nu\lambda}g^{\mu\rho} - i\epsilon^{\mu\nu\lambda\rho}\big) \bar t_{\lambda\rho}~, \nonumber\\
 t^{\mu\nu\dagger} = \frac{1}{4} \big(g^{\mu\lambda} g^{\nu\rho} - g^{\nu\lambda}g^{\mu\rho} + i\epsilon^{\mu\nu\lambda\rho}\big) \bar t_{\lambda\rho}~,
\end{eqnarray}
and the convention $\epsilon^{0123}=1$ has been used for the Levi-Civita tensor. 
 
At NLO, there exist quite a number of corresponding operators, among which the ones contributing to
tensor form factors are given as 
 \begin{eqnarray}
 {\cal L}_{T}^{(6)} &=& i C_{34} \langle t_{+}^{\mu\nu}\{\chi_{+}, u_{\mu} u_{\nu}\}\rangle+i C_{35} \langle t_{+}^{\mu\nu} u_{\mu}\chi_{+}u_{\nu}\rangle+i C_{36} \langle \chi_{+}\rangle\langle t_{+}^{\mu\nu} u_{\mu} u_{\nu}\rangle \nonumber\\
 && +i C_{37} \langle t_{+}^{\mu\nu} \rangle \langle\chi_{+}u_{\mu} u_{\nu}\rangle + i C_{88} \langle\nabla^{\rho} t_{+}^{\mu\nu} [h_{\mu\rho}, u_{\nu}]\rangle+ i C_{89} \langle\nabla_{\mu} t_{+}^{\mu\nu} [h_{\nu\rho}, u^{\rho}]\rangle\nonumber\\
 && + C_{106} \langle t_{+}^{\mu\nu} [\chi_{-\mu}, u_{\nu}]\rangle + iC_{107} \langle t_{+_{\mu\nu}} h^{\mu\alpha}h^{\nu}_{\:\:\alpha}\rangle,
 \end{eqnarray}
with the covariant derivative defined as 
  \begin{equation}
 \nabla_{\mu} X=\partial_{\mu} X+\left[\Gamma_{\mu}, X\right],~~~~\Gamma_{\mu}=\frac{1}{2}\left(u^{\dagger} \partial_{\mu} u+u \partial_{\mu} u^{\dagger}\right).
 \end{equation}
They are used to cancel the divergence that occurs in the one-loop corrections to the tensor form factors.
The renormalized LECs $C_i^r$ and divergence coefficients $\gamma_i^T$ are defined as:
\begin{equation}
  C_i^r=C_i-\frac{\Lambda_2}{F_0^2}\lambda\gamma_i^T.
\end{equation}
In fact, it turns out that the requirement to cancel all divergences in tensor form factors does not fix
all the $\{\gamma_i^T\}$ independently, but only a subset of them, which are determined to be the following
constraints:
\begin{align}
&&\gamma_{34}^T+\frac{1}{2}\left(\gamma_{89}^T+\gamma_{106}^T+\gamma_{107}^T\right)&=-\frac{1}{24}, & 
 \gamma_{35}^T+\gamma_{89}^T+\gamma_{106}^T+\gamma_{107}^T&=-\frac{3}{4},\nonumber\\
 &&\gamma_{36}^T=-\frac{11}{18}, \qquad\qquad  \gamma_{37}^T&=-\frac{1}{2}, &\gamma_{88}^T-\gamma_{89}^T-\gamma_{107}^T&=-\frac{1}{4}.
\end{align}

The numerical values of $\Lambda_2$ and $C_i^r$ at a given renormalization scale $\mu$ are obviously
required to make definite predictions on tensor form factors. Unfortunately, as far as we know, no lattice data
are available for these LECs. We therefore make use of the results in Ref.~\cite{Jiang:2012ir} that
attempted to evaluate the effective action from first principles (under certain uncontrollable approximations).
However, a critical issue in that approach is that one cannot study the scale dependence of the renormalized
LECs, and therefore the issue of how one should match their results to the renormalized LECs with the standard
Gasser--Leutwyler subtraction scheme at a given scale, say $\mu=M_\rho$, remains ambiguous. To account for this
issue, we assume that the LECs in Ref.~\cite{Jiang:2012ir} are given at some unknown scale
$\tilde{\mu}$, and could be run to $\mu=M_\rho$ by a renormalization group (RG) running
\begin{equation}
C_i^r(M_\rho)=C_i^r(\tilde{\mu})-\frac{\gamma_i^T}{32\pi^2}\frac{\Lambda_2}{F_0^2}\ln\frac{M_\rho^2}{\tilde{\mu}^2}~.
\end{equation}
which still involves an unknown coefficient $\ln(M_\rho^2/\tilde{\mu}^2)$. To fix this coefficient,
we refer to Ref.~\cite{Jiang:2015dba}, in which the $\mathcal{O}(p^4)$ LECs $L_i$ are calculated within the same
formalism. We perform a RG running of those LECs and compare them to $\{L_i^r\}$ at $\mu=M_\rho$ that
are fitted to experimental data~\cite{Bijnens:2014lea}. That allows us to get a best-fit value of
$\ln(M_\rho^2/\tilde{\mu}^2)$ by minimizing the $\chi^2$. We find that, with this best-fit value, the changes
in the numerical values of $\{L_i\}$ are relatively small. Therefore, for the case of tensor LECs, we
shall simply cite the results in Ref.~\cite{Jiang:2012ir}, assuming that systematic errors due to the
ambiguity in the renormalization scale are much smaller than the other theoretical errors quoted in the paper.
Within such a framework, the coefficients $C_{36}^r$ and $C_{37}^r$ vanish in the large-$N_c$ limit, where $N_c$
refers to the number of colors in the QCD Lagrangian,
while other nonzero coefficients are collected in Table~\ref{tab:NLOLECTensor} (readers should be alerted
to certain differences in the definition of operators between Refs.~\cite{Cata:2007ns,Jiang:2012ir}
that lead to changes in numerical values of LECs and have been properly taken into account in
Table~\ref{tab:NLOLECTensor}). 
 
The tensor currents are defined as
\begin{equation}
T_{ij}^{\mu\nu}\equiv\bar{q}_i\sigma^{\mu\nu}q_j.
\end{equation}
Using the $\mathcal{O}(p^4)$ and $\mathcal{O}(p^6)$ chiral Lagrangian with tensor sources, we are able to derive the
tensor currents $T^{(4)}_{\mu\nu}$ and $T^{(6)}_{\mu\nu}$ in ChPT, respectively, which are
\begin{eqnarray}
T_{ij}^{(4)\mu\nu}&=&-\frac{i\Lambda_{2}}{4}\left[u^{\dagger}[u^{\mu},u^{\nu}]u^{\dagger}-i\varepsilon^{\mu\nu\lambda\rho}u^{\dagger}u_{\lambda}u_{\rho}u^{\dagger}+u[u^{\mu},u^{\nu}]u+i\varepsilon^{\mu\nu\lambda\rho}uu_{\lambda}u_{\rho}u\right]_{ji}~,\nonumber\\
T_{ij}^{(6)\mu\nu} & = & \frac{i}{F_0^2}\left(\partial^{[\mu}\phi_{a}\right)\left(\partial^{\nu]}\phi_{b}\right)\biggl[C_{34}\{\chi,\lambda^{a}\lambda^{b}\}+C_{35}\lambda^{a}\chi\lambda^{b}+C_{36}\langle\chi\rangle\lambda^{a}\lambda^{b}\nonumber \\
&  & \left.+C_{37}\langle\chi\lambda^{a}\lambda^{b}\rangle+\frac{1}{2}C_{106}[\{\chi,\lambda^{a}\},\lambda^{b}]\right]_{ji} +\frac{2i}{F_0^2}C_{107}\left(\partial^{\alpha}\partial^{[\mu}\phi_{a}\right)\left(\partial^{\nu]}\partial_{\alpha}\phi_{b}\right)\left(\lambda^{a}\lambda^{b}\right)_{ji}\nonumber \\
&  & -\frac{i}{F_0^2}[\lambda^{a},\lambda^{b}]_{ji}\left[C_{88}\partial^{\rho}\left((\partial^{[\mu}\partial_{\rho}\phi_{a})\partial^{\nu]}\phi_{b}\right)+C_{89}\partial^{[\mu}\left((\partial^{\nu]}\partial_{\rho}\phi_{a})\partial^{\rho}\phi_{b}\right)\right]+O(\phi^{3})~,\nonumber\\
\end{eqnarray}
where $A^{[\mu}B^{\nu]}\equiv A^{\mu}B^{\nu}-A^{\nu}B^{\mu}$. In particular, the components of $T^{(4)\mu\nu}$ read
\begin{eqnarray}
\bar{u}\sigma_{\mu\nu}u & = & \frac{i\Lambda_{2}}{F_0^2}(-\partial_{\mu}K^{+}\partial_{\nu}K^{-}+\partial_{\mu}K^{-}\partial_{\nu}K^{+}-\partial_{\mu}\pi^{+}\partial_{\nu}\pi^{-}+\partial_{\mu}\pi^{-}\partial_{\nu}\pi^{+})+\ldots,\nonumber \\
\bar{d}\sigma_{\mu\nu}d & = & \frac{i\Lambda_{2}}{F_0^2}(\partial_{\mu}\bar{K}^{0}\partial_{\nu}K^{0}-\partial_{\mu}K^{0}\partial_{\nu}\bar{K}^{0}+\partial_{\mu}\pi^{+}\partial_{\nu}\pi^{-}-\partial_{\mu}\pi^{-}\partial_{\nu}\pi^{+})+\ldots,\nonumber \\
\bar{s}\sigma_{\mu\nu}s & = & \frac{i\Lambda_{2}}{F_0^2}(-\partial_{\mu}\bar{K}^{0}\partial_{\nu}K^{0}+\partial_{\mu}K^{0}\partial_{\nu}\bar{K}^{0}+\partial_{\mu}K^{+}\partial_{\nu}K^{-}-\partial_{\mu}K^{-}\partial_{\nu}K^{+})+\ldots,\nonumber \\
\bar{u}\sigma_{\mu\nu}d & = & \frac{i\Lambda_{2}}{F_0^2}(\partial_{\mu}K^{-}\partial_{\nu}K^{0}-\partial_{\mu}K^{0}\partial_{\nu}K^{-}+\sqrt{2}\partial_{\mu}\pi^{0}\partial_{\nu}\pi^{-}-\sqrt{2}\partial_{\mu}\pi^{-}\partial_{\nu}\pi^{0})+\ldots,\nonumber \\
\bar{u}\sigma_{\mu\nu}s & = & \frac{i\Lambda_{2}}{2F_0^2}(-2\partial_{\mu}\bar{K}^{0}\partial_{\nu}\pi^{-}+2\partial_{\mu}\pi^{-}\partial_{\nu}\bar{K}^{0}+\sqrt{6}\partial_{\mu}\eta\partial_{\nu}K^{-}+\sqrt{2}\partial_{\mu}\pi^{0}\partial_{\nu}K^{-}\nonumber \\
&  & -\sqrt{6}\partial_{\mu}K^{-}\partial_{\nu}\eta-\sqrt{2}\partial_{\mu}K^{-}\partial_{\nu}\pi^{0})+\ldots,\nonumber \\
\bar{d}\sigma_{\mu\nu}s & = & \frac{i\Lambda_{2}}{2F_0^2}(\sqrt{6}\partial_{\mu}\eta\partial_{\nu}\bar{K}^{0}-\sqrt{2}\partial_{\mu}\pi^{0}\partial_{\nu}\bar{K}^{0}-\sqrt{6}\partial_{\mu}\bar{K}^{0}\partial_{\nu}\eta+\sqrt{2}\partial_{\mu}\bar{K}^{0}\partial_{\nu}\pi^{0}\nonumber \\
&  & +2\partial_{\mu}\pi^{+}\partial_{\nu}K^{-}-2\partial_{\mu}K^{-}\partial_{\nu}\pi^{+})+\ldots.
\end{eqnarray}

The one-loop ChPT results for the tensor form factors are given similarly in Appendix~\ref{sec:FFoneloop}.

\begin{table}
\begin{center}
  \caption{Low-energy constants for the chiral Lagrangian with tensor sources derived from Ref.~\cite{Jiang:2012ir}.
    $\Lambda_2$ is given in units of $10^{-3}b_0$ while the remainders are given in units of $10^{-3}\text{GeV}^{-2}b_0$,
    where $b_0=1.32\ \text{GeV}$. The renormalization scale is assumed to be $\mu=M_\rho$ (see the discussion in the text).}
  \label{tab:NLOLECTensor}
\begin{tabular}{ccccccc}
\hline \hline 
	$\Lambda_{2}$	&   $C_{34}^r$ &  $C_{35}^r$ &  $C_{88}^r$&  $C_{89}^r$ &  $C_{106}^r$ &  $C_{107}^r$  \\  \hline
  $13.79$ 	&  $0.01$ &       $-4.14$    	& $-1.44$ 	& $10.26$	& $-9.04$ & $-0.09$		 \\
\hline
\end{tabular}
\end{center}
\end{table}

 \subsection{Unitarization, Dispersive Improvement and Numerical Results}
 
\begin{figure}
\includegraphics[width=0.49\columnwidth]{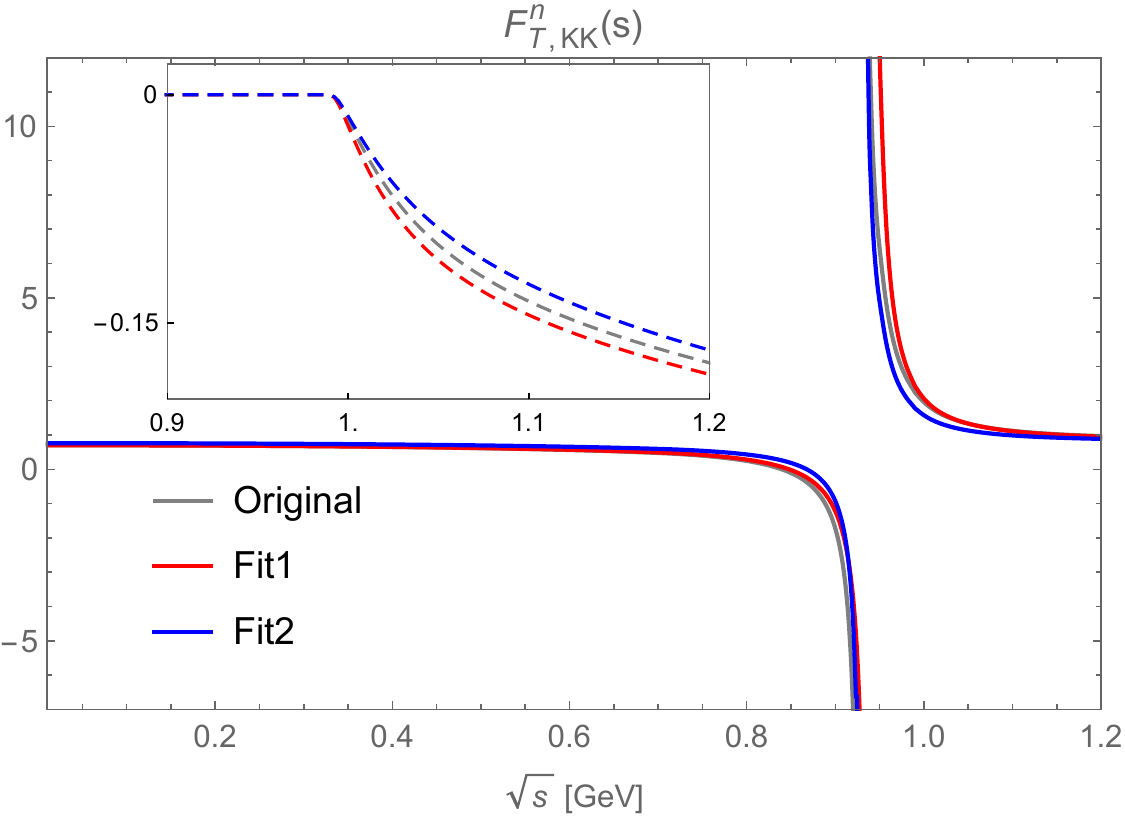}
\includegraphics[width=0.49\columnwidth]{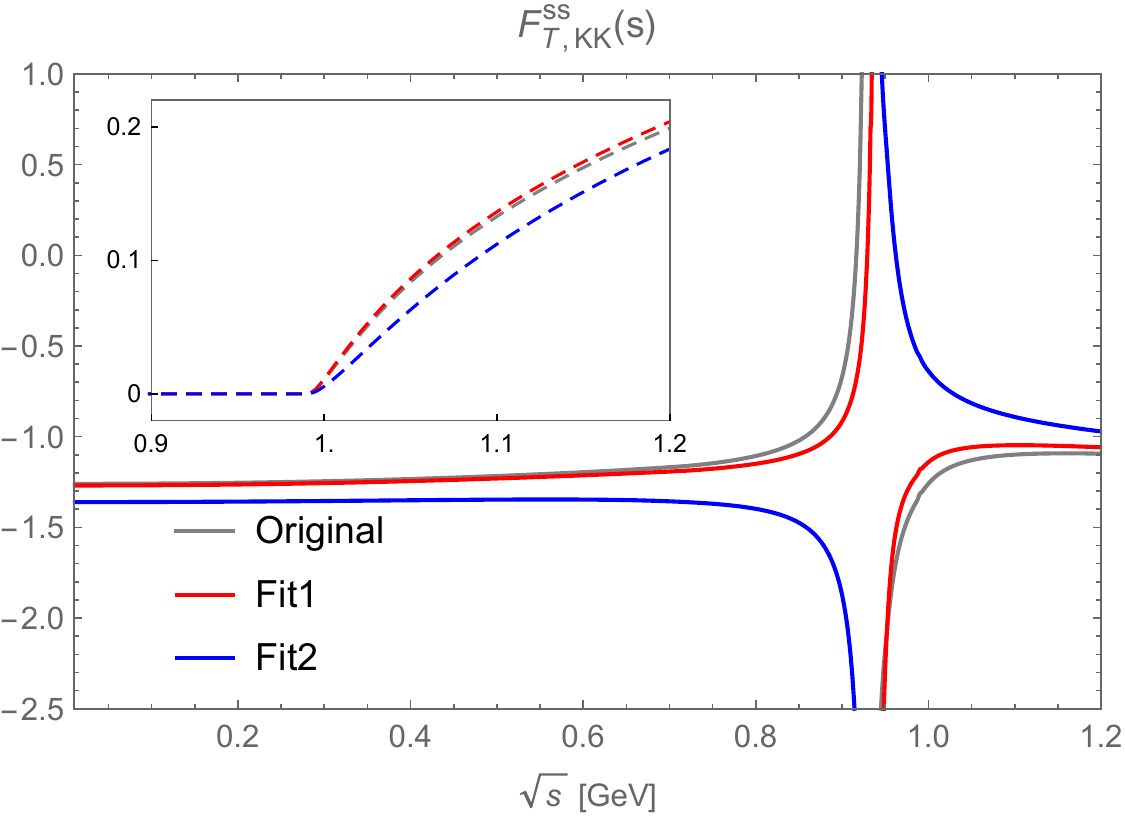} 
\caption{Real and imaginary parts of  $F_{T-,K\bar{K}}^{n},$ respectively (left), and real
  and imaginary parts of  $F_{T-,K\bar{K}}^{\bar s s}$, respectively (right).
  For notations, see Fig.~\ref{fig:Vsbars}.}
\label{fig:Tn} 
\end{figure}

\begin{figure}
\includegraphics[width=0.49\columnwidth]{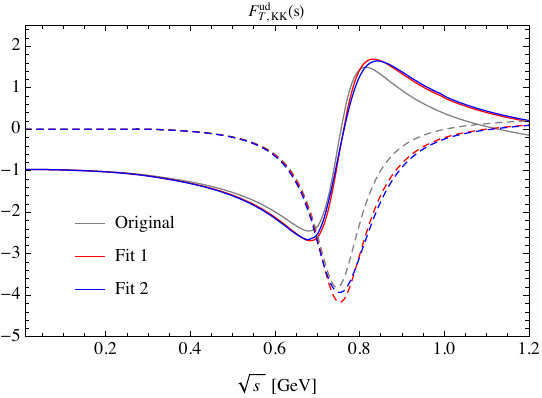} 
\includegraphics[width=0.49\columnwidth]{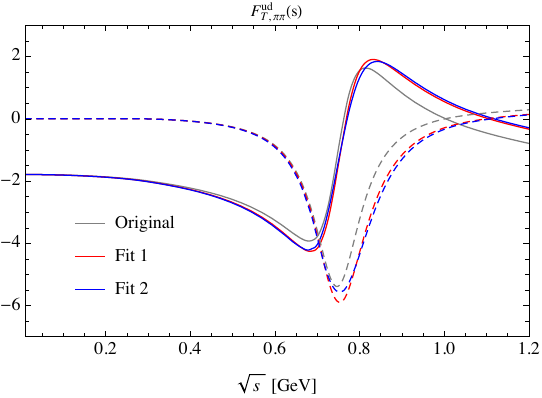} 
\caption{Real and imaginary parts of  $F_{T-,KK}^{\bar{u}d}$ and $F_{T-,\pi\pi}^{\bar{u}d}$, respectively.
  For notations, see Fig.~\ref{fig:Sn}.}  
\label{fig:Tubard} 
\end{figure}

\begin{figure}
\includegraphics[width=0.49\columnwidth]{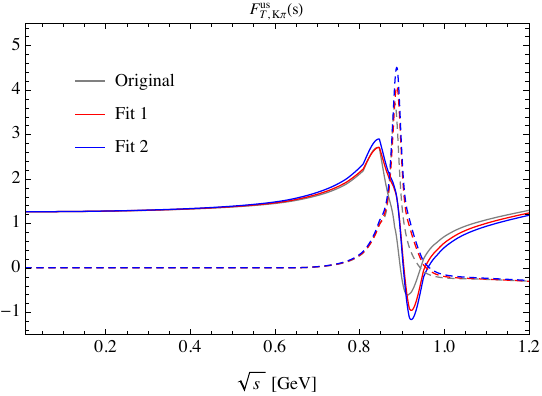} 
\includegraphics[width=0.49\columnwidth]{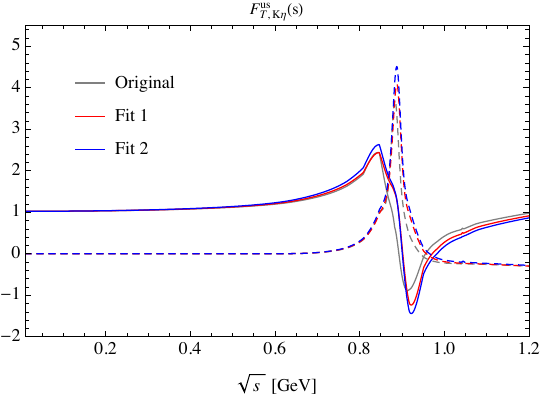} 
\caption{Real and imaginary parts of  $F_{T-,K\pi}^{\bar{u}s}$ and $F_{T-,K\eta}^{\bar{u}s}$, respectively.
  For notations, see Fig.~\ref{fig:Sn}.} 
\label{fig:Tubars} 
\end{figure}

The unitarity relation for tensor form factors $F_{T}$ is identical to that of vector form factors,
so their IAM formulae should also take the same form
\begin{eqnarray}
  F_{T}&=&F_{T}^{(0)}+\mathbb{P}^{-1}T_1^{(2)}(T_1^{(2)}-T_1^{(4)})^{-1}\mathbb{P}F_{T}^{(2)}.
  \label{eq:TFFIAM}
\end{eqnarray}
The results are given in Figs.~\ref{fig:Tn}--\ref{fig:Tubars}, where due to the same reason as that for $F_V^{\bar s s}$,
$F_T^{n}$ and $F_T^{\bar s s}$ are also calculated by cutting off the left-hand cuts and are not improved by dispersive iteration. The iteration procedures for $F_T^{\bar u d}$
and $F_T^{\bar u s}$ are exactly the same as the one for the corresponding vector form factors, so we shall not repeat all
the details here. 

\begin{figure}
\begin{center}
\includegraphics[width=0.6\columnwidth]{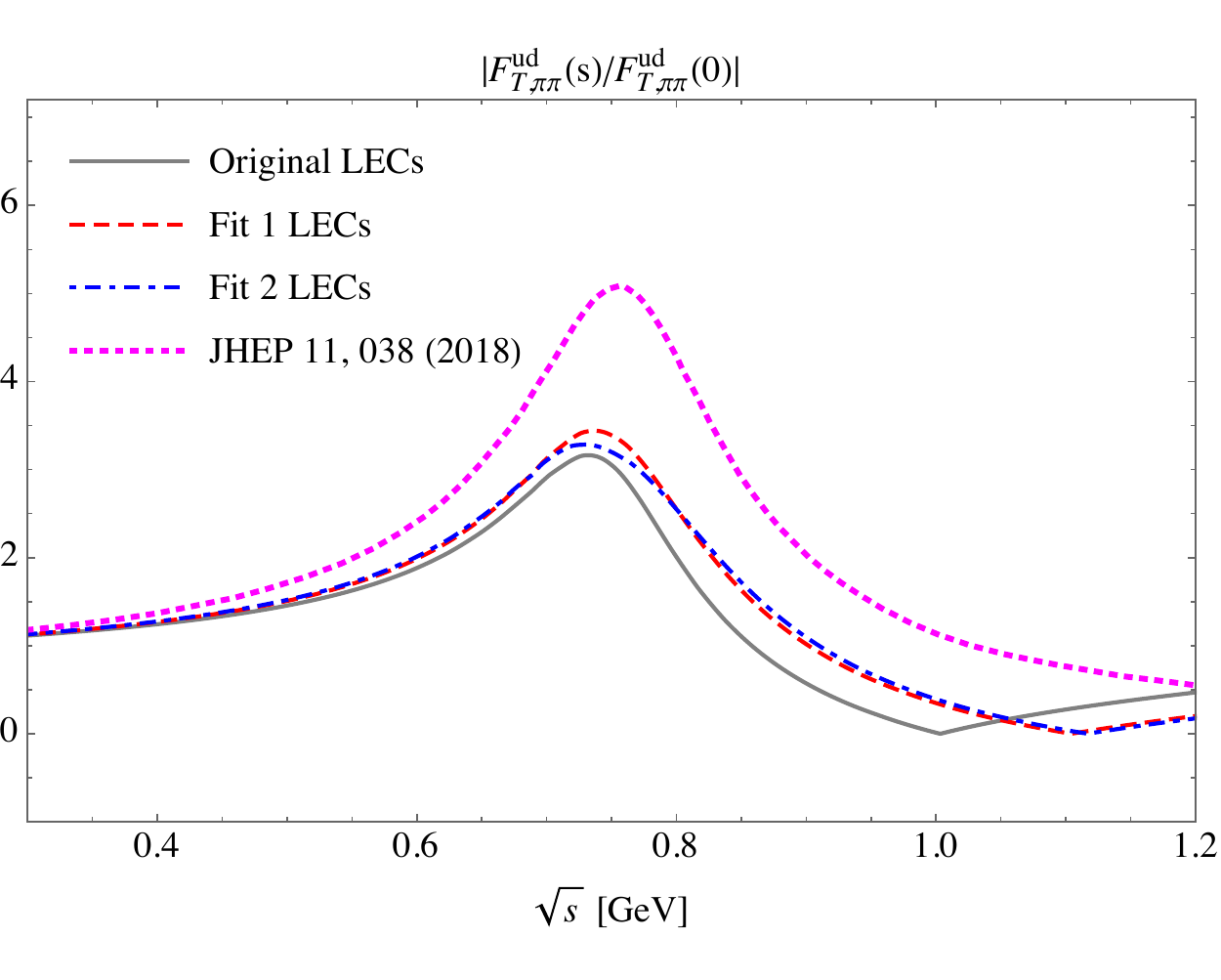} 
\caption{Comparison of $F_{T,\pi\pi}^{\bar u d}$ derived in this work (gray solid, red dashed and blue dot-dashed lines
  corresponding to the three sets of LECs) and that
  from Ref.~\cite{Miranda:2018cpf} by phase dispersive representation (magenta dotted line). }
\label{fig:CompareTFFudpipi} 
\end{center}
\end{figure}

For the form factor $F_{T,\pi\pi}^{ud}$, we compare it with that derived in Ref.~\cite{Miranda:2018cpf},
where $F_{T,\pi\pi}^{ud}$ was obtained using the Omn\`{e}s representation.
Since according to the Watson--Migdal theorem, in the elastic region, the phases of the tensor form factors equal
those of the vector form factors, $\delta_T(s)=\delta_{+}(s)$, one can use the dispersion relation
\begin{align}
  \frac{F_{T}(s)}{F_{T}(0)}=\exp \left\{\frac{s}{\pi} \int_{4 M_{\pi}^{2}}^{\infty} d s^{\prime}
  \frac{\delta_{T}\left(s^{\prime}\right)}{s^{\prime}\left(s^{\prime}-s-i \epsilon\right)}\right\}
\end{align}
to obtain the normalized tensor form factor.
The comparison is shown in Fig.~\ref{fig:CompareTFFudpipi}, and we observe a significant difference.
For instance, the sizes of the peak at $s=M_\rho^2$ are quite different in the two calculations, and there
exists a zero point in our curve above $1\ \text{GeV}$ that does not occur in the phase dispersive representation.
The differences are exclusively due to the SU(3)-breaking LECs in the tensor form factors at NLO, and therefore probably a rather large uncertainty should be associated with them.
An independent cross-check is therefore highly desirable, and in Appendix~\ref{sec:VMD} we argue that
this is in principle doable through a comparison with future lattice QCD calculations of the tensor
charge of the $\rho$-meson.

\subsection{Applications of the Tensor Form Factors}

In the previous section, we have discussed the application of the two-meson vector form factors
that characterize the SM contribution to the hadronic charged-lepton decay. At the same time, these
processes also provide a suitable platform for searching for the BSM physics due to
their large phase space. In the literature, the BSM physics effects are included by
introducing a general set of higher-dimensional operators beyond the SM. For instance, the dimension-6
effective operators (with only left-handed neutrinos) responsible for $\tau\rightarrow
\pi^{-}\pi^{0}\nu_{\tau}$ are given by~\cite{Miranda:2018cpf}
\begin{align}
\mathcal{L}_{C C}=&-\frac{4 G_{F}}{\sqrt{2}}\left[\left(1+\left[v_{L}\right]_{\ell \ell}\right) \bar{\ell}_{L} \gamma_{\mu} \nu_{\ell L} \bar{u}_{L} \gamma^{\mu} d_{L}+\left[v_{R}\right]_{\ell \ell} \bar{\ell}_{L} \gamma_{\mu} \nu_{\ell L} \bar{u}_{R} \gamma^{\mu} d_{R}\right.\nonumber\\
&+\left[s_{L}\right]_{\ell \ell} \bar{\ell}_{R} \nu_{\ell L} \bar{u}_{R} d_{L}+\left[s_{R}\right]_{\ell \ell} \bar{\ell}_{R} \nu_{\ell L} \bar{u}_{L} d_{R} \left.+\left[t_{L}\right]_{\ell \ell} \bar{\ell}_{R} \sigma_{\mu \nu} \nu_{\ell L} \bar{u}_{R} \sigma^{\mu \nu} d_{L}\right]+\text { h.c.},
\end{align}
where the SM Lagrangian is recovered by setting $v_{L}=v_{R}=s_{L}=s_{R}=t_{L}=0$. In particular, the
$t_L$ term contains the tensor interaction. The decay amplitude for $\tau^{-}(P) \rightarrow \pi^{-}
\left(P_{\pi^{-}}\right) \pi^{0}\left(P_{\pi^{0}}\right) \nu_{\tau}\left(P^{\prime}\right)$ reads
\begin{align}
\mathcal{M} &=\mathcal{M}_{V}+\mathcal{M}_{S}+\mathcal{M}_{T} \nonumber\\
&=\frac{G_{F} V_{u d} \sqrt{S_{E W}}}{\sqrt{2}}\left(1+\epsilon_{L}+\epsilon_{R}\right)\left[L_{\mu} H^{\mu}+\hat{\epsilon}_{S} L H+2 \hat{\epsilon}_{T} L_{\mu \nu} H^{\mu \nu}\right],
\end{align}
with the leptonic and hadronic sectors, respectively, given by
\begin{align}
L_{\mu} &=\bar{u}_{\nu_\tau}\left(P^{\prime}\right) \gamma^{\mu}\left(1-\gamma^{5}\right) u_\tau(P)~, \nonumber\\
L &=\bar{u}_{\nu_\tau}\left(P^{\prime}\right)\left(1+\gamma^{5}\right) u_\tau(P)~, \nonumber\\
L_{\mu \nu} &=\bar{u}_{\nu_\tau}\left(P^{\prime}\right) \sigma_{\mu \nu}\left(1+\gamma^{5}\right) u_\tau(P)~,
\end{align}
and
\begin{align}
H &=\left\langle\pi^{0} \pi^{-}|\bar{d} u| 0\right\rangle~, \nonumber\\
H^{\mu} &=\left\langle\pi^{0} \pi^{-}\left|\bar{d} \gamma^{\mu} u\right| 0\right\rangle~, \nonumber\\
H^{\mu \nu} &=\left\langle\pi^{0} \pi^{-}\left|\bar{d} \sigma^{\mu \nu} u\right| 0\right\rangle~.
\end{align}
Therefore an improved understanding of the tensor form factors (as well as the scalar and
vector form factors) will better constrain the strength of the BSM physics interactions. The
same applies to other decay processes such as $\tau^{-} \to (K\pi)^{-} \nu_{\tau}$~\cite{Cirigliano:2017tqn,Rendon:2019awg,Chen:2019vbr}
and $\tau^{-} \to (\pi\eta)^{-} \nu_{\tau}$~\cite{Garces:2017jpz,Castro:2018cot}.
 
\section{Conclusions}
\label{sec:conclusions}
In this work, we have performed a complete study of the scalar, vector, and tensor two-meson
form factors based on uChPT improved by dispersion relations that remove the unphysical sub-threshold
singularities. The slight unitarity violation in the coupled-channel IAM as in Ref.~\cite{GomezNicola:2001as}
is thus removed. The low-energy constants used for the uChPT calculation are fixed by a global fit to the
available data of two-meson scattering observables. The resulting form factors are expected to be
applicable in a wide energy region $0\ldots 1.2\ \text{GeV}$.  For the scalar and vector form factors, our results
may be regarded as an update to those given in  the existing literature, but with a much simpler realization of the
dispersion relation improvement. On the other hand, this work provides the first-ever systematic
analysis of the full two-meson tensor form factors. These form factors play important roles in
precision tests of SM and searches for the BSM physics.

\section*{Acknowledgements}

The authors are very grateful to Jos\'e R.~Pel\'aez and Michael D\"oring for providing data and
for useful discussions. We also thank the anonymous referee for pointing out a defect in the first version of this manuscript.
BK thanks Gilberto Colangelo for helpful e-mail communication.
This work is
supported in part by Natural Science Foundation of China under Grant Nos. 11735010, 11835015, 12047503, 
11911530088, 11961141012  and U2032102, by Natural Science Foundation of Shanghai under Grant No.~15DZ2272100, by
the DFG and the NSFC through funds provided to the Sino-German Collaborative Research Center TRR110 ``Symmetries and the Emergence of Structure in QCD'' (NSFC Grant No.~12070131001, DFG Project-ID 196253076), by the Chinese Academy of Sciences (CAS)
under Grant Nos.~XDB34030000 and QYZDB-SSW-SYS013, and by the CAS Center for Excellence in Particle Physics (CCEPP).
The work of UGM was also supported by the CAS President's International
Fellowship Initiative (PIFI)
(Grant No.~2018DM0034), by the VolkswagenStiftung (Grant No.~93562) and by the EU (Strong2020).
CYS is  also supported by the Alexander von Humboldt Foundation through a Humboldt Research Fellowship.
 

\begin{appendix}
	
	\section{Loop Functions\label{sec:loopfunc}}
	
	In this appendix we summarize the relevant loop functions that appear in the one-loop calculations
        within ChPT. First, from the tadpole diagrams we encounter the following integral:
	\begin{eqnarray}
	\int\frac{d^d q}{(2\pi)^d}\frac{i}{q^2-M_i^2}=2 M_i^2\lambda
	+\frac{M_i^2}{16\pi^2}\ln\frac{M_i^2}{\mu^2} \equiv 2M_{i}^{2}\lambda + 2 F_0^{2}\mu_{i}~,
        \label{deltai} 
	\end{eqnarray}
	where $\mu$ is the renormalization scale and $\mu_i=(M_i^2/32\pi^2F_0^2)\ln(M_i^2/\mu^2)$,
        with $i=\pi,K,\eta$, and $F_0$ is the pion decay constant in the three-flavor chiral limit.
     Next, in a loop integral with two propagators we encounter the following two-point function:
	\begin{eqnarray} 
	-i\int\frac{d^d q}{(2\pi)^d}
	\frac{1}{[q^2-M_P^2][(q-p)^2-M_Q^2]}   =  -2 \lambda +J_{PQ}^{r}(s) = J_{PQ} (0)+{\bar J}_{PQ} (s)~,
	\end{eqnarray}
	where $P,Q=\pi$, $K$, and $\eta$, $s=p^2$, and 
	\begin{eqnarray}
	J_{PQ}(0) &=& -2\lambda-\frac{2F_0^2}{\Delta}(\mu_P-\mu_Q)~,\nonumber\\
	\bar{J}_{PQ}(s)&=&\frac{1}{32\pi^2}\left[2+\left(\frac{\Delta}{s}-\frac{\Sigma}{\Delta}\right)\ln\frac{M_Q^2}{M_P^2}+\frac{\nu}{s}\left\{\ln\left(\frac{(s-\Sigma-\nu-i\varepsilon)^2}{\Delta^2-\Sigma^2}\right)+i\pi\right\}\right]~,
	\end{eqnarray} 
	with \begin{eqnarray} 
	\Delta&=&M_P^2-M_Q^2~,\;\; 
	\Sigma=M_P^2+M_Q^2~,\nonumber\\\nu(s)&=&
	\sqrt{\left(s-\left(M_P+M_Q\right)^2\right)
		\left(s-\left(M_P-M_Q\right)^2\right)}~. 
	\end{eqnarray}
	For the case of a single mass $M_P=M_Q$, the $\bar{J}$-function reads
	\begin{eqnarray}
	J_{PP}(0)&=&-2\lambda-\frac{1}{16\pi^2}-\frac{2F_0^2}{M_P^2}\mu_P~,\nonumber\\
	{\bar J}_{PP}
	(s)&=&\frac{1}{32\pi^2}\left[4+\frac{\sqrt{s(s-4M_P^2)}}{s}\left\{\ln\left(-\frac{(s-2M_P^2-\sqrt{s(s-4M_P^2)}-i\varepsilon)^2}{4M_P^2}\right)+i\pi\right\}\right]~.\nonumber\\
	\end{eqnarray}
	Note that the above integrals have the correct unitarity structure
	along the right-hand cut, which extends on the real axis from
	$s=(M_P+M_Q)^2$ to infinity.

\section{Isospin Decomposition of the Scattering Amplitudes and Form Factors}\label{sec:ScatAmp}
In this appendix we state our conventions in defining one- and two-particle isospin eigenstates,
and show how to construct scattering amplitudes of definite isospin. 

\subsection{One-Particle Isospin Eigenstates}
The one-particle isospin eigenstate is generically denoted as $\left|\phi,I,I_3\right\rangle$. The phases
of such states are chosen such that they satisfy standard results when acted on by isospin-raising
and -lowering operators:
\begin{eqnarray}
\hat{J}_{+}\left|I,I_{3}\right\rangle  & = & \sqrt{(I-I_{3})(I+I_{3}+1)}\left|I,I_{3}+1\right\rangle , \nonumber \\
\hat{J}_{-}\left|I,I_{3}\right\rangle  & = & \sqrt{(I+I_{3})(I-I_{3}+1)}\left|I,I_{3}-1\right\rangle .
\end{eqnarray}
For the pion triplet, we choose:
\begin{eqnarray}
\left|\pi,1,+1\right\rangle  = -\left|\pi^{+}\right\rangle, ~~~~
\left|\pi,1,0\right\rangle  = \left|\pi^{0}\right\rangle, ~~~~
\left|\pi,1,-1\right\rangle  = \left|\pi^{-}\right\rangle .
\end{eqnarray}
For the kaon doublet $K^{+},K^{0}$, we choose:
\begin{eqnarray}
\left|K,\frac{1}{2},+\frac{1}{2}\right\rangle  = \left|K^{+}\right\rangle,~~~~
\left|K,\frac{1}{2},-\frac{1}{2}\right\rangle  = \left|K^{0}\right\rangle .
\end{eqnarray}
For the anti-kaon doublet $\bar{K}^{0}$, $K^{-}$, we choose:
\begin{eqnarray}
\left|\bar{K},\frac{1}{2},+\frac{1}{2}\right\rangle  = \left|\bar{K}^{0}\right\rangle,~~~~
\left|\bar{K},\frac{1}{2},-\frac{1}{2}\right\rangle  = -\left|K^{-}\right\rangle. 
\end{eqnarray}
Finally, the $\eta$-particle is simply an isospin singlet: $\left|\eta,0,0\right\rangle =\left|\eta\right\rangle $. 

\subsection{Two-Particle Isospin Eigenstates}
Two-particle isospin eigenstates, denoted generically as $\left|\phi\phi',I,I_3\right\rangle$, are simply obtained by combining one-particle isospin eigenstates with appropriate Clebsch--Gordan coefficients. There is one complication, namely isospin eigenstates that are constructed by two particles in the same isospin multiplet should be multiplied by a factor $1/\sqrt{2}$ so that the completeness relation they satisfied is properly normalized as $\sum_{I,I_3}\left|I,I_3\right\rangle\left\langle I,I_3\right|=1$.\footnote{Such a factor is sometimes defined into the partial-wave expansion, see, e.g., Ref.~\cite{GomezNicola:2001as}.} In the following we present the two-particle isospin eigenstates that are relevant to this work:
\begin{eqnarray}
\left|\pi\pi,0,0\right\rangle
& = & -\frac{1}{\sqrt{6}}\left[\left|\pi^+\pi^-\right\rangle +\left|\pi^0\pi^0\right\rangle+\left|\pi^-\pi^+\right\rangle\right]~, \nonumber\\
\left|\pi\pi,1,0\right\rangle  & = &  -\frac{1}{2}\left[\left|\pi^+\pi^-\right\rangle -\left|\pi^-\pi^+\right\rangle\right]~, \nonumber \\
\left|\pi\pi,2,2\right\rangle  & = & \frac{1}{\sqrt{2}}\left|\pi^{+}\pi^{+}\right\rangle~,
\nonumber\\
\left|K\pi,\frac{3}{2},\frac{3}{2}\right\rangle&=&-\left|K^{+}\pi^{+}\right\rangle~,
 \nonumber \\
\left|K\pi,\frac{1}{2},-\frac{1}{2}\right\rangle
& = & \sqrt{\frac{2}{3}}\left|K^+\pi^-\right\rangle -\sqrt{\frac{1}{3}}\left|K^0\pi^0\right\rangle~,\nonumber\\
\left|K\bar{K},1,0\right\rangle
& = & -\sqrt{\frac{1}{2}}\left|K^+K^-\right\rangle +\sqrt{\frac{1}{2}}\left|K^0\bar{K}^0\right\rangle~, \nonumber \\
\left|K\bar{K},0,0\right\rangle
& = & -\sqrt{\frac{1}{2}}\left|K^+K^-\right\rangle -\sqrt{\frac{1}{2}}\left|K^0\bar{K}^0\right\rangle~,\nonumber\\
\left|K\eta,\frac{1}{2},-\frac{1}{2}\right\rangle&=&\left|K^{0}\eta\right\rangle~, \nonumber\\
\left|\pi\eta,1,0\right\rangle &=&\left|\pi^0\eta\right\rangle~, \nonumber\\
\left|\eta\eta,0,0\right\rangle &=&\frac1{\sqrt{2}} \left|\eta\eta\right\rangle~.
\end{eqnarray}

\subsection{Scattering Amplitudes of Definite Isospin}

Following Ref.~\cite{GomezNicola:2001as}, one may choose the eight independent $\phi\phi$ scattering amplitudes as follows:
\begin{eqnarray}
T_1(s,t,u)&\equiv&T_{\pi^+\pi^-\rightarrow\pi^0\pi^0}(s,t,u)~,\nonumber\\
T_2(s,t,u)&\equiv&T_{K^+\pi^+\rightarrow K^+\pi^+}(s,t,u)~,\nonumber\\
T_3(s,t,u)&\equiv&T_{K^+K^-\rightarrow K^+K^-}(s,t,u)~,\nonumber\\
T_4(s,t,u)&\equiv&T_{K^0\bar{K}^0\rightarrow K^+K^-}(s,t,u)~,\nonumber\\
T_5(s,t,u)&\equiv&T_{K^0\eta\rightarrow K^0\eta}(s,t,u)~,\nonumber\\
T_6(s,t,u)&\equiv&T_{K^0\pi^0\rightarrow K^0\eta}(s,t,u)~,\nonumber\\
T_7(s,t,u)&\equiv&T_{\pi^0\eta\rightarrow \pi^0\eta}(s,t,u)~,\nonumber\\
T_8(s,t,u)&\equiv&T_{\eta\eta\rightarrow \eta\eta}(s,t,u)~.
\end{eqnarray}
All scattering amplitudes of definite isospin can be expressed in terms of these eight amplitudes and their crossings. The results are as follows. For $I=0$, we have
\begin{eqnarray}
T^{I=0}_{\pi\pi\rightarrow\pi\pi}(s,t,u)&=&\frac{1}{2}\left[3T_{\pi^+\pi^-\rightarrow\pi^0\pi^0}(s,t,u) +T_{\pi^+\pi^-\rightarrow\pi^0\pi^0}(t,s,u)+T_{\pi^+\pi^-\rightarrow\pi^0\pi^0}(u,t,s)\right], \nonumber\\
T^{I=0}_{\pi\pi\rightarrow K\bar{K}}(s,t,u)&=&\frac{\sqrt{3}}{2}\left[T_{K^+\pi^+\rightarrow K^+\pi^+}(u,s,t)+T_{K^+\pi^+\rightarrow K^+\pi^+}(t,s,u)\right]~,\nonumber\\
T^{I=0}_{\pi\pi\rightarrow\eta\eta}(s,t,u)&=&-\frac{\sqrt{3}}{2}T_{\pi^0\eta\rightarrow \pi^0\eta}(t,s,u)~,\nonumber\\
T^{I=0}_{K\bar{K}\rightarrow K\bar{K}}(s,t,u)&=&T_{K^+K^-\rightarrow K^+K^-}(s,t,u)+T_{K^0\bar K^0\rightarrow K^+K^-}(s,t,u)~,\nonumber\\
T^{I=0}_{K\bar{K}\rightarrow\eta\eta}(s,t,u)&=&-T_{K^0\eta\rightarrow K^0\eta}(t,s,u)~,\nonumber\\
T^{I=0}_{\eta\eta\rightarrow\eta\eta}(s,t,u)&=&\frac{1}{2}T_{\eta\eta\rightarrow \eta\eta}(s,t,u)~.
\end{eqnarray}
For $I=1/2$, we have
\begin{eqnarray}
T^{I=1/2}_{K\pi\rightarrow K\pi}(s,t,u)&=&\frac{3}{2}T_{K^+\pi^+\rightarrow K^+\pi^+}(u,t,s)-\frac{1}{2}T_{K^+\pi^+\rightarrow K^+\pi^+}(s,t,u)~,\nonumber\\
T^{I=1/2}_{K\pi\rightarrow K\eta}(s,t,u)&=&-\sqrt{3}T_{K^0\pi^0\rightarrow K^0\eta}(s,t,u)~,\nonumber\\
T^{I=1/2}_{K\eta\rightarrow K\eta}(s,t,u)&=&T_{K^0\eta\rightarrow K^0\eta}(s,t,u)~.
\end{eqnarray}
For $I=1$, we have
\begin{eqnarray}
T^{I=1}_{\pi\pi\rightarrow\pi\pi}(s,t,u)&=&\frac{1}{2}(T_{\pi^+\pi^-\rightarrow\pi^0\pi^0}(t,s,u)-T_{\pi^+\pi^-\rightarrow\pi^0\pi^0}(u,t,s))~,\nonumber\\
T^{I=1}_{\pi\pi\rightarrow K\bar{K}}(s,t,u)&=&\frac{1}{\sqrt{2}}(T_{K^+\pi^+\rightarrow K^+\pi^+}(u,s,t)-T_{K^+\pi^+\rightarrow K^+\pi^+}(t,s,u)~,\nonumber\\
T^{I=1}_{\pi\eta\rightarrow\pi\eta}(s,t,u)&=&T_{\pi^0\eta\rightarrow \pi^0\eta}(s,t,u)~,\nonumber\\
T^{I=1}_{K\bar{K}\rightarrow K\bar{K}}(s,t,u)&=&T_{K^+K^-\rightarrow K^+K^-}(s,t,u)-T_{K^0\bar K^0\rightarrow K^+K^-}(s,t,u)~,\nonumber\\
T^{I=1}_{K\bar{K}\rightarrow\pi\eta}(s,t,u)&=&\sqrt{2}T_{K^0\pi^0\rightarrow K^0\eta}(t,s,u)~.
\end{eqnarray}
For $I=3/2$, we have
\begin{eqnarray}
T^{I=3/2}_{K\pi\rightarrow K\pi}(s,t,u)&=&T_{K^+\pi^+\rightarrow K^+\pi^+}(s,t,u)~. 
\end{eqnarray}
Finally, for $I=2$, we have
\begin{eqnarray}
T^{I=2}_{\pi\pi\rightarrow\pi\pi}(s,t,u)&=&\frac{1}{2}(T_{\pi^+\pi^-\rightarrow\pi^0\pi^0}(t,s,u)-T_{\pi^+\pi^-\rightarrow\pi^0\pi^0}(u,t,s))~.
\end{eqnarray}

\subsection{Form Factors}

Finally, we discuss how the two-meson form factors are classified according to the isospin eigenstates. The form factors of interest have the following general form
\begin{equation}
\left\langle \phi\phi'\right|\bar{q}\Gamma q'\left|0\right\rangle,
\end{equation}
where $\Gamma$ is any matrix in the non-flavor space. It is obvious
that with different choices of $\bar{q}q'$ and $\phi\phi'$ there
will be different form factors. However, not all of them are independent
because some of them are related via charge conjugation and isospin
symmetry. In this appendix, we will extract all the independent form
factors in order to minimize the calculation. 

We shall start from the classification of $\phi\phi'$ states according
to isospin. There are 10 groups of them: $\pi\pi$, $\pi K$, $\pi\bar{K}$,
$\pi\eta$, $KK$, $K\bar{K}$, $K\eta$, $\bar{K}\bar{K}$, $\bar{K}\eta$,
and $\eta\eta$. However, $KK$ and $\bar{K}\bar{K}$ states have
no nonvanishing form factors because they have strangeness  $\pm2$,
which cannot be obtained from a quark bilinear. Next, $\pi K$ and
$\pi\bar{K}$ are related by charge conjugation (the same holds for $K\eta$
and $\bar{K}\eta$), so we just need to choose one of them. Therefore,
there are only 6 groups of independent $\phi\phi'$ states that
will give nonvanishing form factors. They can be chosen as
\begin{equation}
\pi\pi, \quad K\pi, \quad \pi\eta, \quad K\bar{K},\quad K\eta, \quad \text{and} \quad \eta\eta.
\end{equation}

Next we study the quark bilinear operators (we shall ignore the matrix
$\Gamma$ for notational simplicity). At the first sight, there are
9 possible combinations: $\bar{u}u$, $\bar{d}d$, $\bar{s}s$,
$\bar{u}d$, $\bar{u}s$, $\bar{d}u$, $\bar{d}s$, $\bar{s}u$, and
$\bar{s}d$. However, some of them are related by isospin and charge
conjugation (C.C). For example,
\begin{eqnarray}
&  & \bar{u}d ~\stackrel{\mathrm{isospin}}{\longleftrightarrow}~\bar{u}u-\bar{d}d~\stackrel{\mathrm{isospin}}{\longleftrightarrow}~\bar{d}u ~,\nonumber \\
&  & \bar{u}s~\stackrel{\mathrm{isospin}}{\longleftrightarrow}~\bar{d}s~\stackrel{\mathrm{C.C}}{\longleftrightarrow}~\bar{s}d~\stackrel{\mathrm{isospin}}{\longleftrightarrow}~\bar{s}u ~.
\end{eqnarray}
For two bilinears related by isospin, their matrix elements are
related to each other by the Wigner--Eckart theorem. For bilinears
related by charge conjugation, their matrix elements are related
by charge-conjugating the outcoming mesons. Therefore, there are only
four independent quark bilinears, which can be chosen as
\begin{eqnarray}
I=0  : ~  n\equiv\frac{\bar{u}u+\bar{d}d}{\sqrt{2}}~, \quad 
I=0  : ~  \bar{s}s~, \quad 
I=1  : ~ \bar{u}d~, \quad 
I=\frac{1}{2}  : ~ \bar{u}s~.
\end{eqnarray}
Now, 
\begin{table}
\begin{center}
	\caption{\label{tab:independentFF}Quark bilinears and meson fields that form independent form factors.}
		\begin{tabular}{cc}
			\hline 	\hline 
			Quark bilinear & Two-particle isospin eigenstates\tabularnewline
			\hline 
			
			$n=({\bar{u}u+\bar{d}d})/{\sqrt{2}}$ & $\pi\pi$, $K\bar{K}$, $\eta\eta$\tabularnewline
		
			$\bar{s}s$ & $\pi\pi$, $K\bar{K}$, $\eta\eta$\tabularnewline
			
			$\bar{u}d$ & $\pi\pi$, $\pi\eta$, $K\bar{K}$\tabularnewline
			
			$\bar{u}s$ & $K\pi$, $K\eta$\tabularnewline
			\hline
		\end{tabular}
\end{center}
\end{table}
with each independent quark bilinear, we just need to compute
one matrix element for each independent $\phi\phi'$ group; the others are related by Wigner--Eckart theorem. Therefore, the independent form
factors can be chosen as the matrix elements of the quark bilinears between the vacuum
and two-particle isospin eigenstates given in Table~\ref{tab:independentFF}.

\section{Subtraction of Sub-Threshold Poles\label{sec:Adler}}

It is well known that the na\"ive application of the IAM will lead to spurious poles in the sub-threshold
region for quantities in the $J=0$ channel, which is related to the existence of the Adler zero in
the $S$-wave~\cite{Hannah:1997sm,GomezNicola:2007qj}. In this section, we briefly outline the
method one can adopt to subtract such spurious pole in both unitarized partial waves and form factors.
One may refer to Ref.~\cite{GomezNicola:2007qj} for detailed discussions of the topic. 

Let us restrict ourselves to a single-channel unitarization of partial waves and form factors. Also, we shall simply use $T_2$ and $T_4$ to denote the $\mathcal{O}(p^2)$ and $\mathcal{O}(p^4)$ $J=0$ partial wave for notational simplicity. First, let us recall the na\"ive IAM formula for partial waves,
\begin{equation}
T=\frac{T_2^2}{T_2-T_4}\label{eq:naiveIAM}~.
\end{equation}
The $\mathcal{O}(p^2)$ amplitude $T_2$ has a zero at $s=s_2$ below the production threshold: $T_2(s_2)=0$. This is nothing but the Adler zero at the $\mathcal{O}(p^2)$ level for the $S$-wave. Meanwhile, the combination $T_2-T_4$ has a different zero at $s=s'$. Since in general $s'\neq s_2$, the na\"ive IAM formula~\eqref{eq:naiveIAM} has a spurious pole at $s=s'$. Similarly, the na\"ive IAM formula for the scalar form factor
\begin{equation}
F_S=F_S^{(0)}+\frac{T_2}{T_2-T_4}F_S^{(2)}
\end{equation}
suffers from a pole at $s=s'$. 

From the dispersive point of view, the existence of such spurious poles is due to the negligence of the pole contribution of various inverse amplitudes in the derivation of the IAM formula through a dispersion relation. Therefore, the problem can be resolved by appropriately adding back these contributions. Unfortunately, a dispersive derivation of the multi-channel IAM is still missing, so we can only stick to the single-channel case in this discussion.   

To derive the single-channel IAM formula for partial waves, we shall consider the dispersion relation of $1/T$, $1/T_2$ and $T_4/T_2^2$ respectively. Also, here we shall simplify our discussion by considering an unsubtracted dispersion relation of each quantity, the outcome turns out to be equivalent to choosing the subtraction point at the Adler zero of the full amplitude~\cite{GomezNicola:2007qj}. The dispersion relations we obtain are
\begin{eqnarray}
\frac{1}{T(s)}&=&\frac{1}{\pi}\int^\infty_{s_{\rm th}}dz\frac{\mathrm{Im}T^{-1}(z)}{z-s}+LC(1/T)+PC(1/T)~,\nonumber\\
\frac{1}{T_2(s)}&=&PC(1/T_2)~,\nonumber\\
\frac{T_4(s)}{T_2^2(s)}&=&\frac{1}{\pi}\int^\infty_{s_{\rm th}}dz\frac{\mathrm{Im}\{T_4(z)/T_2^2(z)\}}{z-s}+LC(T_4/T_2^2)+PC(T_4/T_2^2)~.
\end{eqnarray}
Here $LC$ and $PC$ denote the left-hand cut and the pole contributions to the dispersion integral, respectively. Since the full and perturbative amplitude satisfies the following unitarity relation on the right-hand cut:
\begin{equation}
-\mathrm{Im}T^{-1}(s)=\mathrm{Im}\{T_4(s)/T_2^2(s)\}=\frac{|\vec{p}_\text{cm}|}{8\pi\sqrt{s}}~,
\end{equation}
the right-hand cut contributions to $1/T$ and $T_4/T_2^2$ simply differ by a sign. Furthermore, one approximates $LC(1/T)\approx LC(T_4/T_2^2)$ by arguing that the left-hand cut contribution is weighted at low energies where the usual ChPT expansion is appropriate. With these, we can write
\begin{equation}
\frac{1}{T(s)}\approx\frac{1}{T_2(s)}-\frac{T_4(s)}{T_2^2(s)}-PC(1/T_2)+PC(T_4/T_2^2)+PC(1/T)~,\label{eq:mIAM}
\end{equation}
where the explicit expressions for the pole contributions are given by
\begin{eqnarray}
PC(1/T)&=&\frac{1}{(s-s_A)T'(s_A)}~,\nonumber\\
PC(1/T_2)&=&\frac{1}{(s-s_2)T'_2(s_2)}~,\nonumber\\
PC(T_4/T_2^2)&=&\frac{T'_4(s_2)}{(s-s_2)(T'_2(s_2))^2}+\frac{T_4(s_2)}{(s-s_2)^2(T'_2(s_2))^2}-\frac{T_4(s_2)T''_2(s_2)}{(s-s_2)(T'_2(s_2))^3}~.\label{eq:polecont}
\end{eqnarray}
Here $s_A$ is the Adler zero of the full partial-wave $T$-matrix. Of course its exact value is unknown, but we can approximate it by the Adler zero of $T_2+T_4$, which is
$s_A\approx s_2+s_4$ where $s_4\approx -T_4(s_2)/T_2'(s_2)$. If one neglects the pole contributions then the na\"ive IAM formula~\eqref{eq:naiveIAM} is recovered, the inclusion of the pole contributions will eliminate the spurious pole in the sub-threshold region. 

The pole subtraction for the scalar form factor follows a similar logic. Let us start by considering the dispersion relations of $(F_S-F^{(0)}_S)/T$ and $F^{(2)}_S/T_2$, respectively,
\begin{eqnarray}
\frac{F_S(s)-F_S^{(0)}(s)}{T}&=&\frac{1}{\pi}\int^\infty_{s_{\rm th}}dz\frac{\mathrm{Im}\{(F_S(z)-F_S^{(0)}(z))/T(z)\}}{z-s}+LC((F_S-F_S^{(0)})/T)\nonumber\\
&&+PC((F_S-F_S^{(0)})/T) ,\nonumber\\
\frac{F_S^{(2)}(s)}{T_2}&=&\frac{1}{\pi}\int^\infty_{s_{\rm th}}dz\frac{\mathrm{Im}\{F_S^{(2)}(z)/T_2(z)\}}{z-s}+LC(F_S^{(2)}/T_2)+PC(F_S^{(2)}/T_2)~.
\end{eqnarray} 
Along the right-hand cut, we have 
\begin{equation}
\mathrm{Im}\left\{\frac{F_S(s)-F^{(0)}_S(s)}{T(s)}\right\}=\mathrm{Im}\left\{\frac{F_S^{(2)}(s)}{T_2(s)}\right\}=F_S^{(0)}(s)\frac{|\vec{p}_{\rm cm}|}{8\pi\sqrt{s}}~.
\end{equation}
Therefore, the right-hand-cut contributions for $(F_S-F_S^{(0)})/T$ and $F_S^{(2)}/T_2$ are the same. Furthermore, we assume that the two left-hand-cut contributions are also approximately the same, following an argument similar to the case of the partial-wave scattering amplitude. With this we obtain
\begin{equation}
F_S(s)\approx F_S^{(0)}(s)+\frac{T(s)}{T_2(s)}F_S^{(2)}(s)+T(s)\left\{PC((F_S-F_S^{(0)})/T)-PC(F_S^{(2)}/T_2)\right\}~,\label{eq:mFF}
\end{equation}
where the explicit expressions for the pole contributions are given by
\begin{eqnarray}
PC((F_S-F_S^{(0)})/T)&=&\frac{F_S(s_A)-F_S^{(0)}(s_A)}{(s-s_A)T'(s_A)}~,\nonumber\\
PC(F_S^{(2)}/T)&=&\frac{F^{(2)}_S(s_2)}{(s-s_2)T_2'(s_2)}~.
\end{eqnarray}
Again, we may approximate $F_S$ and $T$ in the formulae above by their respective ChPT expressions up to NLO. 

Notice that if we neglect the pole contributions in Eq.~\eqref{eq:mFF} and substitute $T(s)$ by the na\"ive IAM formula for the partial-wave $T$-matrix, then we re-obtain the na\"ive IAM unitarized scalar form factor. This expression works fine above threshold but suffers from a spurious pole below threshold. On the other hand, if we use Eq.~\eqref{eq:mFF} with the expression of $T(s)$ given in Eq.~\eqref{eq:mIAM}, then the spurious pole will be smoothly eliminated.


\section{Form Factors in ChPT to One Loop\label{sec:FFoneloop}}

In this section,  we list the NLO ChPT results for scalar, vector, and tensor form factors. Notice that we express our result in terms of the physical pion decay constant $F_\pi$, which is related to $F_0$ by
\begin{equation}
F_\pi=F_0\left\{1-2\mu_\pi-\mu_K+\frac{4}{F_0^2}\left[2L_4^r M_K^2+(L_4^r+L_5^r)M_\pi^2\right]\right\}.
\end{equation}

\subsection[Scalar Form Factors for the $I=0$ System]{Scalar Form Factors for the \boldmath{$I=0$} System} 

The two-meson states are chosen to be exact isospin eigenstates. At NLO, calculating the Feynman diagrams shown in Fig.~\ref{fig:feynmanFormFactor} gives

\begin{eqnarray}
-\frac{F^n_{S,\pi\pi}(s)}{\sqrt{3}}&=&1+\mu_\pi-\frac{1}{3}\mu_\eta-\frac{8(2M_K^2+3M_\pi^2-s)}{F_\pi^2}L_4^r+\frac{4(s-4M_\pi^2)}{F_\pi^2}L_5^r+\frac{16(2M_K^2+3M_\pi^2)}{F_\pi^2}L_6^r\nonumber\\
&&+\frac{32M_\pi^2}{F_\pi^2}L_8^r+\frac{2s-M_\pi^2}{2F_\pi^2}J^r_{\pi\pi}(s)+\frac{s}{4F_\pi^2}J^r_{KK}(s)+\frac{M_\pi^2}{18F_\pi^2}J^r_{\eta\eta}(s)~,\nonumber\\
-F^n_{S,K\bar{K}}(s)&=&1+\frac{2}{3}\mu_\eta-\frac{8(6M_K^2+M_\pi^2-2s)}{F_\pi^2}L_4^r+\frac{4(s-4M_K^2)}{F_\pi^2}L_5^r+\frac{16(6M_K^2+M_\pi^2)}{F_\pi^2}L_6^r\nonumber\\
&&+\frac{32M_K^2}{F_\pi^2}L_8^r+\frac{3s}{4F_\pi^2}J_{\pi\pi}^r(s)+\frac{3s}{4F_\pi^2}J^r_{KK}(s)+\frac{9s-2M_\pi^2-6M_\eta^2}{36F_\pi^2}J^r_{\eta\eta}(s)~,\nonumber\\
3F^n_{S,\eta\eta}(s)&=&1-3\mu_\pi+4\mu_K-\frac{1}{3}\mu_\eta+\frac{8(-10M_K^2+M_\pi^2+3s)}{F_\pi^2}L_4^r-\frac{4(16M_K^2-4M_\pi^2-3s)}{3F_\pi^2}L_5^r\nonumber\\
&&+\frac{16(10M_K^2-M_\pi^2)}{F_\pi^2}L_6^r-\frac{128(M_K^2-M_\pi^2)}{F_\pi^2}L_7^r+\frac{32M_\pi^2}{F_\pi^2}L_8^r+\frac{3M_\pi^2}{2F_\pi^2}J^r_{\pi\pi}(s)\nonumber\\
&&+\frac{9s-2M_\pi^2-6M_\eta^2}{4F_\pi^2}J^r_{KK}(s)+\frac{16M_K^2-7M_\pi^2}{18F_\pi^2}J^r_{\eta\eta}(s)
\end{eqnarray}
for the nonstrange form factors and 
\begin{eqnarray}
-\sqrt{\frac{2}{3}}F^{\bar{s}s}_{S,\pi\pi}(s)&=&\frac{8(s-2M_\pi^2)}{F_\pi^2}L_4^r+\frac{32M_\pi^2}{F_\pi^2}L_6^r+\frac{s}{2F_\pi^2}J^r_{KK}(s)+\frac{2M_\pi^2}{9F_\pi^2}J^r_{\eta\eta}(s)~,\nonumber\\
-\frac{F^{\bar{s}s}_{S,K\bar{K}}(s)}{\sqrt{2}}&=&1+\frac{2}{3}\mu_\eta-\frac{8(4M_K^2+M_\pi^2-s)}{F_\pi^2}L_4^r+\frac{4(s-4M_K^2)}{F_\pi^2}L_5^r+\frac{16(4M_K^2+M_\pi^2)}{F_\pi^2}L_6^r\nonumber\\
&&+\frac{32M_K^2}{F_\pi^2}L_8^r+\frac{3s}{4F_\pi^2}J_{KK}^r(s)+\frac{9s-2M_\pi^2-6M_\eta^2}{18F_\pi^2}J^r_{\eta\eta}(s)~,\nonumber\\
\frac{3\sqrt{2}F^{\bar{s}s}_{S,\eta\eta}(s)}{4}&=&1+2\mu_K-\frac{4}{3}\mu_\eta+\frac{2(-16M_K^2-2M_\pi^2+3s)}{F_\pi^2}L_4^r-\frac{4(16M_K^2-4M_\pi^2-3s)}{3F_\pi^2}L_5^r\nonumber\\
&&+\frac{8(8M_K^2+M_\pi^2)}{F_\pi^2}L_6^r+\frac{64(M_K^2-M_\pi^2)}{F_\pi^2}L_7^r+\frac{32(2M_K^2-M_\pi^2)}{F_\pi^2}L_8^r\nonumber\\
&&+\frac{9s-2M_\pi^2-6M_\eta^2}{8F_\pi^2}J^r_{KK}(s)+\frac{16M_K^2-7M_\pi^2}{18F_\pi^2}J^r_{\eta\eta}(s)
\end{eqnarray}
for the hidden-strangeness form factors.

\subsection[Scalar Form Factors for the $I=\frac{1}{2}$ System]{Scalar Form Factors for the \boldmath{$I=\frac{1}{2}$} System} 

The scalar form factors for the $I=1/2$ meson--meson systems up to NLO in ChPT are given by
\begin{eqnarray}
\sqrt{\frac{2}{3}}F^{\bar{u}s}_{S,K\pi}(s)&=&1+\frac{5s-4M_\pi^2}{4(M_K^2-M_\pi^2)}\mu_\pi-\frac{s}{2(M_K^2-M_\pi^2)}\mu_K+\frac{8M_K^2+4M_\pi^2-9s}{12(M_K^2-M_\pi^2)}\mu_\eta\nonumber\\
&&-\frac{8(2M_K^2+M_\pi^2)}{F_\pi^2}\left(L_4^r-2L_6^r\right)+\frac{4(-2M_K^2-2M_\pi^2+s)}{F_\pi^2}L_5^r+\frac{16(M_K^2+M_\pi^2)}{F_\pi^2}L_8^r\nonumber\\
&&-\frac{3M_\eta^2(3M_K^2-3M_\pi^2+s)-9M_K^4+M_K^2(9M_\pi^2+2s)+s(7M_\pi^2-9s)}{72F_\pi^2s}\bar{J}_{K\eta}(s)\nonumber\\
&&-\frac{3M_K^4+M_K^2(2s-6M_\pi^2)+3M_\pi^4+2M_\pi^2s-5s^2}{8F_\pi^2s}\bar{J}_{\pi K}(s)~,\nonumber\\
-\sqrt{6}F^{\bar{u}s}_{S,K\eta}(s)&=&1-\frac{3(4M_\pi^2-3s)}{4(M_K^2-M_\pi^2)}\mu_\pi+\frac{16M_K^2-9s}{2(M_K^2-M_\pi^2)}\mu_K+\frac{-88M_K^2+28M_\pi^2+27s}{12(M_K^2-M_\pi^2)}\mu_\eta\nonumber\\
&&-\frac{8(2M_K^2+M_\pi^2)}{F_\pi^2}\left(L_4^r-2L_6^r\right)+\frac{4(-14M_K^2+2M_\pi^2+3s)}{3F_\pi^2}L_5^r+\frac{128(M_K^2-M_\pi^2)}{F_\pi^2}L_7^r\nonumber\\
&&-\frac{3M_\eta^2(3M_K^2-3M_\pi^2+s)-9M_K^4+M_K^2(9M_\pi^2+2s)+s(7M_\pi^2-9s)}{8F_\pi^2s}\bar{J}_{\pi K}(s)\nonumber\\
&&-\frac{9M_\eta^4-6M_\eta^2(3M_K^2+s)+9M_K^4-18M_K^2s+s(4M_\pi^2+9s)}{24F_\pi^2s}\bar{J}_{K\eta}(s)\nonumber\\
&&+\frac{16(5M_K^2-3M_\pi^2)}{F_\pi^2}L_8^r~.
\end{eqnarray}

\subsection[Scalar Form Factors for the $I=1$ System]{Scalar Form Factors for the \boldmath{$I=1$} System} 

The scalar form factors for the $I=1$ meson--meson systems up to NLO in ChPT are given by
\begin{eqnarray}
F^{\bar{u}d}_{S,K\bar{K}}(s)&=&1+\frac{2}{3}\mu_\eta-\frac{8(2M_K^2+M_\pi^2)}{F_\pi^2}\left(L_4^r-2L_6^r\right)+\frac{4(s-4M_K^2)}{F_\pi^2}L_5^r+\frac{32M_K^2}{F_\pi^2}L_8^r\nonumber\\
&&+\frac{s}{4F_\pi^2}J^r_{KK}(s)-\frac{8M_K^2+M_\pi^2+3M_\eta^2-9s}{18F_\pi^2}J^r_{\pi\eta}(s)~,\nonumber\\
-\sqrt{\frac{3}{2}}F^{\bar{u}d}_{S,\pi\eta}(s)&=&1-\mu_\pi+2\mu_K-\frac{1}{3}\mu_\eta-\frac{8(2M_K^2+M_\pi^2)}{F_\pi^2}\left(L_4^r-2L_6^r\right)-\frac{4(8M_K^2+4M_\pi^2-3s)}{3F_\pi^2}L_5^r\nonumber\\
&&-\frac{64(M_K^2-M_\pi^2)}{F_\pi^2}L_7^r+\frac{32M_\pi^2}{F_\pi^2}L_8^r+\frac{9s-8M_K^2-M_\pi^2-3M_\eta^2}{12F_\pi^2}J^r_{KK}(s)\nonumber\\
&&+\frac{M_\pi^2}{3F_\pi^2}J^r_{\pi\eta}(s)~,\nonumber\\
F^{\bar{u}d}_{S,\pi\pi}(s)&=&0 \quad (\text{by isospin and Bose symmetry}).
\end{eqnarray}

\subsection[Vector Form Factors for the $I=0$ System]{Vector Form Factors for the \boldmath{$I=0$} System}

As mentioned in Sect.~\ref{sec:vector},
the SU(3) singlet vector form factor of $\bar{u}\gamma^\mu u+\bar{d}\gamma^\mu d+\bar{s}\gamma^\mu s$ is zero up to $\mathcal{O}(p^4)$. Therefore, it is sufficient to present the form factors of $\bar{s}\gamma^\mu s$ only:
\begin{eqnarray}
\frac{F^{\bar{s}s}_{V+,K\bar{K}}(s)}{\sqrt{2}}&=&1-2\mu_K+\frac{s-6M_K^2}{96\pi^2F_\pi^2}+\frac{2s}{F_\pi^2}L_9^r+\frac{s-4M_K^2}{4F_\pi^2}J^r_{KK}(s)~,\nonumber\\
F^{\bar{s}s}_{V+,\pi\pi}(s)&=&0 \quad (\text{by isospin and Bose symmetry}),\nonumber\\
F^{\bar{s}s}_{V+,\eta\eta}(s)&=&0 \quad (\text{by Bose symmetry}).
\end{eqnarray}

\subsection[Vector Form Factors for the $I=\frac{1}{2}$ System]{Vector Form Factors for the \boldmath{$I=\frac{1}{2}$} System} 

The vector form factors for the $I=1/2$ meson--meson systems up to NLO in ChPT are given by
\begin{eqnarray}
-\sqrt{\frac{2}{3}}F^{\bar{u}s}_{V+,K\pi}(s)&=&1+\frac{s-4M_K^2}{4(M_K^2-M_\pi^2)}\mu_\pi+\frac{-8M_K^2+4M_\pi^2+s}{2(M_K^2-M_\pi^2)}\mu_K+\frac{3(4M_K^2-s)}{4(M_K^2-M_\pi^2)}\mu_\eta\nonumber\\
&&-\frac{5M_K^2+M_\pi^2-s}{96\pi^2F_\pi^2}+\frac{2s}{F_\pi^2}L_9^r+\frac{M_\eta^4-2M_\eta^2(M_K^2+s)+(M_K^2-s)^2}{8F_\pi^2s}\bar{J}_{K\eta}(s)\nonumber\\
&&+\frac{M_K^4-2M_K^2(M_\pi^2+s)+(M_\pi^2-s)^2}{8F_\pi^2s}\bar{J}_{\pi K}(s)~\nonumber\\
&=&-\sqrt{\frac{2}{3}}F^{\bar{u}s}_{V+,K\eta}(s)~.
\end{eqnarray}

\subsection[Vector Form Factors for the $I=1$ System]{Vector Form Factors for the \boldmath{$I=1$} System}

 The vector form factors for the $I=1$ meson--meson systems up to NLO in ChPT are given by 
 \begin{eqnarray}
 \frac{F^{\bar{u}d}_{V+,\pi\pi}(s)}{\sqrt{2}}&=&1-\frac{4}{3}\mu_\pi-\frac{2}{3}\mu_K+\frac{s-2M_K^2-4M_\pi^2}{96\pi^2F_\pi^2}+\frac{2s}{F_\pi^2}L_9^r+\frac{s-4M_\pi^2}{6F_\pi^2}J^r_{\pi\pi}(s)+\frac{s-4M_K^2}{12F_\pi^2}J^r_{KK}(s)\nonumber\\
 &=&F^{\bar{u}d}_{V+,K\bar{K}}(s)~,\nonumber\\
 F^{\bar{u}d}_{V+,\pi\eta}(s)&=&0 \quad (\text{by C parity}).
 \end{eqnarray}

\subsection[Tensor Form Factors for the $I=0$ System]{Tensor Form Factors for the \boldmath{$I=0$} System} 
 
 For simplicity of notation, we will define $\tilde{C}_i^r\equiv C_i^r/\Lambda_2$. The tensor form factors for the isoscalar systems up to NLO in ChPT are given by 
 \begin{eqnarray}
 F^n_{T,K\bar{K}}(s)&=&1-7\mu_\pi-\frac{1}{3}\mu_\eta+\frac{s-6M_K^2}{96\pi^2F_\pi^2}-\frac{8(M_K^2-M_\pi^2)}{F_\pi^2}L^r_5-4M_\pi^2\tilde{C}^r_{34}+2(M_\pi^2-2M_K^2)\tilde{C}^r_{35}\nonumber\\
 &&-2(2M_K^2+M_\pi^2)\tilde{C}^r_{36}+8(M_K^2-M_\pi^2)\tilde{C}^r_{37}-2s\tilde{C}^{r}_{88}+2(s-2M_K^2)\tilde{C}^r_{89}-4M_K^2\tilde{C}^r_{106}\nonumber\\
 &&+2(s-2M_K^2)\tilde{C}^r_{107}+\frac{s-4M_K^2}{4F_\pi^2}J^r_{KK}(s)~,\nonumber\\
 F^n_{T,\pi\pi}(s)&=&0 \quad (\text{by isospin and Bose symmetry}),\nonumber\\
 F^n_{T,\eta\eta}(s)&=&0 \quad (\text{by isospin and Bose symmetry}), \\
 -\frac{F^{\bar{s}s}_{T,K\bar{K}}(s)}{\sqrt{2}}&=&1-4\mu_\pi-2\mu_K-\frac{4}{3}\mu_\eta+\frac{s-6M_K^2}{96\pi^2F_\pi^2}-\frac{8(M_K^2-M_\pi^2)}{F_\pi^2}L^r_5+4(M_\pi^2-2M_K^2)\tilde{C}^r_{34}\nonumber\\
 &&-2M_\pi^2\tilde{C}^r_{35}-2(2M_K^2+M_\pi^2)\tilde{C}^r_{36}-4(M_K^2-M_\pi^2)\tilde{C}^r_{37}-2s\tilde{C}^{r}_{88}+2(s-2M_K^2)\tilde{C}^r_{89}\nonumber\\
 &&-4M_K^2\tilde{C}^r_{106}+2(s-2M_K^2)\tilde{C}^r_{107}+\frac{s-4M_K^2}{4F_\pi^2}J^r_{KK}(s)~,\nonumber\\
 F^{\bar{s}s}_{T,\pi\pi}(s)&=&0 \quad (\text{by isospin and Bose symmetry}),\nonumber\\
 F^{\bar{s}s}_{T,\eta\eta}(s)&=&0 \quad (\text{by Bose symmetry}).
 \end{eqnarray}

\subsection[Tensor Form Factors for the $I=\frac{1}{2}$ System]{Tensor Form Factors for the \boldmath{$I=\frac{1}{2}$} System} 

 The tensor form factors for the $I=1/2$ systems up to NLO in ChPT are given by 
 \begin{eqnarray}
 \sqrt{\frac{2}{3}}F^{\bar{u}s}_{T,K\pi}(s)&=&1+\frac{-20M_K^2+16M_\pi^2+s}{4(M_K^2-M_\pi^2)}\mu_\pi+\frac{-12M_K^2+8M_\pi^2+s}{2(M_K^2-M_\pi^2)}\mu_K+\frac{44M_K^2-8M_\pi^2-9s}{12(M_K^2-M_\pi^2)}\mu_\eta\nonumber\\
 &&-\frac{5M_K^2+M_\pi^2-s}{96\pi^2F_\pi^2}-\frac{4(M_K^2-M_\pi^2)}{F_\pi^2}L_5^r-4M_K^2\tilde{C}^r_{34}-2M_\pi^2\tilde{C}^r_{35}-2(2M_K^2+M_\pi^2)\tilde{C}^r_{36}\nonumber\\
 &&-2s\tilde{C}^r_{88}-2(M_K^2+M_\pi^2-s)\tilde{C}^r_{89}-2(M_K^2+M_\pi^2)\tilde{C}^r_{106}-2(M_K^2+M_\pi^2-s)\tilde{C}^r_{107}\nonumber\\
 &&+\frac{M_\eta^4-2M_\eta^2(M_K^2+s)+(M_K^2-s)^2}{8F_\pi^2s}\bar{J}_{K\eta}(s) \nonumber\\
 && +\frac{M_K^4-2M_K^2(M_\pi^2+s)+(M_\pi^2-s)^2}{8F_\pi^2s}\bar{J}_{\pi K}(s)~,\nonumber\\
 \sqrt{\frac{2}{3}}F^{\bar{u}s}_{T,K\eta}(s)&=&1+\frac{-20M_K^2+16M_\pi^2+s}{4(M_K^2-M_\pi^2)}\mu_\pi+\frac{-12M_K^2+8M_\pi^2+s}{2(M_K^2-M_\pi^2)}\mu_K+\frac{44M_K^2-8M_\pi^2-9s}{12(M_K^2-M_\pi^2)}\mu_\eta\nonumber\\
 &&-\frac{5M_K^2+M_\pi^2-s}{96\pi^2F_\pi^2}-\frac{28(M_K^2-M_\pi^2)}{3F_\pi^2}L_5^r-4M_K^2\tilde{C}^r_{34}+\frac{2(M_\pi^2-4M_K^2)}{3}\tilde{C}^r_{35}\nonumber\\
 &&-2(2M_K^2+M_\pi^2)\tilde{C}^r_{36}-2s\tilde{C}^r_{88}+\frac{2(-7M_K^2+M_\pi^2+3s)}{3}\tilde{C}^r_{89}+\frac{2(M_\pi^2-7M_K^2)}{3}\tilde{C}^r_{106}\nonumber\\
 &&+\frac{2(-7M_K^2+M_\pi^2+3s)}{3}\tilde{C}^r_{107}+\frac{M_\eta^4-2M_\eta^2(M_K^2+s)+(M_K^2-s)^2}{8F_\pi^2s}\bar{J}_{K\eta}(s)\nonumber\\
 &&+\frac{M_K^4-2M_K^2(M_\pi^2+s)+(M_\pi^2-s)^2}{8F_\pi^2s}\bar{J}_{\pi K}(s)~.
 \end{eqnarray}

\subsection[Tensor Form Factors for the $I=1$ System]{Tensor Form Factors for the \boldmath{$I=1$} System}
 
 The tensor form factors for the isovector systems up to NLO in ChPT are given by
 \begin{eqnarray}
 -\frac{F^{\bar{u}d}_{T,\pi\pi}(s)}{\sqrt{2}}&=&1-\frac{13}{3}\mu_\pi-\frac{8}{3}\mu_K-\frac{1}{3}\mu_\eta+\frac{-2M_K^2-4M_\pi^2+s}{96\pi^2F_\pi^2}-4M_\pi^2\tilde{C}^r_{34}-2M_\pi^2\tilde{C}^r_{35}\nonumber\\
 &&-2(2M_K^2+M_\pi^2)\tilde{C}^r_{36}-2s\tilde{C}^r_{88}+2(s-2M_\pi^2)\tilde{C}^r_{89}-4M_\pi^2\tilde{C}^r_{106}+2(s-2M_\pi^2)\tilde{C}^r_{107}\nonumber\\
 &&+\frac{s-4M_\pi^2}{6F_\pi^2}J^r_{\pi\pi}(s)+\frac{s-4M_K^2}{12F_\pi^2}J^r_{KK}(s)~,\nonumber\\
 -F^{\bar{u}d}_{T,K\bar{K}}(s)&=&1-\frac{13}{3}\mu_\pi-\frac{8}{3}\mu_K-\frac{1}{3}\mu_\eta+\frac{-2M_K^2-4M_\pi^2+s}{96\pi^2F_\pi^2}-\frac{8(M_K^2-M_\pi^2)}{F_\pi^2}L^r_5-4M_\pi^2\tilde{C}^r_{34}\nonumber\\
 &&+2(M_\pi^2-2M_K^2)\tilde{C}^r_{35}-2(2M_K^2+M_\pi^2)\tilde{C}^r_{36}-2s\tilde{C}^r_{88}+2(s-2M_K^2)\tilde{C}^r_{89}-4M_K^2\tilde{C}^r_{106}\nonumber\\
 &&+2(s-2M_K^2)\tilde{C}^r_{107}+\frac{s-4M_\pi^2}{6F_\pi^2}J^r_{\pi\pi}(s)+\frac{s-4M_K^2}{12F_\pi^2}J^r_{KK}(s)~,\nonumber\\
 F^{\bar{u}d}_{T,\pi\eta}(s)&=&0 \quad (\text{by C parity}).
 \end{eqnarray}
 
 \section{\boldmath{$B_s\to \pi^+\pi^-$} Form Factors}\label{sec:LCDAs}

Normalized by the scalar form factor, the $S$-wave $\pi^+\pi^-$  LCDAs   are defined as~\cite{Diehl:1998dk,Polyakov:1998ze,Kivel:1999sd,Diehl:2003ny,Meissner:2013hya}
\begin{eqnarray}
 \langle {(\pi^+\pi^-)_S}|\bar s (x)\gamma_\mu s(0)|0\rangle &=&  F_{S,\pi\pi}^{\bar s s}(m_{\pi\pi}^2)
 p_{{\pi\pi},\mu}
 \int_0^1 du\,e^{i up_{\pi\pi}\cdot x}\phi_{\pi\pi}(u) \,,
 \nonumber\\
 \langle {(\pi^+\pi^-)_S}|\bar s (x)  s(0)|0\rangle &=&F_{S,\pi\pi}^{\bar s s}(m_{\pi\pi}^2) B_0
 \int_0^1 du\,e^{iup_{\pi\pi}\cdot x}\phi_{\pi\pi}^s(u) \,,
 \nonumber \\
 \langle {(\pi^+\pi^-)_S}\bar s(x)\sigma_{\mu\nu}   s(0)|0\rangle &=&
 -F_{S,\pi\pi}^{\bar s s}(m_{\pi\pi}^2)   B_0\frac{1}{6}  (p_{{\pi\pi}\mu} x_\nu -p_{{\pi\pi}\nu} x_\mu) \int_0^1 du\,e^{i up_{\pi\pi}\cdot x} {\phi_{\pi\pi}^\sigma(u)} \,.
\end{eqnarray}
$\phi_{\pi\pi}$ and  $\phi_{\pi\pi}^s, \phi_{\pi\pi}^{\sigma}$ are twist-2 and twist-3 LCDAs, respectively. They are normalized as
\begin{eqnarray}
 \int_0^1 du \phi_{\pi\pi}^s(u)=\int_0^1
 du\phi_{\pi\pi}^\sigma(u)=1 \,.
\end{eqnarray}
According to the  conformal symmetry in QCD~\cite{Braun:2003rp},  the twist-3 LCDAs  have the asymptotic form~\cite{Diehl:1998dk,Polyakov:1998ze,Kivel:1999sd,Diehl:2003ny}:
\begin{eqnarray}
 \phi_{\pi\pi}^s(u)=1, \;\;\; \phi_{\pi\pi}^\sigma(u)=  6u(1-u) \,,
\end{eqnarray}
while the twist-2 LCDA can be expanded in terms of the Gegenbauer moments as
\begin{eqnarray}
 \phi_{\pi\pi}(u)=  6u(1-u) \sum_{n} a_n C_n^{3/2}(2u-1) \,.
\end{eqnarray}
Since the contributions from higher Gegenbauer moments are suppressed, it is enough to only consider the lowest moment $a_1$.
The first Gegenbauer moment for the $f_0(980)$ is~\cite{Cheng:2005nb}
\begin{eqnarray}
a_1=-1.35\, .  \label{eq:a_1_cheng}
\end{eqnarray}
The $B_s\to \pi^+\pi^-$ form factors  in terms of  the $\pi^+\pi^-$ LCDAs read as~\cite{Meissner:2013hya},
\begin{align}
  &{\cal F}^{B_s\to \pi\pi}_1(M_{\pi\pi}^2, q^2)\nonumber\\
  &=  N_F \bigg\{\int_{u_0}^1\frac{du}{u}{\rm exp}\left[-\frac{m_b^2+u\bar u M_{\pi\pi}^2-\bar uq^2}{uM^2}\right]  \bigg[-\frac{m_b}{B_0}\Phi_{\pi\pi}(u)+u \Phi_{\pi\pi}^s(u)+\frac{1}{3} \Phi_{\pi\pi}^\sigma(u) \nonumber\\
  &   +\frac{m_b^2+q^2-u^2M_{\pi\pi}^2}{uM^2}\frac{\Phi_{\pi\pi}^\sigma(u)}{6}\bigg]
 +\exp{\left[-\frac{s_0}{M^2}\right]}\frac{\Phi_{\pi\pi}^\sigma(u_0)}{6}\frac{m_b^2-u_0^2M_{\pi\pi}^2+q^2}
 {m_b^2+u_0^2M_{\pi\pi}^2-q^2}\bigg\},
 \label{eq:fplus}\\
  &{\cal F}^{B_s\to \pi\pi}_-(M_{\pi\pi}^2, q^2)\nonumber\\
  &=  N_F\left\{\int_{u_0}^1\frac{du}{u}{\rm
 exp}\left[-\frac{m_b^2+u\bar u M_{\pi\pi}^2-\bar uq^2}{uM^2}\right]
 \bigg[ \frac{m_b}{B_0}\Phi_{\pi\pi}(u)+(2-u)  \Phi_{\pi\pi}^s(u)\right.
 \nonumber\\
 &\;\;\;\left. +\frac{1-u}{3u}\Phi_{\pi\pi}^\sigma(u) -\frac{u({m_b^2+q^2-u^2M_{\pi\pi}^2})+2(
 m_b^2-q^2+u^2M_{\pi\pi}^2)}{u^2M^2}\frac{ \Phi_{\pi\pi}^\sigma(u)}{6}
 \bigg]\right.\nonumber\\
 &\;\;\;\left. -\frac{u_0({m_b^2+q^2-u_0^2M_{\pi\pi}^2})+2(
 m_b^2-q^2+u_0^2M_{\pi\pi}^2) }{u_0(m_b^2+u_0^2M_{\pi\pi}^2-q^2)}
 \exp{\left[-\frac{s_0}{M^2}\right]}\frac{ \Phi_{\pi\pi}^\sigma(u_0)}{6}\right\},
  \label{eq:fminus}\\
   &{\cal F}^{B_s\to \pi\pi}_0(M_{\pi\pi}^2, q^2)=  {\cal F}^{B_s\to \pi\pi}_1(M_{\pi\pi}^2, q^2) + \frac{q^2}{M_{B_s}^2 -M_{\pi\pi}^2}   {\cal F}^{B_s\to \pi\pi}_-(M_{\pi\pi}^2, q^2),\\
  &{\cal F}^{B_s\to \pi\pi}_T(M_{\pi\pi}^2, q^2)\nonumber\\
  &=2   N_F (M_{B_s}+M_{\pi\pi})
 \bigg\{\int_{u_0}^1\frac{du}{u} {\rm exp}\left[-\frac{(m_b^2-\bar uq^2+u\bar
 uM_{\pi\pi}^2)}{uM^2}\right]
 \left[-\frac{\Phi_{\pi\pi}(u)}{2B_0}+m_b\frac{ \Phi_{\pi\pi}^\sigma(u)}{6uM^2}\right] \nonumber\\
 & \;\;\; +m_b\frac{ \Phi_{\pi\pi}^\sigma(u_0)}{6}
   \frac{\exp[-s_0/M^2]}{m_b^2-q^2+u_0^2M_{\pi\pi}^2}\bigg\},
   \label{eq:ftensor}
\end{align}

\section{Vector Meson Dominance and Tensor Form Factors at the Vector Pole\label{sec:VMD}}

The antisymmetric tensor interaction is not a part of the elementary SM interaction structure, so it is quite difficult to compare theoretical results of tensor form factor calculations with data. Therefore, for an independent cross-check one should rely on future lattice QCD calculations. One could of course calculate the two-meson tensor form factors directly on the lattice, but since the results of different theory predictions of the tensor form factors differ mainly around the $\rho$-peak (see Fig.~\ref{fig:CompareTFFudpipi}), a more straightforward cross-check is to calculate the tensor charge of the $\rho$-meson. 
Combining its value and the vector meson dominance (VMD) picture, one is able to compare the two-meson tensor form factors with lattice predictions around the vector-meson mass pole. We outline the method below.

The tensor charge $f_\mathcal{V}^T$ of a vector meson $\mathcal{V}$ is defined through
\begin{equation}
\left\langle\mathcal{V}(p)|\bar{q}'\sigma^{\mu\nu}q|0\right\rangle=-if^T_{\mathcal{V},\bar{q}'q}(\varepsilon^{\mu*}p^\nu-\varepsilon^{\nu*}p^\mu) \,,
\end{equation} 
where $\varepsilon^\mu$ is the polarization vector of $\mathcal{V}$. Suppose now we are interested in the tensor form factor $F_{T,\phi\phi'}^{\bar{q}'q}(s)$ defined in Eq.~\eqref{eq:tensorFFdef} at $s\approx M_\mathcal{V}^2$. If one is able to parameterize the $\mathcal{V}\rightarrow\phi\phi'$ amplitude as
\begin{equation}
iM_{\mathcal{V}\rightarrow\phi\phi'}=if_{\mathcal{V}\phi\phi'}(p_\phi-p_{\phi'})_\mu\varepsilon^\mu \,,
\end{equation}
then the VMD picture provides an approximate expression of $F_{T,\phi\phi'}^{\bar{q}'q}(s)$ at $s\approx M_\mathcal{V}^2$,
\begin{equation}
\frac{\Lambda_2}{F_\pi^2}F_{T,\phi\phi'}^{\bar{q}'q}(s)\approx-\frac{2f_{\mathcal{V}\phi\phi'}f_{\mathcal{V},\bar{q}'q}^T}{s-M_\mathcal{V}^2+i\Theta(s-4M_\pi^2)M_\mathcal{V}\Gamma_\mathcal{V}}~,
\end{equation}
where $\Gamma_\mathcal{V}$ is the total decay width of $\mathcal{V}$. An interesting consequence of this formula is that one expects $\mathrm{Re}F_{T,\phi\phi'}^{\bar{q}'q}(s)$ to vanish and $\mathrm{Im}F_{T,\phi\phi'}^{\bar{q}'q}(s)$ to peak at $s=M_\mathcal{V}^2$. Furthermore, with future lattice inputs of $f_{\mathcal{V},\bar{q}'q}^T$, the equation above serves as a consistency check of the theoretical result for two-meson tensor form factors at the vector-meson pole.

\end{appendix}

\providecommand{\href}[2]{#2}\begingroup\raggedright

\endgroup

\end{document}